\documentstyle[12pt]{article}

\renewcommand{\theequation}{\thesection.\arabic{equation}}

\newcommand \beq{\begin{eqnarray}}
\newcommand \eeq{\end{eqnarray}}
\begin{document}
\begin{titlepage}
\begin{center}
{\Large SOFT COLLECTIVE EXCITATIONS IN HOT\\
GAUGE THEORIES}
\vskip 1cm
Jean-Paul BLAIZOT\footnote{CNRS}  and
Edmond IANCU\\
Service de Physique Th\'eorique\footnote{Laboratoire de la Direction des
Sciences de la Mati\`ere du Commissariat \`a l'Energie
Atomique}, CE-Saclay \\ 91191 Gif-sur-Yvette, France\\
\vskip0.5cm
18 June 1993
\end{center}

\vskip 2cm \begin{abstract}
The longwavelength excitations of a quark-gluon plasma at high temperature
can be described as collective oscillations of gauge and fermionic average
fields. We show that, at leading order in the coupling strength, the
Dyson-Schwinger equations for the $N$-point functions reduce to a set of
coupled equations for these average fields and their induced sources which
involve only 2-point functions. The equations for the 2-point functions
describe the dynamics of the hard plasma particles in the presence of soft
background fields. They may be given the form of simple kinetic equations.
Both the wavelength and the amplitude of the collective modes are controlled
by the coupling strength, and we show that it is important to take
this properly into account in order to obtain consistent equations of motion
which are
covariant under gauge transformations. By solving the kinetic equations
for the 2-point functions with well defined (i.e. retarded or advanced)
boundary
conditions, we obtain the induced currents in closed forms. These act as
generating
functionals for the one-particle irreducible amplitudes with soft external
lines,
  and yield in particular all the so called
``hard thermal loops'' identified in  the diagrammatic analysis.
\end{abstract}
\vskip 1cm
\begin{flushright}
SPhT/93-064
\end{flushright}
\begin{flushleft}
Submitted to Nuclear Physics B\\
PACS No: 12.38.Mh, 12.38.Bx, 52.25.Dg, 11.15.Kc
\end{flushleft}
\end{titlepage}


\def\square{\hbox{{$\sqcup$}\llap{$\sqcap$}}}   
\def\grad{\nabla}                               
\def\del{\partial}                              

\def\frac#1#2{{#1 \over #2}}
\def\smallfrac#1#2{{\scriptstyle {#1 \over #2}}}
\def\half{\ifinner {\scriptstyle {1 \over 2}}
   \else {1 \over 2} \fi}

\def\bra#1{\langle#1\vert}              
\def\ket#1{\vert#1\rangle}              

\def\simge{\mathrel{%
   \rlap{\raise 0.511ex \hbox{$>$}}{\lower 0.511ex \hbox{$\sim$}}}}
\def\simle{\mathrel{
   \rlap{\raise 0.511ex \hbox{$<$}}{\lower 0.511ex \hbox{$\sim$}}}}


\def\parenbar#1{{\null\!                        
   \mathop#1\limits^{\hbox{\fiverm (--)}}       
   \!\null}}                                    
\def\nunubar{\parenbar{\nu}}
\def\ppbar{\parenbar{p}}


\def\buildchar#1#2#3{{\null\!                   
   \mathop#1\limits^{#2}_{#3}                   
   \!\null}}                                    
\def\overcirc#1{\buildchar{#1}{\circ}{}}


\def\slashchar#1{\setbox0=\hbox{$#1$}           
   \dimen0=\wd0                                 
   \setbox1=\hbox{/} \dimen1=\wd1               
   \ifdim\dimen0>\dimen1                        
      \rlap{\hbox to \dimen0{\hfil/\hfil}}      
      #1                                        
   \else                                        
      \rlap{\hbox to \dimen1{\hfil$#1$\hfil}}   
      /                                         
   \fi}                                         %


\def\subrightarrow#1{
  \setbox0=\hbox{
    $\displaystyle\mathop{}
    \limits_{#1}$}
  \dimen0=\wd0
  \advance \dimen0 by .5em
  \mathrel{
    \mathop{\hbox to \dimen0{\rightarrowfill}}
       \limits_{#1}}}                           

\def\real{\mathop{\rm Re}\nolimits}     
\def\imag{\mathop{\rm Im}\nolimits}     

\def\tr{\mathop{\rm tr}\nolimits}       
\def\Tr{\mathop{\rm Tr}\nolimits}       
\def\Det{\mathop{\rm Det}\nolimits}     

\def\mod{\mathop{\rm mod}\nolimits}     
\def\wrt{\mathop{\rm wrt}\nolimits}     


\def\TeV{{\rm TeV}}                     
\def\GeV{{\rm GeV}}                     
\def\MeV{{\rm MeV}}                     
\def\KeV{{\rm KeV}}                     
\def\eV{{\rm eV}}                       

\def\mb{{\rm mb}}                       
\def\mub{\hbox{$\mu$b}}                 
\def\nb{{\rm nb}}                       
\def\pb{{\rm pb}}                       

%
\def\journal#1#2#3#4{\ {#1}{\bf #2} ({#3})\  {#4}}

\def\AdvPhys{\journal{Adv.\ Phys.}}
\def\AnnPhys{\journal{Ann.\ Phys.}}
\def\EurophysLett{\journal{Europhys.\ Lett.}}
\def\JApplPhys{\journal{J.\ Appl.\ Phys.}}
\def\JMathPhys{\journal{J.\ Math.\ Phys.}}
\def\LettNuovoCimento{\journal{Lett.\ Nuovo Cimento}}
\def\Nature{\journal{Nature}}
\def\NPA{\journal{Nucl.\ Phys.\ {\bf A}}}
\def\NPB{\journal{Nucl.\ Phys.\ {\bf B}}}
\def\NuovoCimento{\journal{Nuovo Cimento}}
\def\Physica{\journal{Physica}}
\def\PLA{\journal{Phys.\ Lett.\ {\bf A}}}
\def\PLB{\journal{Phys.\ Lett.\ {\bf B}}}
\def\PhysRev{\journal{Phys.\ Rev.}}
\def\PRC{\journal{Phys.\ Rev.\ {\bf C}}}
\def\PRD{\journal{Phys.\ Rev.\ {\bf D}}}
\def\PRL{\journal{Phys.\ Rev.\ Lett.}}
\def\PhysRept{\journal{Phys.\ Repts.}}
\def\ProcNatlAcadSci{\journal{Proc.\ Natl.\ Acad.\ Sci.}}
\def\ProcRoySoc{\journal{Proc.\ Roy.\ Soc.\ London Ser.\ A}}
\def\RevModPhys{\journal{Rev.\ Mod.\ Phys.}}
\def\Science{\journal{Science}}
\def\SovPhysJETP{\journal{Sov.\ Phys.\ JETP}}
\def\SovPhysJETPLett{\journal{Sov.\ Phys.\ JETP Lett.}}
\def\SovJNuclPhys{\journal{Sov.\ J.\ Nucl.\ Phys.}}
\def\SovPhysDoklady{\journal{Sov.\ Phys.\ Doklady}}
\def\ZPhys{\journal{Z.\ Phys.}}
\def\ZPhysA{\journal{Z.\ Phys.\ A}}
\def\ZPhysB{\journal{Z.\ Phys.\ B}}
\def\ZPhysC{\journal{Z.\ Phys.\ C}}


\setcounter{equation}{0}


\section{Introduction}

The study of the long wavelength excitations of the quark-gluon plasma is
interesting in several respects. In particular, it  provides clues to basic
features of the plasma off-equilibrium dynamics, and it is  a first step
towards a full
description of its transport properties. Interesting
progress in this area has been made recently, and consistent schemes to perform
calculations of amplitudes in hot gauge theories have begun to emerge [1-6],
giving rise to various applications [7-11].

We consider in this paper  a quark-gluon plasma in thermal equilibrum at high
temperature
$T$ well beyond the critical temperature for the deconfining phase
transition, and neglect quark masses. Then, the only energy scale in
 the problem is $T$.
We also assume a regime where the effective coupling constant $g$ is
small.  In the plasma, quarks and gluons have typical energies and momenta of
order $T$, and will be referred to as ``hard'' particles; their number
density is proportional to $T^3$. Thus, the typical interparticle distance is
of the order $1/T$, that is, of the order of the thermal wavelength
$\lambda_T$. This means, for instance, that the Pauli principle cannot be
ignored for the hard particles. However,
when coupled to weak and slowly varying disturbances, the plasma may develop
a collective behaviour on a typical length/time scale of order $1/gT$. A
familiar example of such a collective behaviour is Debye's screening.
Collectivity arises because any motion taking place over a distance scale
$1/gT$ may involve coherently a large number of hard particles. We shall
see later that such collective motions can be treated as classical degrees of
freedom.

The existence of such possible excitations manifests itself in specific
difficulties encountered in perturbative calculations. In particular,
diagrammatic analyses have revealed that amplitudes with ``soft'' external
lines, i.e. with all external momenta of order $gT$, may receive contributions
from  an
infinite number of loop diagrams which are as large
 as the tree amplitudes
 \cite{Shuryak78,Kalashnikov80,Weldon82a,Frenkel90,Braaten90b}.
These contributions  arise from one-loop integrals in which the loop momentum
is hard, and
for this reason they have been named ``hard thermal loops'' (HTL).
As they are of the same order of magnitude
as the tree amplitudes, they have to be resummed
in consistent higher order calculations \cite{Pisarski89,Braaten90b}.

The need for resummation in a perturbative expansion usually points to simple
underlying physics, as shown, for instance, by the classic example of the
 electron gas. There, the resummation of ring diagrams
performed by Brueckner and Gell-Mann \cite{GMB} reflects the simple physical
phenomenon of Debye's screening. In order to illuminate the simple physics
underlying the resummation needed in QCD, we have initiated \cite{QED,PRL} a
new approach to the problem, by working directly at the level of the
equation of motion rather than on the Feynman diagrams. The crucial remark
is that the collective phenomena that
we are trying to describe involve {\it mean fields} which represent the average
motion of hard particles over distances large compared to their mean
separation. Such mean fields may be induced by external disturbances, but they
are also the natural degrees of freedom describing
the soft normal modes of the quark-gluon plasma. Our main objective
will be therefore to obtain consistent equations of motions for these mean
fields. In order
to allow for collective excitations with arbitrary quantum numbers,
we shall consider both fermionic and bosonic average fields. The physical
picture which will emerge from our analysis is that the dominant interactions
in
the plasma are those between the mean fields and the hard particles. Direct
interactions among hard particles can be neglected, but since hard particles
can
contribute coherently, they strongly affect the dynamics of the mean fields.
Our final equations will take the form of a coupled set of mean field
equations and kinetic equations describing the motion of hard particles in
the presence of the mean fields. These equations generalize the familiar set
of Vlasov and field equations of ordinary electromagnetic plasmas
\cite{LLPhysKin,Silin60}.

When dealing with this type of problems one is usually facing three kinds
of approximations, involving, respectively, the strength of the coupling, the
amplitude of the oscillation, and the wavelength or period of the modes that
one is studying. In the present case, there is no other scale than the
temperature, and it turns out that all three
approximations are controlled by the single small parameter, $g$. They should
be viewed,
then, as three facets of the same approximation, namely an expansion in powers
of $g$.
Thus, the   collective modes with  wavelength $\sim 1/gT$, have, as we shall
see,
amplitudes limited by
 $\bar\psi\psi\simle gT^3$ and $F_{\mu\nu}\simle gT^2$,
respectively. In order to preserve  gauge
covariance, it is essential that these three features of the
approximation scheme be implemented together consistently.
The naive loop expansion fails precisely because it takes care only of the
first
facet, namely the expansion in powers of the interaction vertices.

As alluded to before, one  outcome of the present analysis is a set of
gauge covariant  kinetic equations for hard quarks and gluons
 moving in soft gauge and fermionic fields. We thus make contact with
previous works on the kinetic theory of non abelian plasmas [21-25]. However,
the special
role of the scale $gT$, as a unifying parameter controlling the various
approximations being done in deriving kinetic equations, has never been
properly recognized.  Because they include terms which are found to be of
different
orders in our expansion scheme, most kinetic equations proposed so far are not
only more complicated, they are also inconsistent. Besides,
although quite some effort was spent in using  gauge invariant Wigner
functions, specific
issues related to gauge invariance in a quantum theory, such as gauge fixing
independence,
or the elimination of ghost particles, have never been properly addressed. The
resulting
inconsistencies  can be seen in calculating one-particle irreducible
amplitudes which, in leading order,  can be deduced from the kinetic
equations. As well known, these amplitudes  satisfy Ward  identities imposed by
 gauge
symmetry. This  puts  constraints on the  structure of the non-linear terms
which enter
the  kinetic equations, and such  constraints do not appear to be satisfied by
the
transport equations proposed so far.

Our strategy for deriving the relevant equations of motion
is similar to that already used in
connection with the abelian case \cite {QED}.
However, the structure of the non abelian theory is much richer, and the
present work
contains many new technical developments. It is organized
as follows. In Sec.2, after summarizing the conventions and the
notations used in this paper, we write the equations of motion
relating  the average fields to the off-equilibrium 2-point functions. The
connected
parts of these 2-point functions play the role of induced sources for
the average fields. From them, one-particle
irreducible amplitudes for the soft fields can be calculated.

The equations
for the average fields constitute the first level in the hierarchy of
Dyson-Schwinger  equations for the $N$-point functions. In Sec.3, we describe
all the approximations which allow us to calculate the 2-point functions. In
leading order, these functions concern only hard particles, for
which radiative corrections can be neglected. This implies that the
hierarchy truncates at the level of 2-point
functions. We analytically continue the 2-point functions to real time, and we
implement the condition that the average fields are slowly varying with the
help of Wigner functions and gradient expansions. We also take into account
the fact that the strength of the mean fields have to be constrained so as to
obtain consistent equations. We emphasize that this does not result in a
trivial
linearization of the equations since, for example, gauge fields involved in
covariant
derivatives are to be kept to all orders.
At the end of Sec.3, we have a set of gauge
covariant equations for the induced pieces of the 2-point functions. These
equations
incorporate consistently all leading order effects in $g$.

 In Sec.4, the equations for the 2-point functions  are
transformed into equations for the induced sources. These
describe the polarizability of the plasma at leading order in $g$. They
involve only the physical hard particles from the thermal bath, i.e. the
quarks, the
antiquarks and  the transverse gluons. The equations are
covariant under gauge transformations of
 the mean fields. They reduce to simple kinetic equations for density matrices
which, in general, are non diagonal in color space, or which can mix boson and
fermion
degrees of freedom. These  density matrices  play the role of generalized
classical
distribution functions for the hard particles. The kinetic  equations
are solved for retarded boundary conditions. We get then explicit expressions
for the induced sources in terms of the average fields.

The knowledge of these induced sources entirely determines the dynamics of
the average fields, which is discussed in Sec.5. There we obtain all
 one-particle irreducible amplitudes for soft fields by
functional differentiation of the induced sources. We  recover in this way all
the ``hard thermal loops'', with the boundary conditions
appropriate to the problem at
hand, i.e. the causal propagation of the mean fields. The
 Ward identities satisfied by the HTL  are seen to follow simply from the
conservation laws obeyed by the induced currents. In
cases where Landau damping is inoperative, a generating
functional for the induced currents is obtained.
It may be regarded  as the effective action
which describes
 the propagation of long wavelength
fields in a polarizable non abelian plasma, and coincides with the
generating functional for hard thermal loops previously derived in
\cite{Taylor90,Braaten91,Frenkel91}. Properties of this generating functional
are also
discussed in \cite{Nair92,Weldon92}.

The
main conclusions are  summarized in Sec.6.
Technical details related to the use of a general covariant gauge in our
main discussion are gathered in Appendices A and B.  In App.A we show that,
in leading order, ghost particles do not undergo collective motion. This  is in
line with our conclusions concerning the semiclassical character of the leading
order polarization phenomena. This same conclusion is supported by the
independence of our results with respect to the choice of the gauge in the
original quantum theory. This is proved in App.B.
 In App.C, we clarify the conditions  under which the effective
action proposed in Sec.5 generates indeed the renormalized mean fields
equations.

\vskip 2cm
\setcounter{equation}{0}


\section{Mean fields and induced sources}

In this section, after specifying some of the conventions and the notations
 used in this work (in subsection 2.1), we derive the first set of equations
in the hierarchy of Dyson-Schwinger equations for $N$-point functions, $i.e.$,
 the equations of motion for the average gauge and fermionic fields (in
subsection 2.2). These equations relate the mean fields to 2 and 3-point
functions which are interpreted here as induced sources for the mean
fields. Functional relations between these induced sources and the one-particle
 irreducible amplitudes are recalled in subsection 2.3. The relation between
induced sources and 2-point functions is further analysed in the last
subsection. Most of the material presented in this section is standard,
and the equations written are exact; the  approximations to be  made will be
presented in the next section.

\subsection{Generalities}

  We consider an ultrarelativistic plasma in, or close to, thermal equilibrum,
at a temperature $T=1/\beta$. The equations of motion which describe its
long wavelength excitations will be
derived using the imaginary time formalism, and their real time versions
will be obtained through an analytic continuation.
Since the only quantities  to be continued in the present work
are at most 2-point functions, the procedure is non ambiguous. It allows us
to avoid the intricacies of real time formalisms and to preserve, at least
in the first stage of our approximation scheme, the familiar structure
of perturbation theory for systems in equilibrium. In doing the analytic
 continuation, the relevant 2-point functions  may be chosen to
obey retarded boundary conditions,
appropriate to the calculation of the causal response functions that one
is eventually interested in.
We use Minkovski metric and we note  $x=(x_0,\vec x)=(-i\tau,\vec x)$,
 $\del_0=i\del_\tau$ and $\del^2=-\del_\tau^2-\vec{\grad }^2$.
In the analytic continuation, $x_0$ becomes a complex variable ($\tau$
acquires an imaginary part) and the metric, the gamma matrices, etc...,
are untouched.
 Our units are such that $\hbar=c=1$. We consider a
SU($N$) gauge theory with $N_f$ flavors of quarks .
 The
color indices for the adjoint representation, $a$,$b$,... run from 1 to
$N^2$-1,
while those for the fundamental representation, $i$,$j$,... run from 1 to $N$.
We
work in four space-time dimensions, $d=4$, but on occasion we shall
 keep $d$ as an explicit parameter in our equations in order to
exhibit the contribution of the various gauge degrees of freedom.
Minkovski indices $\mu,\,\nu,...$ run from 0 to $d-1$.
The generators of the gauge group in different representations are taken to
 be Hermitian and traceless. They are denoted by
$t^a$ and $T^a$, respectively, for the fundamental and the adjoint
representations,
and are normalized such that
\beq
\label{gen}
tr(t^at^b)=\frac {1}{2}\delta^{ab},\qquad\qquad
tr(T^aT^b)=N\delta^{ab}.
\eeq
Furthermore, ($T^a)_{bc}
=-if^{abc}$, and $t^at^a=C_f$, where $C_f=(N^2-1)/(2N$) is the Casimir
of the fundamental representation and $f^{abc}$ are the structure
constants of the group:
\beq
\label{lie}
[t^a,t^b]=if^{abc}t^c.
\eeq
We use, without distinction, upper and lower positions for the color
indices.

The
thermal averages of fields operators in the
presence of external sources can be obtained from the generating functional
\beq
\label{Z}
Z[j_\mu^a,\eta,\bar\eta,C^a,\bar C^a]&=&\int[ d\bar\psi d\psi dA_\mu^a
 d\bar\zeta^b d\zeta^c]\qquad\nonumber\\
&\times&\exp-\left\{
S+(j,A)
+(\bar\eta,\psi)+(\bar\psi,\eta)+(\bar C,\zeta)+(\bar\zeta,C) \right\},
\eeq
where $S$ is the usual QCD action in a covariant gauge
\beq
\label{QCD}
S=\int_0^\beta d\tau d^3x\left\{
\frac{1}{4}F_{\mu\nu}^aF^{\mu\nu\,a}+\frac{\lambda}{2}
(\del\cdot A^a)^2+\bar\psi(-i\slashchar{D})\psi+\bar\zeta^a\del_\mu(\tilde
D^\mu\zeta)^a
 \right\},
\eeq
and
 $\psi$, $\bar\psi$, $A$, $\zeta$ and $\bar\zeta$ denote, respectively, the
quark fields, the gauge fields and the Fadeev-Popov ghosts. They obey
antiperiodic  (for $\psi$, $\bar\psi$) and
periodic (for $A$, $\zeta$ and $\bar\zeta$) boundary conditions,
with period $\beta$.
  In the formula above, $\lambda$  is the gauge fixing parameter, and
$\slashchar{D}\equiv \gamma^\mu D_\mu$, with the usual
 notations for the covariant derivatives in the fundamental ($D_\mu$)
 and in the adjoint representations ($\tilde D_\mu$):
\beq
\label{cder}
D_\mu\equiv \del_\mu+igA_\mu(x),\qquad\tilde D_\mu\equiv \del_\mu+ig\tilde
A_\mu(x). \eeq
Here $A_\mu\equiv A_\mu^at^a$ and $\tilde A_\mu\equiv A_\mu^aT^a$ are the
gauge field matrices in the two representations.
The source terms in eq.~(\ref{Z}) are
\beq
(j,A)\equiv \int_0^\beta d\tau d^3x j_\mu^a(x)A_a^\mu(x),
\eeq
and similarly for $(\bar\eta,\psi)$, $(\bar\psi,\eta)$, $(\bar C,\zeta)$ and
$(\bar\zeta,C)$.

The gauge field strength tensor is $F_{\mu\nu}\equiv [D_\mu,
D_\nu]/(ig) \equiv F_{\mu\nu}^at^a$.
For any operator of the form $O(x)=O^a(x)t^a$, we have
\beq \label{adder} [D_\mu,
O(x)]\equiv \del_\mu O(x)+ig[A_\mu(x),O(x)], \eeq
and similarly in the adjoint representation.

\subsection{Equations of motion for the mean fields}

In the absence of the external sources, the plasma is in
thermal equilibrium and the expectation values of the fields vanish.
For nonvanishing sources, the fields acquire nontrivial
 expectation values which obey  equations of motion that
are readily derived from eq.~(\ref{Z}):
\beq
\label{Acl}
\langle \left [D^\nu, F_{\nu\mu}(x) \right ]
 \rangle +\lambda\del_\mu\del_\nu\langle A^\nu(x) \rangle
-gt^a \langle \bar\psi(x) \gamma_\mu t^a\psi(x)\rangle
+gt^af_{abc}\langle (\del_\mu\bar\zeta ^b(x))\zeta^c(x)\rangle = j_\mu(x),
\nonumber\\\eeq
\beq
\label{psicl}
i\slashchar{\del}\langle{\psi(x)}\rangle
-g\langle\slashchar{A}(x)\psi(x)\rangle=\eta(x),
\eeq
\beq
\label{zcl}
-\del^2 \left\langle\zeta^a(x)\right\rangle+ gf^{abc}\left\langle\del^\mu\left
(A_\mu^b(x)\zeta^c(x)\right )\right\rangle = C^a(x), \eeq
\beq
\label{bzcl}
-\del^2 \left\langle\bar\zeta^a(x)\right\rangle+ gf^{abc}\langle A_\mu^b(x)
\del^\mu\bar\zeta^c(x)\rangle =\bar C^a(x), \eeq
together with the Hermitian conjugate equation for $\langle\bar\psi\rangle$.
The explicit
form of the commutator term in eq.~(\ref{Acl}) is
 \beq\label{ave}
\langle\left [D^\nu, F_{\nu\mu}\right ]^a\rangle=\left
(g_{\mu\nu}\del^2-\del_\mu\del_\nu\right )\langle A_a^\nu\rangle
-gf^{abc}\Gamma_{\mu\rho\lambda\nu}\left\langle A_b^\nu\left
 (\del^\lambda A_c^\rho\right
)\right\rangle -g^2f^{abc}f^{cde}\left\langle A_\nu^bA_\mu^dA^{\nu}_e
\,\right\rangle,\nonumber\\&&
\eeq
where
\beq
\label{Gamma}
\Gamma_{\mu\nu\rho\lambda}\equiv\,2g_{\mu\nu}g_{\rho\lambda}-g_{\mu\rho}g_{\nu\lambda}-g_{\mu\lambda}g_{\nu\rho}.
\eeq
Note the following identities which will be used later:
\beq
\label{kGk}
k^\mu\Gamma_{\mu\nu\rho\lambda}k^\lambda=k_\nu k_\rho -g_{\nu\rho}k^2,
\qquad\qquad\qquad\qquad\nonumber\\
 \Gamma_{\mu\nu\rho\lambda}=\Gamma_{\nu\mu\rho\lambda}
=\Gamma_{\mu\nu\lambda\rho}=\Gamma_{\rho\lambda\mu\nu},\qquad
\Gamma_{\mu\nu\rho\lambda}+\Gamma_{\mu\rho\lambda\nu}
+\Gamma_{\mu\lambda\nu\rho}=0.\eeq
We shall also use the following shorthand notation for the four gluon vertex:
\beq
\label{4g}
\hat \Gamma_{\mu\nu\lambda\rho}^{abcd}\equiv\,\frac {1}{2}
\biggl\{\,f^{eab}f^{ecd}(g_{\mu\lambda}g_{\nu\rho}-g_{\mu\rho}g_{\nu\lambda})
&+&f^{eac}f^{edb}(g_{\mu\rho}g_{\lambda\nu}-g_{\mu\nu}g_{\lambda\rho})\nonumber\\
&+&f^{ead}f^{ebc}(g_{\mu\nu}g_{\rho\lambda}-g_{\mu\lambda}
g_{\rho\nu})\,\biggr\}.\qquad\qquad\qquad
\eeq
Note that
$\hat \Gamma_{\mu\nu\rho\lambda}^{abcd}$ and $\Gamma_{\mu\nu\rho\lambda}$
are related by
\beq\label{tGG}
\hat\Gamma_{\mu\nu\rho\lambda}^{abcd}A_c^\rho A_d^\lambda\equiv
-f^{ade}f^{bce}
\Gamma_{\mu\rho\nu\lambda}A_c^\rho A_d^\lambda.
\eeq

{\parindent 30pt   We separate the 2-point functions into connected parts,
denoted by the
symbol $\langle\cdots\rangle_c$, and disconnected parts. }
For example,
$\langle\slashchar{A}(x)\psi(x)\rangle=\langle\slashchar{A}(x)\psi(x)\rangle_c+
\langle\slashchar{A}(x)
\rangle\langle\psi(x)\rangle$, and $\langle\bar\psi\gamma_\mu t^a\psi\rangle=
\langle\bar\psi\gamma_\mu t^a\psi\rangle_c+
\langle\bar\psi\rangle\gamma_\mu t^a\langle\psi\rangle$.
Then eq.~(\ref{ave}) can be written compactly as follows:
\beq
\label{Fconn}
\langle\, \left [\,D^\nu, F_{\nu\mu}\, \right ]^a\, \rangle&=&\left [\,{\cal
D}^\nu,\,{\cal F}_{\nu\mu}\,\right
]^a-\,gf^{abc}\Gamma_{\mu\rho\lambda\nu}\left\langle\, A_b^\nu\left
 (\del^\lambda A_c^\rho\right
)\,\right\rangle_c\nonumber\\
&-&g^2f^{abc}f^{cde}\left\langle\,A_\nu^bA_\mu^dA^{\nu}_e\,
\right\rangle_c-g^2\hat
\Gamma_{\mu\nu\rho\lambda}^{abcd}\left\langle\,A_b^\nu\right\rangle\,
\left\langle
\,A_c^\rho A_d^\lambda\,\right\rangle_c, \eeq
where ${\cal D}_\mu\equiv \del_\mu +ig\langle A_\mu(x)\rangle$ and
${\cal F}_{\mu\nu}\equiv [{\cal D}_\mu, {\cal
D}_\nu]/(ig)$.

After the separation of the connected parts in the 2 and 3-point functions,
the mean field equations (\ref{Acl})--(\ref{bzcl}) can be  written as:
\beq
\label{avA}
\left [{\cal D}^\nu,{\cal F}_{\nu\mu}(x)\right
]^a+\lambda\del_\mu\del_\nu\langle A_a^\nu(x) \rangle -g \langle \bar\psi(x)
\rangle \gamma_\mu t^a\langle \psi(x)\rangle
+gf^{abc}\left (\del_\mu\langle\bar\zeta ^b(x)\rangle
\right)\langle\zeta^c(x)\rangle\nonumber\\
=\,j_\mu^a(x)+j_\mu^{ind\,a}(x),\qquad\qquad\eeq
\beq
\label{avpsi}
\left (i\slashchar{\del}
-g\langle\slashchar{A}(x)\rangle\right )\langle\psi(x)\rangle=\eta(x)+
\eta^{ind}(x),
\eeq
\beq
\label{avz}
-\del^2 \left\langle\zeta^a(x)\right\rangle+ gf^{abc}\del^\mu\left (\langle
A_\mu^b(x)\rangle\,\langle\zeta^c(x)\rangle \right )= C^a(x)+C^{ind\,a}(x),
 \eeq
\beq
\label{avbz}
-\del^2 \left\langle\bar\zeta^a(x)\right\rangle+ gf^{abc}\langle A_\mu^b(x)
\rangle\del^\mu\langle\bar\zeta^c(x)\rangle =\bar C^a(x) + \bar C^{ind\,a}(x).
 \eeq
In the left hand sides of these equations
 we have isolated the terms corresponding to the tree level equations of motion
 generated by the QCD action
 ~(\ref{QCD}). In their right hand sides, we have,
 next to the external sources, connected 2 and 3-point Green's functions which
we refer as ``induced sources'', because they play a role similar
to that of $\eta$, $j_\mu$, or $C$, that is, they act as extra sources
for the mean fields. Explicitly, they are
\beq\label{1jind}
j_\mu^{ind\,a}(x)\equiv g \langle \bar\psi
\gamma_\mu t^a \psi\rangle_c -gf^{abc}\left\langle (\del_\mu\bar\zeta
^b)\zeta^c\right\rangle_c+gf^{abc}\Gamma_{\mu\rho\lambda\nu}\left
\langle\,A_b^\nu\left (\del^\lambda A_c^\rho\right
)\,\right\rangle_c\nonumber\\\qquad\qquad
+g^2\hat
\Gamma_{\mu\nu\rho\lambda}^{abcd}\left\langle\,A_b^\nu\right\rangle\,
\left\langle \,A_c^\rho A_d^\lambda\,\right\rangle_c
+g^2f^{abc}f^{cde}\left\langle\,A_\nu^bA_\mu^d
A^{\nu}_e\,\right\rangle_c, \eeq
\beq
\label{eta1}
\eta^{ind}(x)\equiv\,g\gamma^\nu t^a\,\left\langle A_\nu^a(x)\psi (x)
\right\rangle_c,
\eeq
\beq
\label{C1}
C^{a\,ind}(x)\equiv\,-
gf^{abc}\left\langle\del^\mu\left (A_\mu^b(x)\zeta^c(x)\right )\right\rangle_c,
\eeq\beq\label{bC1}
\bar C^{a\,ind}(x)\equiv\,-
gf^{abc}\left\langle A_\mu^b(x)\del^\mu\bar\zeta^c(x)\right\rangle_c.\eeq
In equilibrium, when the external sources vanish,
 the induced sources are also vanishing. This follows from symmetry.
In equilibrium,
the expectation values involve thermal averages over color singlet states,
and elementary group theory can then be used to prove that, in this case,
all terms on the r.h.s. of eqs.~(\ref{1jind})--(\ref{bC1}) indeed vanish.

At this point, it is
perhaps useful to illustrate the structure of the equations of motion
 by means of diagrams. This is done in Figs.1 and 2,
which represent schematically the various contributions to the
expectation values of the fields, as obtained from the equations of motion
(\ref{avA}) and (\ref{avpsi}).

\subsection{Induced sources and one particle irreducible amplitudes}

It will be convenient later to express the induced sources as functionals
 of the mean fields.
Following a standard procedure, this may be achieved in a systematic fashion
by introducing the effective action, that is, the Legendre transform of ln$Z$,
with $Z$ given by eq.~(\ref{Z}).
The external sources are then considered as the first derivatives of the
effective action with respect to the mean fields, which are treated as
independent variables. Further derivations yield the one particle
irreducible amplitudes.
Thus, by differentiating eqs.~(\ref{avA})
and (\ref{avpsi}) with respect to the mean fields,
one gets, using the usual definitions of self energies,
   \beq\label{pisia}
-\frac{\delta j_\mu^a(x)}{\delta\langle A_b^\nu(y)\rangle}
=\delta^{ab}D^{-1}_{0\,\mu\nu}+\Pi_{\mu\nu}^{ab}(x,y),\qquad\qquad\Pi_{\mu\nu}^{ab}(x,y)=\frac{\delta j_\mu^{ind\,a}(x)}{\delta\langle A_b^\nu(y)\rangle}
,\eeq  \beq\label{pisib}
-\frac{\delta\eta(x)}{\delta\langle\psi(y)\rangle}=
\left (-i\slashchar{\del}
+g\langle\slashchar{A}(x)\rangle\right )\delta (x-y)+
\Sigma(x,y),\qquad\qquad\Sigma(x,y)=\frac{\delta\eta^{ind}(x)}{\delta\langle\psi(y)\rangle}. \eeq
In eq.~(\ref{pisia})
\beq
\label{D}
D^{-1}_{0\,\mu\nu}(x,y)\equiv \left (-g_{\mu\nu}\del^2+(1-\lambda)
\del_\mu\del_\nu\right )\delta (x-y)
\eeq
is the free inverse propagator for gluons.
A further differentiation gives the 3-point functions. The quark-gluon vertex
is
\beq\label{gamma}
g\gamma^\mu
t^a+g\Gamma_a^\mu(x,y,z)=-\frac{\delta^2j_a^\mu(x)}{\delta\langle\psi(z)
\rangle\delta\langle\bar\psi(y)\rangle}
=-\frac{\delta^2\eta(y)}{\delta\langle
A_\mu^a (x)\rangle\delta\langle\psi(z)
\rangle},\eeq
that is,
\beq\label{Vqg}
g\Gamma^a_\mu(x,y,z)=\frac{\delta\Sigma(y,z)}{\delta\langle
A^\mu_a(x)\rangle}=\frac{\delta^2 j^{ind\,a}_{\mu}
(x)}{\delta\langle\psi(z)\rangle
\delta\langle\bar\psi(y)\rangle}
 .\eeq
Similarly, we get for the vertex correction to the three-gluons amplitude:
\beq
\label{V3g}
g\Gamma_{\mu\nu\rho}^{abc}(x,y,z)=\frac{\delta^2
j_\mu^{ind\,a}(x)}{\delta\langle A_b^\nu(y)\rangle\delta\langle A_c^\rho
(z)\rangle}=\frac{\delta \Pi_{\mu\nu}^{ab}(x,y)}{\delta\langle
A_c^\rho(z)\rangle}.
\eeq
We do not write here the  expressions corresponding to amplitudes with
ghost external lines since we shall not need them.

\subsection{Induced sources and two point functions}

 By inspecting  eqs.~(\ref{1jind})--(\ref{bC1}),
one sees that the induced sources are given by off-equilibrum 2-point
functions.
The only exception is the induced current $j^{ind}$ which also contains
 the 3-point function $\left\langle\,A_\nu^bA_\mu^dA^{\nu}_e\,\right\rangle_c$.
However, in the present work, we shall be interested only in the
contributions to the induced sources which appear at leading order in $g$.
Since the connected 3-point function contains at least two powers of $g$
more than the other terms, it can be safely ignored. This allows us
to express the induced current $j^{ind}$ (\ref{1jind}) in terms of 2-point
functions only. We write
\beq
\label{jind1}
j_\mu^{ind\,a}(x)\equiv\,j_{{\rm f}\,\mu}^a(x)+\,j_{{\rm G}\,\mu}^a(x)
+j_{{\rm g}\,\mu}^a(x)
+j_{\Omega\,\mu}^a(x),
\eeq
where the subscripts (${\rm f,\,\,G,\,\,g,}\,\,\Omega$) refer
to which hard particles contribute to the induced current.
Thus $j_{{\rm f}}$ is the current associated with the polarization of the
 hard fermions induced by the average quark and gauge fields
\beq
\label{jf1}
j_{{\rm f}\,\mu}^a(x)\equiv\,g \langle \bar\psi (x) \gamma_\mu t^a
\psi(x)\rangle_c.
\eeq
Similarly, $j_{{\rm G}}$ results from the polarization
 of the hard ghost particles induced by the gauge  mean fields
\beq
\label{jG1}
j_{G\,\mu}^a(x)\equiv\,
-gf^{abc}\left\langle (\del_\mu\bar\zeta ^b(x))\zeta^c(x)\right\rangle_c.
\eeq
The role of this current is to eliminate spurious polarization effects which,
 in a covariant gauge, are due to non-physical gluonic degrees of freedom.
The induced current resulting from the polarization of the hard
gluons by the quark and gauge mean fields is
\beq
\label{jg1}
j_{{\rm g}\,\mu}^a(x)\equiv\,gf^{abc}\Gamma_{\mu\rho\lambda\nu}\left\langle\,
A_b^\nu(x)\left
 (\del^\lambda A_c^\rho(x)\right )\,\right\rangle_c.
\eeq
The last term in eq.~(\ref{jind1}) is the ``tadpole'' contribution
\beq
\label{jt1}
j_{\Omega\,\mu}^a(x)\equiv g^2\hat\Gamma^{abcd}_{\mu\nu\rho\lambda}\langle
A_b^\nu(x)
\rangle\,\langle A^\rho_c(x) A_d^\lambda(x)\rangle_c.
\eeq

It is convenient to express the induced sources in terms of time-ordered
propagators, for which we shall derive equations of motion in the next
section. Let us first introduce specific notations.
Aside from the usual quark, gluon
and ghost propagators,
\beq\label{nor}
S(x,y)&\equiv&\langle{\rm T}\psi(x)\bar\psi(y)\rangle_c=-\frac {\delta
\langle\psi(x)\rangle}{\delta\eta(y)}, \nonumber\\
D_{\mu\nu}^{ab}(x,y)&\equiv&\langle{\rm T}A_\mu^a(x)A_\nu^b(y)\rangle_c
=-\frac
{\delta\langle A_\mu^a(x)\rangle}{\delta j_b^\nu(y)},\nonumber\\
G^{ab}(x,y)&\equiv&\langle{\rm
T}\zeta^a(x)\bar\zeta^b(y)\rangle_c=-\frac{\delta\langle\zeta^a(x)\rangle}{\delta
C^b(y)},\eeq
we shall also need the following ``abnormal'' propagators
\beq\label{abn}
K_\nu^a(x,y)&\equiv&\langle{\rm T}\psi(x)A_\nu^a(y)\rangle
_c=-\frac{\delta\langle\psi(x)\rangle}{\delta j_a^\nu(y)}
=-\frac{\delta\langle A_\nu^a(y)\rangle}{\delta\bar\eta(x)},
\nonumber\\
H_\nu^a(x,y)&\equiv&\langle{\rm T}
A_\nu^a(x)\bar\psi(y)\rangle_c=-\frac{\delta\langle\bar\psi(y)\rangle}{\delta
j_a^\nu(x)}=\frac{\delta\langle A_\nu^a(x)\rangle}{\delta\eta(y)},\nonumber\\
O_{\,\,\nu}^{ab}(x,y)&\equiv&\langle{\rm T}\zeta^a(x)
A_\nu^b(y)\rangle_c=-\frac{\delta\langle\zeta^a(x)\rangle}{\delta
j_b^\nu(y)}=-\frac{\delta\langle A_\nu^b(y)\rangle}{\delta\bar C^a(x)},
\nonumber\\
\bar O_{\nu\,}^{ab}(x,y)&\equiv&\langle{\rm
T}A_\nu^a(x)\bar\zeta^b(y)\rangle_c
=-\frac{\delta\langle\bar\zeta^b(y)\rangle}{\delta
j_a^\nu(x)}=\frac{\delta\langle A_\nu^a(x)\rangle}{\delta C^b(y)},
\eeq
which, of course,  vanish in equilibrium.
In these definitions, the symbol ${\rm T}$
denotes time ordering with respect to
imaginary time. The identities listed in the right hand sides of the
previous expressions will be useful later, when we derive the
equations of motion for 2-point functions (see Sec.3.3).

The time ordered propagators can be separated
into components
 which are  analytic functions of their time arguments
 \cite{KB62,Landsman87}. For example,
for the fermion propagator, we write      \beq\label{analytic}
S(x,y)\equiv\theta(\tau_x-\tau_y)S^>(x,y)-\theta(\tau_y-\tau_x)S^<(x,y),   \eeq
with similar definitions  for the other 2-point functions $D, G, K_\nu^a,
H_\nu^a, O_{\,\,\nu}^{ab}$ and
$\bar O_\nu^{ab}$, but without the minus sign (remember that the ghosts fields,
while anticommuting,
are submitted to periodic boundary conditions in imaginary time). The two
functions  $S^>(x,y)$ and
$S^<(x,y)$ are analytic functions of $x_0$ and $y_0$ for  $-\beta<{\rm Im}
\,(x_0- y_0)<0$ and
$0<{\rm Im} \,(x_0- y_0)<\beta$  respectively. This is the case at equilibrium;
we assume that the
sources satisfy the the conditions necessary to preserve this property
off-equilibrium\cite{KB62}.
We may assume, to be specific, that the
(analytic) sources  are adiabatically switched on at (real)
time $ -\infty$. Then,  when (real) time goes to $-\infty$, the various
 2-point functions go
over to their equilibrium values  in the absence of sources, and, for arbitrary
time, they represent correlation functions averaged over the initial
 equilibrium grand canonical ensemble.

The induced sources eqs.~(\ref{eta1})--(\ref{bC1}) and
 (\ref{jf1})--(\ref{jt1})  involve
products of fields with equal time arguments $\tau_x=\tau_y$. They may be
calculated from  the time ordered propagators (\ref{nor})--(\ref{abn}) taking
consistently  $\tau_y - \tau_x=\eta\to 0^+$.
Alternatively, they may be expressed in terms of the analytic pieces
of these propagators, as defined in (\ref{analytic}):
\beq
\label{eind}
\eta^{ind}(x)=g\gamma^\nu t^aK^<_{a\,\nu}(x,x), \eeq
\beq
\label{jf}
j_{{\rm f}\,\mu}^a(x)=g\,{\rm Tr}\left (\gamma_\mu t^aS^<(x,x)\right ),
\eeq
where the trace acts both on spinor and color indices,
\beq\label{jG}
j_{{\rm G}\,\mu}^a(x)=-gf^{abc}\left [\del_\mu ^y G_{bc}^<(x,y)\right ]\Big
|_{y\to x^+},
\eeq
\beq\label{jg}
j_{{\rm g}\,\mu}^a(x)=gf^{abc}\Gamma_{\mu\rho\lambda\nu}\left
[\del_x^\lambda\,D_{cb}^{\rho\nu\,<}(x,y)\right ]\Big |_{y\to x^+},
\eeq
\beq\label{Ci}
C_a^{ind}(x)=gf^{abc}\left [\,(\del_y^\mu+\del_x^\mu)\,O_{\,\,\mu}^{bc\,<}
(x,y)\right ]\Big |_{y\to x^+},
\eeq
\beq\label{bCi}
\bar C^{ind}_a(x)=-gf^{abc}\left [\,\del_y^\mu{\bar O}_{\mu}^{bc\,<}
(x,y)\right ]\Big |_{y\to x^+},
\eeq
where $y\to x^+$ stands for $\tau_y-\tau_x\to 0^+$.
These expressions hold after analytic continuation to real time. Then
$y\to x^+$ should be understood as $y_0-x_0 =i\eta$ and $\eta\to 0^+$.

Because the induced sources involve products of fields at the same point,
one could expect to encounter ultraviolet divergences when calculating them.
However, this will not be the case here (with the exception of the tadpole
term to be discussed in the next section). Indeed, as we have seen, in
equilibrum all the induced sources vanish, for symmetry reasons.
Off-equilibrum, no divergence occurs at leading order, as we shall verify
later.

For further reference, we list here expressions
for analytic components of the free, equilibrum propagators. For the fermions,
we have
\beq\label{S0}
S_0^<(x,y)=\int\frac{d^4k}{(2\pi)^4}e^{-ik\cdot (x-y)}
S_0^<(k),\qquad S_0^<(k)=\slashchar{k}\rho_0(k)n(k_0)\equiv\slashchar{k}
\tilde\Delta(k),
\eeq
where $\rho_0(k)$ is the  spectral function \beq
\label{rho0}\rho_0(k)=2\pi\epsilon(k_0)\delta(k^2),
\eeq
$n(k_0)=1/(\exp(\beta k_0)+1)$
is the fermion occupation factor,  and $\epsilon(k_0)=k_0/|k_0|$. Similarly
the equilibrium free gluon 2-point function $D^<_{0\,\mu\nu}(k)$ is, in Feynman
gauge ($\lambda =1$),
\beq
\label{Do}
D^{<\,ab}_{0\,\mu\nu}(k)=-\delta^{ab} g_{\mu\nu}\Delta(k),\eeq
where now
$\Delta(k)\equiv \rho_0(k)N(k_0)$, and  $N(k_0)=1/(\exp(\beta k_0)-1)$ denotes
the
boson occupation factor. Finally, for the free ghost propagator, we have
\beq
\label{G0}
G^<_{0\,ab}(k)=\delta ^{ab}\Delta(k),
\eeq
because of the periodic boundary conditions imposed on the
 ghost fields.
We note that in the description of the plasma using a covariant gauge,
the equilibrum state contains ghost particles obeying Bose statistics.
These are unphysical degrees of freedom, whose role is to cancel, when
computing physical observables, the contribution of the longitudinal
gluons, which are also present in the state,
 according to the propagator (\ref{Do}).

\vskip 2cm
\setcounter{equation}{0}

\section{Equations for the induced sources}

Up to now all the equations that we have written are exact, but
incomplete. Indeed, these are equations of motion for the mean
fields which depend on induced sources given themselves by
2-point functions. (A possible contribution from a 3-point
function has been seen to be negligible in the previous
section). These 2-point functions are determined by equations
which involve generally 3 and 4-point functions, and so on.
The approximations to be presented in this section will provide us
 with equations for the 2-point functions which involve
 the mean fields, but no higher point functions, thereby closing
the hierarchy.

The various features of our approximation scheme are summarized in
subsection 3.1. Then, in the following subsections, we go on and
derive systematically the equations of motion for the various
2-point functions of interest. This is done in several steps.
In subsection 3.3 we take into account the smallness of the coupling
in order to truncate the hierarchy. In the next subsections,
we implement, for each 2-point function of interest, the approximations which
follow from the fact that both the wavenumber and the amplitude of the mean
field oscillations are also small. The technical problems arising
from the use of a general covariant gauge are discussed in
Appendices A and B.

\subsection{Approximation scheme}

The cornerstone of our approximation scheme is the fact that the long
wavelength excitations that we are studying are {\it collective excitations
which develop at a particular energy scale}, $gT$, well separated,
when $g\ll 1$,
from the typical energies of plasma particles, $\sim T$. This scale controls
both the wavelength and the amplitude of the mean field oscillations.
Collectivity arises because the wavelength of the oscillations, $\sim 1/gT$,
is much larger than the typical interparticle distance, $\sim 1/T$.

The way how the scale $gT$ emerges, is perhaps best seen on the equations
  (\ref{avA})--(\ref{avbz}) for the mean fields. Technically speaking,
one expects collective excitations to develop whenever the
induced sources on the right hand sides of these equations
are of the same order of
magnitude as the first order differential operators which stand
on their l.h.s. Assuming that the wavelength/period of
the field oscillations is $\lambda$ ($i.e.$, the fields carry a typical
momentum $P\sim gT$),  the l.h.s. are then of order $\sim
\lambda^{-1}\phi$, where $\phi$ denotes one of the average
fields, that is either $\langle F_{\mu\nu}\rangle$ or $\langle\psi\rangle$.
{\it
The r.h.s., that is, the induced sources are indeed of order $
\lambda^{-1}\phi$  when $\lambda^{-1}$ is of order $gT$.} For instance,
eq.~(\ref{pisib}) shows that $\eta^{ind}\sim \Sigma\langle\psi\rangle$. The
dominant
contribution to $\Sigma (P)$ at high temperature is $\sim g^2T^2/P$.
Thus, $\eta^{ind}\sim (g^2T^2/P)\langle\psi\rangle$, which
is $\sim P\langle\psi\rangle\sim
\lambda^{-1}\langle\psi\rangle$ when $P\sim gT$. Similarly, $j^{ind}\sim \Pi
\langle A\rangle\sim
g^2T^2\,\langle A\rangle$ at high temperature, while
 $\del \langle F\rangle\sim\del^2 \langle A\rangle\sim P^2 \langle A\rangle$;
again, $j^{ind}\sim\del \langle F\rangle$ if $P\sim gT$.
Furthermore, the requirement that the plasma stay close to
 equilibrium puts  constraints on the
strength of the mean fields which also depend on $\lambda$.
The existence of such constraints can be anticipated on the basis
of the following heuristic argument. The plasma will be weakly perturbed
if, for instance, the color force $\sim g\langle F\rangle$ acting on a typical
particle
with momentum $k\sim T$, produces a small change in its momentum.
Assuming that the force
is acting during a time $\Delta t\sim \lambda$,
we require $\Delta k=g\langle F\rangle\lambda\ll k$.  One sees that, for
$\lambda^{-1}\sim
gT$ and $\Delta k\simle gT$, we have $\langle F\rangle\simle gT^2$. More
generally, {\it for
 $\lambda^{-1}\sim gT$,  there exists a limit in which all terms in the
equations of motion are of the same order of magnitude.}
 This is achieved when $\langle F\rangle\sim gT^2$ and $\langle\bar\psi\rangle
\langle\psi\rangle\sim gT^3$, that is, when these fields are $g$ times a power
of $T$
equal to their canonical
dimension. When the fields take this particular limiting
strength, the coupling constant disappears from the equations of
motion if one expresses all energies and momenta in units of
$gT$, and length and time in units of $1/gT$. Furthermore, in
this limit, all terms in the field strength tensor, or in
covariant derivatives, are of the same order of magnitude
($\partial\sim g\langle A\rangle$,
 $\partial \langle A\rangle\sim g\langle A\rangle^2$), and the same is true for
all the contributions  to the induced sources which are non linear
 in $\langle A\rangle$.
In the rest of this work, we shall work in
this limit, where the equations keep their full non-abelian structure.
 However, one should
emphasize that the equations to be derived are also valid in the
case of weaker fields. In the limit of weak fields, the
equations for $\langle A\rangle$ and $\langle\psi\rangle$ decouple,
and $\langle A\rangle$ behaves as an abelian
gauge field, that is, the equations are linear.

We turn now to the 2-point functions which are needed for
the calculation of the induced sources and explain the general
structure of our approximation scheme.
 For the present discussion, we shall consider as a generic example
the off-equilibrium fermion 2-point function
$S(x,y)$. Most arguments extend to other 2-point functions as well.

The equations of motion for the 2-point functions will be obtained in Sec.3.3,
by differentiation of the mean field equations with respect to the
external sources. In doing this, we make use of the identities listed in
 eqs.~(\ref{nor})--(\ref{abn}). For $S(x,y)$ one obtains
 (in the case $\langle\psi\rangle =\langle\bar\psi\rangle=0$)
\beq\label{dyson}
\left (S_0^{-1}\,+g\langle{\slashchar A}\rangle\,+\Sigma\right )\,S\,=\,1,\eeq
where operator notations have been used, and $S_0^{-1}(x,y)\equiv -i
{\slashchar \del}_x\,\delta(x-y)$. The off-equilibrium
self-energy $\Sigma$ arises from
differentiating the induced source. We argue now that such a term can, in fact,
 be ignored in leading order. To understand why it is so,  let us recall
that we need $S(x,y)$ only to evaluate the induced current $j_{\rm f}$,
according to eq.~(\ref{jf}).
Since the equilibrium propagator $S_{eq}(x-y)$ does not contribute
to $j_{\rm f}$,  we  need, in fact, only the deviation induced
by the gauge fields, $\delta S\equiv S-S_{eq}$. Furthermore, as will be
clear shortly, we only need $\delta S(x,y)$ for small relative distances
$|x-y|\simle 1/T$, or, equivalently, for hard momenta $k\sim T$. The diagrams
contributing to $\delta S$ involve at least one mean field insertion, and
the lowest order ones are shown in Figs.3a and 3b. An elementary analysis
reveals that, indeed, those in Fig.3b, which represent typical corrections
coming from $\Sigma$, are negligible compared to that of Fig.3a when $k\sim T$.
 Consequently, we shall neglect $\Sigma$ in eq.~(\ref{dyson}) and, at the same
 time, replace $S_{eq}$ by its free value $S_0$. We may then write the final
equation as
\beq\label{dyson1}
\left (S_0^{-1}\,+g\langle{\slashchar A}\rangle\right )\,S\,=\,1,
\qquad\qquad S\,=\,S_0+\delta S.\eeq

More generally, in deriving the equations of motion for the 2-point functions
in Sec.3.2, we shall ignore all terms coming from differentiating the induced
sources, to which we refer as radiative corrections, and at same time replace
the equilibrium propagators by their free expressions. This approximation,
equivalent to the familiar one-loop approximation to the induced currents
(see Figs.7 and 8 below),
closes the hierarchy of equations at the level of 2-point functions.
The net result is that only
mean field insertions are to be considered in the off
equilibrium 2-point functions. In physical
terms, this is to say that the dominant interactions
in the plasma are those between the hard particles and the mean fields.

It should be stressed that the one-loop approximation alone is not by
itself a consistent approximation for the calculation of the
induced sources.
It does contain all leading terms
in the weak coupling expansion, but also terms which are subleading,
and which should therefore be discarded. These subleading contributions
will be eliminated after taking carefully into account the other sources of
$g$ dependence, namely, the soft character of the average fields.

In order to implement the approximation that the
 fields are
slowly varying in space and time, we need to
consider the 2-point functions in real time. As
explained in section 1, these functions, as well as the
equations that they satisfy, can be obtained by analytic
continuation of the imaginary time functions and the
corresponding equations.  We can then proceed to a gradient expansion, for
which it is convenient to use the Wigner transforms of the various 2-point
functions. In the case of the fermion propagator, this is
\beq\label{SWIG}
 {\cal S}(k,X)\equiv\int d^4s
\,e^{ik\cdot s} S^<(X+{s\over 2},X-{s\over 2}), \eeq
 where  \beq\label{Rel}
 s^\mu\equiv x^\mu-y^\mu,\qquad\qquad
X^\mu\equiv{x^\mu+y^\mu\over2}.\eeq
 Note that in order to avoid the
proliferation of symbols, we use similar symbols to denote
different functions. Thus, for example, we shall often write $S^<(s,X)$ in
place of $S^<(X+s/2,X-s/2)=S^<(x,y)$.
Which function is meant can be recognized
from its arguments. Wigner functions will be denoted
consistently with calligraphic capitals.
In equilibrium, the non vanishing two point functions are
independent  of $X^\mu$ and sharply peaked about $s^\mu=0$, the
range of variation being fixed by the thermal wavelength
$\lambda_T\sim 1/T$. In the presence of a slowly varying
disturbance, with a wavelength of order $1/gT$, we expect the
$s^\mu$ dependence to be little affected, while the acquired
$X$ dependence will have a long wavelength variation. This
will enable us, for example, to neglect the derivatives with
respect to $X^\mu$, of order $gT$, relative to those with
respect to $s^\mu$, of order $T$. We shall thus transform the
equations of motion for the 2-point functions into
equations for $\delta {\cal S}(k,X)$, valid for $k\sim T$ and $\del_X\sim gT$.

One can use the Wigner function (\ref{SWIG}) to write the induced
current (\ref{jf}) as
\beq\label{fwig}
j^a_{{\rm f}\,\mu}(X)&=& g\int\frac{d^4k}{(2\pi)^4}\,{\rm Tr}\,\gamma_\mu
t^a\,\delta{\cal S}(k,X),
\eeq
where $\delta{\cal S}(k,X)\equiv {\cal S}(k,X)-S_{eq}(k)$.
This expression explains why we need $\delta{\cal S}(k,X)$ only for
$k\sim T$:  soft momenta $k\sim gT$ represent a small integration region and,
 consequently,  give generally a small contribution
to such an integral. We are
assuming that $\delta{\cal S}(k,X)$ is not infrared singular, which is
the case in leading order. Note that $\delta{\cal S}(k,X)$ is
 not  accurately determined by our approximate equation (\ref{dyson1}) if
 $k\sim gT$ (then the one loop
diagrams in Fig.3b contribute at the same order as the tree-level diagram
in Fig.3a). However,   (\ref{dyson1}) provides a
 $\delta{\cal S}(k,X)$ with a smooth
behaviour at small $k$ and the integration in (\ref{fwig}) can
be extended to all momenta without harm.

After solving the equations of motion for the 2-point functions,
we find that, for fields as strong as allowed,
the induced pieces of the propagators $\delta S$,
$\delta D$ and $\delta G$, are, at most, $g$ times
the corresponding equilibrium propagators. It is in this sense that
 the induced effects are perturbatively small, and that the plasma is
weakly disturbed. When isolating the leading
order terms in our equations, we are led to retain terms of
lowest order in the fermionic fields, that is, terms linear in
 $\langle\psi\rangle$
and $\langle\bar\psi\rangle$ for ${\cal K}$ and ${\cal H}$, and, owing
 to baryon number conservation, bilinear expressions
like $\langle\bar\psi\rangle \langle\psi\rangle$ for $\delta{\cal S}$ and
 $\delta{\cal D}$.
However, in order to satisfy
gauge covariance, we have to keep terms to all orders in $\langle A\rangle$.
In fact, the gauge fields enter our equations  in two different
ways. This may be seen by rewriting eq.~(\ref{dyson}) as
\beq\label{dyson2}
\left (-i{\slashchar \del}\,+g\langle{\slashchar A}\rangle\right )
\,\delta S\,=\,-g \langle{\slashchar A}\rangle S_0.\eeq
The gauge field enters linearly on the right hand side, as
a usual perturbation. It also enters non
linearly, as part of a covariant derivative,  on the left hand
side. The existence of these covariant derivatives,
in the equations for both the mean fields and the 2-point functions, will be
seen to be the only source of non linearity in $\langle A\rangle$.
This particular way in which
the mean fields enter our equations is the origin of the fact
that there are hard thermal loops with any number of gluon
lines, but no more than two fermion lines; and also of the fact
that, in QED, there is no hard thermal loop with more than two photon
external lines.


\subsection{The induced sources and their densities}

According to eq.~(\ref{fwig}), the induced current $j^a_{{\rm f}\,\mu}(X)$
can be written as an integral over $k$ of a {\it density}
$ {\cal J}_{{\rm f}\, \mu}^{\, a}(k,X)$:
\beq\label{fcd}
j^a_{{\rm f}\,\mu}(X)\equiv g\int\frac{d^4k}{(2\pi)^4}\,{\cal J}^{\,a}_{{\rm f}
\,\mu}(k,X),\,\,\,\,\qquad {\cal J}_{{\rm f}\, \mu}^{\, a}(k,X)
={\rm Tr}\gamma_\mu t^a \delta{\cal S}(k,X).\eeq
 Once all 2-point functions are expressed in terms of Wigner
functions, this can be generalized to all induced sources. In order to
facilitate the reading of this paper, we now specify the systematics of the
notation used for the various densities of induced sources. All these
densities will be denoted by calligraphic capitals, as the Wigner functions
to which they are related.

According to  eq.~(\ref{eind}),  the induced source $\eta^{ind}(X)$
is given by
 \beq\label{4eind}
\eta^{ind}(X)=g\int\frac{d^4k}{(2\pi)^4}\slashchar{{\cal K}}(k,X),
\eeq
where ${\cal {\slashchar K}}(k,X)\equiv t^a\gamma^\mu {\cal K}_\mu
^a(k,X)$.
The densities of the induced currents will carry, besides Minkowski and color
indices, subscripts (${\rm f}$, ${\rm g}$, $G$, $\Omega$) and upperscripts
($A$, $\psi$). As indicated after eq.~(\ref{jind1}), the subscripts refer
to which hard particles are contributing to the induced sources, ${\rm f}$
standing for fermions, ${\rm g}$ for gluons, $G$ for ghosts; $\Omega$
refers to the tadpole
contribution. The upperscript indicates the nature of the
background field which induces the current ($A$ for a gauge field, and
 $\psi$ for a fermionic field). We shall see that, in the present
approximation, one
can separate the contributions of the fermionic and gauge fields, (up to
 trivial  color ``rotations'' imposed by the gauge transformation properties
of the various densities). Thus, we shall have a collection of induced currents
$\{j^A_{\rm f},\,\,j^A_{\rm g},\,\,j^A_G,\,\,j^A_{\Omega},\,\,
j^\psi_{\rm f},\,\,j^\psi_{\rm g}\}$, together with the corresponding densities
$\{ {\cal J}^A_{\rm f},\,\,{\cal J}^A_{\rm g},\,\,{\cal J}^A_{G},\,\,
{\cal J}^A_{\Omega},\,\,{\cal J}^\psi_{\rm f},\,\,{\cal J}^\psi_{\rm g}\}$.
 In some cases, it will be convenient to operate various grouping of these
contributions. The absence of the upperscript will imply a grouping of all
contributions corresponding to the various upperscripts:
\beq\label{group1}
{\cal J}_{\rm f}={\cal J}^A_{\rm f}+{\cal J}^\psi_{\rm f},\qquad
{\cal J}_{\rm g}={\cal J}^A_{\rm g}+{\cal J}^\psi_{\rm g},\qquad
{\cal J}_{G}={\cal J}^A_{G},\qquad {\cal J}_{\Omega}={\cal J}^A_{\Omega}.
\eeq
Similarly, the absence of a subscript implies a grouping of all contributions
corresponding to the various possible subscripts:
\beq\label{group2}
{\cal J}^A\,=\,{\cal J}^A_{\rm f}+{\cal J}^A_{\rm g}+{\cal J}^A_{G}+
{\cal J}^A_{\Omega},\qquad\,\,{\cal J}^\psi={\cal J}^\psi_{\rm f}+
{\cal J}^\psi_{\rm g}.\eeq
We shall also use the subscript ${\rm b}$  to group together all the
contributions coming from bosonic particles:
\beq\label{group3}
{\cal J}^\psi_{\rm b}\,=\,{\cal J}^\psi_{\rm g},\qquad {\cal J}^A_{\rm b}\,=\,
  {\cal J}^A_{\rm g}+ {\cal J}^A_{G}+ {\cal J}^A_{\Omega},\eeq
and similarly for the corresponding induced currents.

We shall also be led to consider the baryonic vector and axial currents. They
are defined by
\beq\label{3bar}
b_{\mu}(x)\equiv {\rm Tr}\gamma_\mu \delta S^<(x,x),\qquad\qquad
b^5_{\mu}(x)\equiv {\rm Tr}\gamma_5\gamma_\mu \delta S^<(x,x)\,\eeq
and their densities will be denoted by ${\cal B}_\mu(k,X)$ and $
{\cal B}^5_\mu(k,X)$, respectively:
\beq\label{4bari}
{\cal B}_\mu(k,X)={\rm Tr}\gamma_\mu\delta{\cal S}(k,X),\qquad\qquad
{\cal B}^5_\mu(k,X)={\rm Tr}\gamma_5\gamma_\mu\delta{\cal S}(k,X).\eeq
Finally, as we shall see, a fermionic field may also induce oscillations
in the {\it axial} color currents $j^5_{\rm f}$ and $j^5_{\rm g}$, with
\beq\label{jf5}
j_{{\rm f}\,\mu}^{5\,a}(x)\equiv g\,{\rm Tr}\left (\gamma_5\gamma_\mu t^a
\delta S^<(x,x)\right ),\qquad\,\,{\cal J}_{{\rm f}\, \mu}^{5\, a}(k,X)=
{\rm Tr}\left (\gamma_5 \gamma_\mu t^a \delta{\cal S}(k,X)\right ).
\eeq
and similarly for $j^5_{\rm g}$.

We end this subsection by calculating  the induced current $j_\Omega$ which
arises from the four gluon interaction. According to
eq.~(\ref{jt1}), $j_\Omega$
is proportional to the gluon loop $\langle A_\mu^a A_\nu^b
\rangle_c\equiv D_{\mu\nu}^{ab}(x,x^+)$, where $D_{\mu\nu}^{ab}(x,y)$ denotes
 the (exact) gluon propagator. Since this is multiplied by $g^2$, in order to
get the leading order contribution to $j_\Omega$ we can replace it whith the
 free, equilibrum propagator,  given by eq.~(\ref{Do}). Indeed, according
to the previous discussion, the difference $D-D_0$ is expected to be, at most,
of order $gD_0$.
We obtain then
\beq
\label{TP1}
j^a_{\Omega\,\mu}(x)=\,g^2\hat\Gamma_{\mu\nu\rho\lambda}^{abcd}\left\langle\,A_b^\nu(x)\right
\rangle\,D_{0\,cd}^{\rho\lambda}(x,x^+)
= -\Omega^2\,\langle A_\mu^a(x)\rangle,
\eeq
where we have noted
\beq
\label{Omg}
\Omega^2\equiv\, (d-1)Ng^2\int\frac {d^3k}{(2\pi)^3}\,\frac{N(k)}{k}
=N\frac {d-1}{12}g^2 T^2,
\eeq
and eq.~(\ref{Do}) for $D_{0\,\mu\nu}^{<\,ab}$ in Feynman gauge
has been used. (Note that the ultraviolet divergence associated
with the $T=0$ part of $D_0(x,x^+)$ has been removed by the usual
 zero-temperature substraction.)
The momentum space density for $j_\Omega$ is
\beq
\label{4jt}
{\cal J}_{\Omega\,
\mu}^a(k,X)\equiv\,-gN(d-1)\,\Delta(k)\,\langle A_\mu^a(X)\rangle.
  \eeq

\subsection{Equations for the 2-point  functions}

In this section, the imaginary time
equations of motion for the 2-point Green's functions are obtained
 by differentiating the mean field equations (\ref{avA})--(\ref{avbz})
 with respect to the external sources.
 As explained in the previous subsection,
we ignore all radiative corrections, that is, all terms coming from
differentiating the induced sources.

We define
 \beq \label{L} {\cal
L}_{\mu\nu}^{ab}(x)\equiv\left
[g_{\mu\nu}\,\del_x^2       -(1-\lambda)\del_\mu^x\del_\nu^x\right
]\delta^{ab}\qquad\qquad\qquad\nonumber\\+gf^{abc}\left
[\Gamma_{\mu\nu\rho\lambda}\langle A_c^\rho(x)\rangle \del_x^\lambda-
\Gamma_{\mu\rho\lambda\nu}\left (\del_x^\lambda\langle A_c^\rho(x)\rangle\right
)\right ]-g^2\hat \Gamma_{\mu\nu\rho\lambda}^{abcd}\langle
A_c^\rho(x)\rangle\langle A_d^\lambda(x)\rangle. \eeq This operator appears
whenever we differentiate eq.~(\ref{avA}) with respect to an arbitrary source
$J$ \beq \label{LJ} \frac {\del}{\del J(y)}\biggl\{\left [\,{\cal
D}_x^\nu,\,{\cal F}_{\nu\mu}(x)\,\right ]^a+\lambda\del_\mu\del_\nu\langle
A_a^\nu(x) \rangle\biggr\}\,=\,{\cal
L}_{\mu\nu}^{ab}(x)\frac{\delta\langle A_b^\nu(x)\rangle}{\delta J(y)},\eeq
and it may be interpreted as the inverse propagator of a gluon in a
background gauge field (see eq.~(\ref{D1}) below).

In writing the equations of motion for the 2-point functions,
 we shall ignore, whenever they appear, the following
abnormal propagators $\langle{\rm T}\psi(x)\psi(y)\rangle_c$, $\langle{\rm
T}\psi(x)\zeta(y)\rangle_c$,\\ $\langle{\rm T}\psi(x)\bar\zeta(y)\rangle_c$,
together
with their Hermitian conjugates.
These functions are vanishing in equilibrum, while their off-equilibrum
expressions involve at least one pair of weak fields; because of this,
they are found to be negligible relative to the leading terms appearing next
to them in the  equations.
In order to make this statement clear, it is useful to analyze a specific
example. Consider the equation for $K^a_\nu(x,y)$ which follows
from differentiating eq.~(\ref{avA})  with respect to $\bar\eta$
\beq
\label {1K1}
{\cal L}_{\mu\nu}^{ab}(y)\,K_b^\nu(x,y)=\,gS(x,y)\gamma_\mu t^a\langle\psi(y)
\rangle +g\langle\bar\psi(y)\rangle\gamma_\mu t^a \langle{\rm T}\psi(y)
\psi(x)\rangle_c\qquad\qquad\qquad\nonumber\\
\qquad\qquad -g f^{abc}\left (\del_\mu\langle\bar\zeta^b(y)\rangle\right )
\langle{\rm T}\psi(x)\zeta^c(y)\rangle_c
-g f^{abc}\left (\del_\mu^x\langle{\rm T}\psi(x)\bar\zeta^b(y)\rangle_c\right )
\langle\zeta^c(y)\rangle.
\eeq
The first term in the r.h.s. of this equation involves
the off-equilibrum quark propagator $S(x,y)$. We write it as
$S\equiv S_0+\delta S$, where $S_0$ is the free equilibrum propagator,
while $\delta S$ is the correction due to the average fields.
We assume that  $\delta S\simle gS_0$, and this will be verified
explicitly in Sec.3.4.
In leading order, we can therefore neglect the contribution of
 $\delta S$ relative to that of $S_0$.
As mentioned, the other 2-point functions in the r.h.s. of eq.~(\ref{1K1}) are
also expected to be small with respect
to $S_0$. We show how to verify this explicitly for the first of them,
which is proportional to $\langle{\rm T}\psi(x)\psi(y)\rangle_c$.
By differentiating eq.~({\ref{avpsi}) with respect to $\bar\eta$, we obtain
the equation
\beq\label{pp}
\slashchar{{\cal D}}_x\,\langle{\rm T}\psi(x)\psi(y)\rangle_c
=ig\,K_\nu^a(y,x)\gamma^\nu t^a\langle\psi(x)\rangle.\eeq
 As we shall see in Sec.3.4 (after eq.~(\ref{Spsi})), this equation implies
 that  $\langle{\rm T}\psi(x)\psi(y)\rangle_c
\sim(S_0/T^3)\langle\psi\rangle ^2$. Therefore, the second term in
the r.h.s. of eq.~(\ref{1K1}) is
 $g\langle\bar\psi(y)\rangle\gamma_\mu t^a$ $\langle{\rm T}\psi(y)
\psi(x)\rangle_c\sim g S_0 (\langle\bar\psi\rangle \langle\psi\rangle/T^3)
\langle\psi\rangle$. Relative to the first term
$gS_0(x-y)\gamma_\mu t^a\langle\psi(y)\rangle$,
 this is of  order
$\left (\langle\bar\psi\rangle \langle\psi\rangle\right )/T^3$,
that is, at most of order $g$.
Similar considerations apply to the last two terms in eq.~(\ref{1K1}),
which can be also neglected. Our final equation for $K^a_\nu(x,y)$ is therefore
\beq
\label {K1}
{\cal L}_{\mu\nu}^{ab}(y)\,K_b^\nu(x,y)=\,gS_0(x-y)\gamma_\mu t^a\langle\psi(y)
\rangle.\eeq
The previous arguments may be  extended to all the other
equations of interest. We shall not repeat them, but simply list the
equations obtained after the corresponding simplifications have been
done.

By differentiating again eq.~({\ref{avpsi}), but this time
with respect to $j_\nu^b$, we derive a second equation for $K^b_\nu(x,y)$
\beq
\label{K2}
\slashchar{{\cal D}}_x\,K_b^\nu(x,y)=-igt^a\gamma_\mu\langle\psi(x)\rangle
\,D_{0\,ab}^{\mu\nu}(x-y).
\eeq
A diagrammatic interpretation of eqs.~(\ref{K1}) and (\ref{K2}) is given in
Fig.4.
There are, of course, similar equations for $H_\nu^a$, but we do not
write them explicitly, because they are simply related (by
Hermitian conjugation) to eqs.~(\ref{K1}) and (\ref{K2}).

 The
equation for $S(x,y)$ is obtained by
 differentiating eq.~({\ref{avpsi}) with respect to $\eta$
\beq
\label{S1}
\slashchar{{\cal D}}_x\,S(x,y)=i\delta(\tau_x-\tau_y)\delta^3({\bf
x-y})-ig\gamma^\nu t^a\langle\psi(x)\rangle H_\nu^a(x,y).\eeq
The diagrammatic contributions to $S$ are displayed in Fig.5.

By differentiating the equations (\ref{avz}) and (\ref{avbz})
for the ghost fields  with respect $C^b$ and $\bar C^a$, we
generate two equations for $G^{ab}(x,y)$:
\beq
\label{G1}
\lefteqn{\del_x^2\,G^{ab}(x,y)-gf^{acd}
\langle A_\mu^c(x)\rangle\,\del_x^\mu G^{db}(x,y)=}\nonumber\\
& &\delta^{ab}\delta(\tau_x-\tau_y)\delta^3({\bf x-y})-gf^{acd}\,
\del_x^\mu\left [\langle\zeta^c(x)\rangle\,\bar O_\mu^{db}(x,y)\right ],
\eeq and
\beq
\label{bG1}
\lefteqn{\del_y^2\,G^{ab}(x,y)-gf^{bcd}\langle A_\mu^c(y)\rangle\,\del_y^\mu
G^{ad}(x,y
)=\qquad\qquad}\nonumber\\
& &\delta^{ab}\delta(\tau_x-\tau_y)\delta^3({\bf x-y})-gf^{bcd}\,\left (\del^
\mu\langle\bar\zeta^c(y)\rangle\right ) O_{\,\,\mu}^{ad}(x,y).
\eeq

Finally,
by differentiating eq.~(\ref{avA}) with respect to
$j_b^\nu$, we get the following equation for $D^{ab}_{\mu\nu}(x,y)$:
\beq
\label{D1}
{\cal L}_{\mu\sigma}^{ac}(x)D_{c\nu}^{\sigma b}(x,y)&=&g_{\mu\nu}\delta^{ab}
\delta(\tau_x-\tau_y)\delta^3({\bf x-y})\qquad\qquad\nonumber\\
&+&g\left \{\langle\bar\psi(x)\rangle\gamma_\mu t^aK_\nu^b(x,y)
+H_\nu^b(y,x)\gamma_\mu t^a\langle\psi(x)\rangle\right\}\nonumber\\
&-&gf^{acd}\left\{\left (\del_\mu\langle\bar\zeta^c(x)\rangle\right )
\,O_{\,\,\nu}^{db}(x,y)+\left (\del_\mu^x\,\bar O_\nu^{bc}(y,x)\right )
\langle\zeta^d(x)\rangle\right \},
\eeq
Diagrammatic contributions to $D$ are shown in Fig.6.

We have not written here the corresponding equations for the
ghost-gluon abnormal propagators, $O^{ab}_{\,\,\nu}$ and $\bar O^{ab}_\nu$.
  We do this in Appendix A, where we prove that, in
leading order, the equation of motion for the ghost mean fields
coincide with the tree level equations; there is no
collective excitation with ghost quantum numbers.
Accordingly, we shall  not be concerned anymore
in the main text with the propagation of the ghost fields. We set
 the ghost sources $C^a$ and $\bar C^a$ to zero. Then
 $\langle\zeta\rangle$ and
$\langle\bar\zeta\rangle$ vanish, and all terms involving
  $O^{ab}_{\,\,\nu}$ and $\bar O^{ab}_\nu$ disappear from
 eqs.~(\ref{G1})--(\ref{D1}).

 In  what follows, we simplify the writing
and denote the average gauge potential or the average fermionic field
respectively by $A(x)$
and $\psi(x)$, omitting the brackets. We also denote simply by
 $D$ or $F_{\mu\nu}$ the covariant derivative or the field strength tensor
 associated to the average gauge field, in place of ${\cal D}$ or
${\cal F}_{\mu\nu}$. Calligraphic capitals
will therefore be reserved for Wigner functions and for the densities of
the induced sources.
Furthermore, we shall restrict our discussion in the main text
to the  Feynman gauge ($\lambda=1$);
we prove in Appendix B that our final equations are independent of the gauge
fixing parameter.

The calculation of the induced sources from the 2-point functions given
by the previous equations is equivalent to the calculation of
the one-loop Feynman diagrams displayed
in Figs.7 and 8.
The approximations to be implemented
in the rest of this section
will provide us with consistent equations from which the leading
contributions to these diagrams will be directly obtained.

\subsection{Equations for ${\cal K}$ and ${\cal H}$}

The real-time equations which determine the abnormal quark-gluon propagator
$K_b^\nu$ are obtained from the imaginary-time equations
(\ref{K1}) and (\ref{K2}). We consider $\tau_y >\tau_x$ and obtain,
after analytic continuation,
\beq
\label {K3}
{\cal L}_{\mu\nu}^{ab}(y)\,K_b^{\nu\,<}(x,y)=-\,gS_0^<(x-y)\gamma_\mu
t^a\psi(y),
\eeq
and
\beq
\label{K4}
\slashchar{D}_x\,K_b^{\nu\,<}(x,y)=-igt^a\gamma_\mu\psi(x)\,D_{0\,ab}
^{<\,\mu\nu}(x-y).\eeq

By acting on eq.~(\ref{K4}) with the differential operator
$\slashchar{D}_x$ from the left, we obtain
\beq
\label{K5}
\slashchar{D}_x\slashchar{D}_x\,K_a^{\nu\,<}(x,y)=
ig\slashchar{D}_x \left [t^a\gamma^\nu\psi(x)\,\Delta(x-y)\right ].
\eeq
To simplify the l.h.s. of this equation, we use the identity
\beq\label{D^2}
\slashchar{D}_x\,\slashchar{D}_x=\,D_x^2+\frac {g}{2}\sigma^{\mu\nu}\,F_{\mu
\nu}(x),
\eeq
where $\sigma^{\mu\nu}=(i/2)[\gamma^\mu,\gamma^\nu]$ and
\beq\label{d^2}
D_x^2=\del_x^2+2ig\,A(x)\cdot\del_x +ig(\del\cdot A(x))-\,g^2\,A^2(x).
\eeq
Then we replace $x$ and $y$ by the
coordinates $s$ and $X$ (see eq.~(\ref{Rel})) and rewrite the derivatives as
 $\del_x=\del_s+\half\del_X$,
$\del_y=-\del_s+\half\del_X$ and $\del_x^2-\del_y^2=2\del_s\cdot\del_X$.
We recall that
 a derivative $\del_X$ is to be regarded as being of order $gT$, and a
derivative $\del_s$ as being of  order $T$.
 Furthermore,  the mean field is assumed to be of order $T$,
 that is $gA$ is to be considered of the same order as the derivative
$\del_X$, which insures the consistency of the soft covariant derivative,
$D_X=\del_X+igA$. As for $gF_{\mu\nu}$, it is of order $g^2T^2$. Finally,  we
expand $A(x)\equiv A(X+s/2)\approx A(X)+(s/2)\cdot \del_X A(X)$,
 where the second term is $g$ times the first one, and likewise
for the other fields. Finally, eq.~(\ref{K5})  reduces to
\beq
\label{K6}
\left [ \del_s^2+\del_s\cdot\del_X+2igA(X)\cdot\del_s\right ]K_a^{\mu\,<}
(s,X)= ig(\slashchar{\del}_s\Delta)\gamma^\mu t^a\psi(X).
\eeq
Similar manipulations performed on eq.~(\ref{K3}) lead to
\beq
\label{K7}
\left [\delta^{ab} g_{\mu\nu}\left (\del_s^2 -\del_s\cdot\del_X \right )
-gf^{abc}\Gamma_{\mu\nu\rho\lambda} A_c^\rho(X) \del_s^\lambda\right ]K_b^
{\nu\,<}(s,X)\nonumber\\
= -g S_0^<(s)\gamma_\mu t^a\psi(X).\qquad\qquad
\eeq
 We remark that the contribution from the
four gluon  interaction ($i.e.$, the term proportional to $\hat\Gamma$)
 disappears at this order.

In writing eqs.~(\ref{K6}) and (\ref{K7}), we have neglected terms of order
$g^2 T^2$ inside the brackets on the left hand sides. As for the
remaining terms, the first one, $\del_s^2$, is of order $T^2$, while all
the others are of order $gT^2$. Therefore, in order to have
consistent equations
order by order in $g$, we should also expand $K^< = K^{(0)}+K^{(1)}+...$,
with $K^{(1)}\sim gK^{(0)}$, etc... Of course, because of the
approximations which have been done,
 eqs.~(\ref{K6}) and (\ref{K7}) can only be used to determine $K^{(0)}$.
The dominant terms in  the two equations lead to the consistency condition
\beq
\label{ms}
\del_s^2 K^{(0)}(s,X)=0.
\eeq
Next, we note that the ``big'' derivatives $\del_s^2\sim T^2$
give rise to contributions of the form $\del_s^2K^{(1)}$
which are of the same
order as, for example, $ (\del_s\cdot\del_X )K^{(0)}$.
Such terms are inconsistent and, besides,
they cannot be easily evaluated. In order to do so, one would need
 equations for $K^{(1)}$ containing the unknown function $K^{(2)}$,
and so forth. However, such ``spurious'' contributions
cancel in {\it the difference} of the two equations (\ref{K6}) and
(\ref{K7}). By doing this difference, we end up with
 the following equation for $K^{(0)}$, which we denote from
now on simply by $K^<$:
\beq
\label{KsX}
\left [\delta^{ab} g_{\mu\nu}\left (\del_X+igA(X)\right )\cdot\del_s+
\frac{g}{2}f^{abc}\Gamma_{\mu\nu\rho\lambda} A_c^\rho(X) \del_s^\lambda\right
]K_b^{\nu\,<}(s,X)\nonumber\\
 =\,\frac {g}{2}\left [S_0^<(s)+i\slashchar{\del}_s\Delta(s)\right]
\gamma_\mu t^a\psi(X).\qquad\qquad
\eeq
When Fourier transformed with respect to $s$, eq.~(\ref
{KsX}) translates into an equation for
the Wigner function ${\cal K}^a_\mu(k,X)$:
\beq
\label{KXk}
 k\cdot\left (\del_X+igA(X)\right ){\cal K}^a_\mu(k,X)+
\frac{g}{2}f^{abc}\Gamma_{\mu\nu\rho\lambda} A_c^\rho(X) k^\lambda {\cal
K}_b^\nu(k,X)\nonumber\\
= \,i\frac {g}{2}\rho_0(k)[N(k_0)+n(k_0)]
\slashchar{k}\gamma_\mu t^a\psi(X).\qquad\qquad
\eeq
{}From the right hand side of eq.~(\ref{KXk}), one sees that
${\cal K}_\mu^a$ is proportional to $\slashchar {k}\delta(k^2)$.
Consequently
\beq\label{K-prop}
k^2{\cal K}^a_\mu(k,X)=0,\qquad\qquad\slashchar{k}{\cal K}^a_\mu(k,X)=0.
\eeq
The first identity above is the consistency condition (\ref{ms}).
It simply expresses the mass-shell condition for ${\cal K}^a_\mu$. The second
identity is the Dirac equation for the quark-gluon propagator.
It can be also verified that the right hand side of eq.~(\ref{KXk}) is
transverse with respect to $k$, so that
\beq
\label{trK}
k^\mu{\cal K}^a_\mu(k,X)=0.
\eeq
By using these properties, namely,
Eqs.(\ref{K-prop}) and (\ref{trK}), one easily transforms
 eq.~(\ref{KXk})  into an equation
for ${\cal {\slashchar K}}^a(k,X)\equiv\gamma^\mu{\cal K}^a_\mu(k,X)$
\beq
\label{slashK}
 k\cdot \left [\left (\del_X+igA(X)\right )\delta^{ab}
+ ig\tilde A^{ab}(X)\right ] {\cal {\slashchar K}}^b(k,X)\nonumber\\
=-ig\frac{d-2}{2}\rho_0(k)[N(k_0)+n(k_0)]\slashchar{k} t^a\psi(X),
\eeq
where the factor $(d-2)$ results from the identity $\gamma^\mu
{\slashchar k}\gamma_\mu=-(d-2){\slashchar k}$.
Eq.~(\ref{slashK}) is covariant with respect to local gauge
 transformations of the mean fields $\psi$ and $A$.
This is seen by noting that the differential operator from the
left hand side is a covariant derivative acting on vectors in
the direct product of the fundamental and the adjoint representations.
It follows that ${\cal {\slashchar K}}^a$ transforms as $t^a\psi(x)$
under a gauge transformation. Explicitly, when $\psi(x)\to S(x)\psi(x)$,
with $S(x)=\exp(i\theta^a(x) t^a)$, then
\beq
\label{gtK}
{\cal {\slashchar K}}_i^a(k,X)\to\tilde S_{ab}
(X) S_{ij}(X){\cal {\slashchar K}}_j^b(k,X),
\eeq
where $\tilde S(x)=\exp(i\theta^a(x) T^a)$ is the gauge transformation
in the adjoint representation.

The equation satisfied by ${\cal {\slashchar K}}(k,X)\equiv t^a {\cal
{\slashchar K}}^a(k,X)$ results from
eq.~(\ref{slashK}) after multiplication by $t^a$
\beq
\label{kinK}
\left (k\cdot D_X\right ) {\cal {\slashchar K}}(k,X)
=-ig\frac{d-2}{2}C_f\rho_0(k)[N(k_0)+n(k_0)]\slashchar{k} \psi(X),
\eeq
where $C_f$ is the quark Casimir. The covariant derivative
which remains in the left hand side of eq.~(\ref{kinK}) is that of
the fundamental representation, $D_X$,   showing that
 ${\cal {\slashchar K}}$ transforms  as $\psi$. In contrast with what happens
in eq.~(\ref{slashK}) for $ {\cal {\slashchar K}}^a$, the gauge
 field insertions on the
hard $gluon$ lines have disappeared in  eq.~(\ref{kinK}), which coincides
with the corresponding equation for an abelian plasma \cite{QED}.

We end this section by writing the equation satisfied by the Wigner
transform of the quark-gluon propagator $H_\mu^{a\,<}$. This
follows from eq.~(\ref{kinK}),
by noting that  $ {\cal {\slashchar H}}(k,X)= {\cal {\slashchar
K}}^\dagger(k,X)\gamma^0$,
\beq
\label{kinH}
 {\cal {\slashchar H}}(k,X)\left (k\cdot D_X^\dagger\right )
=ig\frac{d-2}{2}C_f\rho_0(k)[N(k_0)+n(k_0)]\bar\psi(X)\slashchar{k}.
\eeq
Here ${D}_\mu^\dagger \equiv {\del}_\mu^\dagger - ig
{A}_\mu$, where  the derivative operator
 $\del_\mu^\dagger\equiv\left(\buildchar{\del}{\leftarrow}{}/\del
x^\mu\right)$ acts on its left.

\subsection{Equations for $\delta{\cal S}_A$ and $\delta{\cal S}_\psi$}

The equations for the off-equilibrium fermion propagator are
 obtained by analytic continuation
 from eq.~(\ref{S1}) and its  conjugate:
 \beq
\label{S2}
\slashchar{D}_xS^<(x,y)=ig\gamma^\nu t^a\psi(x)H^{a\,<}_\nu(x,y),
\eeq
\beq\label{S3}
S^<(x,y)\slashchar{D}^\dagger_y=-igK^{a\,<}_\nu(x,y)\bar\psi(y) t^a\gamma^\nu,
\eeq
with $\slashchar{D}^\dagger \equiv \slashchar{\del}^\dagger - ig\slashchar
{A}$ and $\slashchar{\del}^\dagger\equiv\left(\buildchar{\del}
{\leftarrow}{}/\del
x^\mu\right)\,\gamma^\mu$. We transform them by first
 acting with the covariant derivatives $\slashchar {D}_x$ on eq.~(\ref{S2})
from the left, and with $\slashchar{D}^\dagger_y$ on
eq.~(\ref{S3}) from the right. We subtract the resulting equations and
separate the leading order terms by manipulations similar to those encountered
in the previous subsection. A new feature stems from the fact that
in equilibrium $S^<(x,y)$ does not vanish, but is equal to
 $S_0^<$, eq.~(\ref{S0}).
As explained in Sec.3.1, we write  $S^<(x,y)= S_0^<(x-y)+\delta S(x,y)$ and
assume $\delta S$ to be of order $g\,S_0^<$.
It is then straightforward to derive the  equation satisfied in leading order
by the Wigner function $\delta {\cal S}(k,X)$,
\beq
\label{dS}
\left [k\cdot D_X,\,\delta{\cal S}(k,X) \right ]
=-i\frac{g}{4}F^{\mu\nu}[\sigma_{\mu\nu},\,S_0^<(k)]
-g\left (\del_X^\nu\,A^\mu \right )\,k_\mu\del_\nu^k S_0^<(k)
\nonumber\\ \qquad -i\frac{g}{2}\left\{ {\cal K}_\nu^a(k,X)
\bar\psi(X) t^a\gamma^\nu\slashchar{k}
-\slashchar{k}\gamma^\nu t^a\psi(X){\cal H}^a_\nu(k,X) \right\}.
 \eeq
We call $\delta {\cal S}^A(k,X)$ the solution of this equation when $\psi=
\bar\psi=0$, and $\delta {\cal S}^\psi(k,X)$ the correction which includes
the effects of the fermionic fields. We have $\delta {\cal S}(k,X)=\delta {\cal
S}^A(k,X)+\delta {\cal S}^\psi(k,X)$, and
\beq
\label{Spsi}
\left [k\cdot D_X,\,\delta{\cal S}^\psi(k,X) \right ]=
-i\frac{g}{2}\left\{ {\cal K}_\nu^a(k,X)\bar\psi(X) t^a\gamma^\nu\slashchar{k}
-\slashchar{k}\gamma^\nu t^a\psi(X){\cal H}^a_\nu(k,X) \right\}.
 \eeq
By using this equation, together with eqs.~(\ref{kinK}) and (\ref{kinH}),
we can verify that  $\delta {\cal S}^\psi(k,X)
\sim (S_0/T^3)\bar\psi\,\psi$. Thus $\delta {\cal S}^\psi(k,X)
\simle gS_0$ when  $\bar\psi\,\psi\simle gT^3$, as anticipated in Sec.3.1.
 Remark also the similarity between eqs.~(\ref{S2})--(\ref{S3}) for
$S^<(x,y)$ and eq.~(\ref{pp}) for the abnormal
propagator $\langle{\rm T}\psi(x)\psi(y)\rangle_c$.
By following the same steps as above, one can derive for the Wigner transform
of $\langle{\rm T}\psi(x)\psi(y) \rangle_c$ an equation analogous
 to  eq.~(\ref{Spsi}) for  $\delta {\cal S}^\psi(k,X)$.  From it we
 derive the estimate  $\langle\psi\psi\rangle_c\sim(S_0/T^3)\psi^2$ which
 has been  used in Sec.3.3 in order to discard the contribution of such
abnormal propagators.

 As a color matrix acting in the fundamental representation,
$\delta{\cal S}^\psi$ can be decomposed into a color singlet and a color octet:
 $\delta {\cal S}^\psi\equiv
\delta {\cal S}^\psi_{\rm o}+\delta {\cal S}^\psi_a t^a$.
 Besides, $\delta{\cal S}^\psi$ is a $4\times 4$ matrix in Dirac indices.
We can verify on eq.~(\ref{Spsi})
that ${\rm tr}\delta{\cal S}^\psi={\rm tr}\gamma_5\delta{\cal S}^\psi
={\rm tr}\sigma_{\mu\nu}\delta{\cal S}^\psi=0$, and these equalities
hold for both  the color singlet and the color octet
(the trace acts on Dirac indices only).
The quantities
${\rm tr}\gamma_\mu \delta{\cal S}^\psi$ and ${\rm tr}\gamma_5\gamma_\mu
\delta{\cal S}^\psi$ are nonvanishing. Thus a fermionic field induces
 oscillations in the vector and axial
baryonic currents  $b_\mu$ and $b^5_\mu$, eq.~(\ref{3bar}), as well as in
the vector and axial color currents, $j^a_{{\rm f}\,\mu}$ and
 $j_{{\rm f}\,\mu}^{5\,a}$. The corresponding densities are given in eqs.~(\ref
{4bari}) and (\ref{jf5}), with $\delta {\cal S}$ replaced by $\delta {\cal S}
^\psi(k,X)$.

Let us consider now the equation
satisfied by $\delta {\cal S}^A(k,X)$. By using eq.~(\ref{S0}), we can write
it as follows:
\beq
\label{SA}
\left [k\cdot D_X,\,\delta{\cal S}^A(k,X) \right ]=-g
\left \{ \left [k\cdot D_X,\,{\slashchar A}\right ]+
\slashchar{k}\,k_\mu\left (\del^\nu_X A^\mu\right )\del_\nu \right \}
\tilde \Delta(k).
\eeq
This equation shows that  $\delta{\cal S}^A\sim g(A/T)S_0$, so that
$\delta{\cal S}^A\sim gS_0$ when  $A\sim T$.
  It can also be verified that  $\delta{\cal S}^A\propto\gamma_\mu t^a
\,({\rm Tr}\,\gamma^\mu t^a
\delta{\cal S}^A)$. Thus, a gauge field induces only
fluctuations in the vector color current.

 Eq.~(\ref{SA}) is not covariant under local gauge transformations.
This reflects the lack
of covariance of our definition for the fermionic Wigner function
$\delta{\cal S}(k,X)$. We could, instead, define a gauge covariant Wigner
transform, as for example in ref. \cite{Elze86}. But it is easier to
recover  gauge covariance directly at the level of eq.~(\ref{SA}). This
can be done in a standard way, by replacing the $canonical$
momentum $k$, associated with the gradient operator,
by the $kinetic$ momentum $p\equiv k-gA$, related to the particle velocity.
By applying this operation to the quark Wigner transform ${\cal S}(k,X)$,
we obtain a new function of $p$ and $X$,
$\acute{\cal S}(p,X)\equiv {\cal S}(k=p+gA,\,X)$,
which transforms covariantly. This
can be verified directly by writing (for $\psi=\bar\psi=0$)
$\acute{\cal S}(p,X)\equiv S_0^<(p)+
\delta\acute{\cal S}^A(p,X)$. Then, in leading order,
\beq
\label{delS}
\delta\acute{\cal S}^A(p,X)=\delta{\cal S}^A(p,X)+g(A(X)\cdot\del_p)S_0^<(p),
\eeq
and  $\delta\acute{\cal S}^A(p,X)$ satisfies
 the {\it gauge covariant} equation
\beq
\label{Vlas}
 \left [p\cdot D_X,\, \delta\acute{\cal S}^A(p,X) \right ]=
g\,\slashchar{p}\,p\cdot F(X)\cdot \del_p\tilde\Delta(p),
\eeq deduced from  eq.~(\ref{SA}).
This is the non-abelian generalization of the well-known linearized Vlasov
equation.
Note that although $gA\sim gT \ll k\sim T$,  the change from
$k$ to $p$ in ${\cal S}(k,X)$ is non trivial, as the second
term in eq.~(\ref{delS}) is of the same order of magnitude as the
first one.
However, this term is a total derivative with respect to $p$, so
that it does
not contribute to the induced current $j_{\rm f}$ (see eq.~(\ref{fwig})).

The change from canonical to kinetic momentum
in $\delta {\cal S}^\psi(k,X)$ is irrelevant at this order.
 Indeed, if we replace
$k$ by $p=k-gA$ in eq.~(\ref{Spsi}), we only generate terms of higher order.
And, in fact,
eq.~(\ref{Spsi}) is gauge covariant as it stands.

In summary, the previous discussion shows that all effects induced by the
mean fields on the hard quarks can be described by the
{\it gauge covariant} Wigner function
\beq\label{qwig}
 \acute{\cal S}(k,X)\equiv S_0^<(k)+
\delta {\cal S}^\psi_{\rm o}(k,X)+\left (\delta {\cal S}^\psi_a(k,X)
+ \delta \acute{\cal S}^A_a(k,X)\right )\,t^a,\eeq
and $k$ is to be understood here and from now on as the {\it kinetic}
momentum. In particular, the induced current $j_{\rm f}$ has the gauge
covariant  density
  \beq \label{4jf} {\cal J}_{{\rm f}\, \mu}^{\, a}(k,X)={\rm Tr}\gamma_\mu
t^a \left(N_{\rm f} \delta\acute{\cal S}^A(k,X)+ \delta{\cal S}^
\psi(k,X)\right)\equiv
{\cal J}_{{\rm f}\, \mu}^{A\, a}(k,X)+ {\cal J}_{{\rm f}\, \mu}^{\psi\,
a}(k,X).
\eeq
The factor $N_{\rm f}$ reflects the fact that all quark flavors
contribute identically to the current induced by the  average gauge fields.
Similarly, eq.~(\ref{Spsi}) contained  an implicit sum over the different
flavors of the quark mean fields.

\subsection{Equations for $\delta{\cal G}$}

{}From eqs.~(\ref{G1}) and (\ref{bG1}), one obtains the
following real-time equations of motion for the ghost propagator
\beq
\label{G2}
\del_x^2\,G_{ab}^<(x,y)=gf^{acd} A_\mu^c(x)\,\del_x^\mu G_{db}^<(x,y),
\eeq
and
\beq
\label{G3}
\del_y^2\,G_{ab}^<(x,y)=gf^{bcd} A_\mu^c(y)\,\del_y^\mu G_{ad}^<(x,y).
\eeq
After separating the induced part, by setting
$G_{ab}^<(s,X)=G_{0\,ab}^<(s)+\delta G_{ab}(s,X)$,
with $G_{0\,ab}^<$ given by eq.~(\ref{G0}), we  transform these equations
in a way which should by now be familiar. We obtain
\beq
\label{kinG}
k\cdot \left [\del_X \delta {\cal G}_{ab}(k,X)+
\frac {g}{2}\left(f^{acd}\delta{\cal G}_{cb}+f^{bcd}\delta{\cal G}_{ac}\right)
\,A^d(X)\right ]\nonumber\\
=i\,\frac{g}{2}\,f^{abc}\left(\del_\nu A_\mu^c(X)\right)k^\mu(\del_k^\nu
\Delta),\qquad\qquad
\eeq
where  $\delta {\cal G}(k,X)$ is the Wigner transform of $\delta G (s,X)$.
This equation shows that, as expected, $\delta {\cal G}\sim g(A/T)\Delta$.

For later reference, we note that $\delta {\cal G}_{ab}$ determined
by eq.~(\ref{kinG}) is a color matrix of the adjoint representation,
$i.e.$, it can be written as $\delta {\cal G}_{ab}\equiv (T^c)_{ab}
\delta {\cal G}_c$, where the new
functions $\delta {\cal G}_a(k,X)$ satisfy
\beq
\label{kinGa}
(k\cdot \del_X) \delta {\cal G}_a(k,X)-
\frac {g}{2}f^{abc} (k\cdot A_b(X))\delta{\cal G}_c(k,X)
=-\frac{g}{2}\left(\del_\nu A_\mu^a(X)\right)k^\mu(\del_k^\nu
\Delta).
\eeq
Then the ghost contribution to the induced current density
 may be written as (see eq.~(\ref{jG}))
\beq
\label{4jG} {\cal J}_{G\, \mu}^{\, a}(k,X)=-i\,k_\mu\,f^{abc}\delta{\cal G}
_{bc}(k,X)\,=\,-N\,k_\mu\,\delta{\cal G}^{\, a}(k,X).\eeq

\subsection{Equations for $\delta{\cal D}_A$ and  $\delta{\cal D}_\psi$}

After analytic continuation to real time,  eq.~(\ref{D1}) gives
\beq
\label{D2}
{\cal L}_{\mu\sigma}^{ac}(x)D_{c\nu}^{\sigma b\,<}(x,y)=
g\left \{\bar\psi(x)\gamma_\mu t^aK_\nu^{b\,<}(x,y)
+H_\nu^{b\,>}(y,x)\gamma_\mu t^a\psi(x)\right\}.
\eeq
Another  equation for $D^<$ is obtained by interchanging
 in this equation the
space-time arguments, as well as the color and Minkowski indices,
\beq
\label{D3}
{\cal L}_{\nu\sigma}^{bc}(y)D_{\mu c}^{a\sigma \,<}(x,y)=
g\left \{\bar\psi(y)\gamma_\nu t^bK_\mu^{a\,>}(y,x)
+H_\mu^{a\,<}(x,y)\gamma_\nu t^b\psi(y)\right\}.
\eeq
The symmetry property $D^{ab\,<}_{\mu\nu}(x,y)=D^{ba\,>}_{\nu\mu}(y,x)$
has been used.
We set $ D^{ab\,<}_{\mu\nu}(x,y)\equiv -g_{\mu\nu}\delta^{ab}\Delta(s)+
\delta D_{\mu\nu}^{ab}(s,X)$, with $\delta D$ expected to be
 of order $g\Delta$,
and we separate the leading order terms in
eqs.~(\ref{D2})--(\ref{D3}) in order to get equations for
the  Wigner function $\delta{\cal D}^{ab}_{\mu\nu}(k,X)$.
In order to do this, we need all the analytic pieces of the abnormal
quark-gluon propagators, $i.e.$, not only $K^<$ and $H^<$, for which we have
 obtained equations in Sec.3.3, but also $K^>$ and $H^>$. In fact, by
repeating the calculations of Sec.3.3, one can show that,
in  the present approximation,
$K^>$ and $K^<$ satisfy identical equations, and both vanish in equilibrium,
so that  \beq
{\cal K}_\mu^{a\,>}(k,X)\approx {\cal K}_\mu^{a\,<}(k,X)\equiv {\cal
K}_\mu^a(k,X),
\eeq
and similarly for the ${\cal H}$ functions.

In order to separate
the effects of gauge fields from those of fermionic
fields, we write  $\delta {\cal D}
\equiv\delta{\cal D}^A+\delta{\cal D}^\psi$, where $\delta{\cal D}^A(k,X)$
 is the modification  induced by the  gauge field when $\psi=\bar\psi=0$.
At leading order in $g$,
 eqs.~(\ref{D2}) and (\ref{D3}) imply the consistency conditions
\beq
\label{k2D}
k^2\delta{\cal D}^{A\,ab}_{\,\,\mu\nu}(k,X)=igf^{abc}\Delta(k)\Gamma_{\mu\nu
\rho\lambda}A^\rho_a(X)k^\lambda,\qquad\qquad k^2\delta{\cal D}^{\psi\,ab}
_{\,\,\mu\nu}=0.
\eeq
These will be used later in simplifying some formulae.
At next to leading order, we take the difference of eqs.~(\ref{D2}) and
(\ref{D3}) and obtain, after a lengthy, but straightforward
calculation,
\beq
\label{kinDA}
k\cdot\del_X\,\delta{\cal D}^{A\,ab}_{\,\,\mu\nu} -\frac {g}{2}
\left (f^{acd}\,\Gamma_{\mu\sigma\lambda\rho}\,\delta{\cal D}^
{A\,\sigma b}_{\,\,d\nu}+
f^{bcd}\,\Gamma_{\nu\sigma\lambda\rho}\,\delta{\cal D}^{A\,a\sigma}_{\,\,\mu d}
\right )\,k^\lambda\,A^\rho_c\nonumber\\
=-i\,\frac{g}{2}\,f^{abc}\,\Gamma_{\mu\nu\rho\lambda}\left
(\del^\sigma_X A^\rho_c\right )\,k^\lambda (\del_\sigma ^k\Delta),
\qquad\qquad
\eeq
and
\beq
\label{kinDpsi}
k\cdot\del_X\,\delta{\cal D}^{\psi\,ab}_{\,\,\mu\nu} -\frac {g}{2}
\left (f^{acd}\,\Gamma_{\mu\sigma\lambda\rho}\,\delta{\cal D}^
{\psi\,\sigma b}_{\,\,d\nu}+
f^{bcd}\,\Gamma_{\nu\sigma\lambda\rho}\,\delta{\cal D}^{\psi\,a\sigma}_{\,\,\mu
d}
\right )\,k^\lambda\,A^\rho_c\nonumber\\
=i\,\frac{g}{2}\,\bar\psi\left \{\gamma_\mu t^a{\cal K}_\nu^b-
\gamma_\nu t^b{\cal K}_\mu^a\right \}\,
+i\,\frac{g}{2}\,\left \{{\cal H}_\nu^b\gamma_\mu t^a-{\cal H}_\mu^a
\gamma_\nu t^b\right \}\,\psi.\qquad
\eeq
Note that the four gluons vertex does not contribute to this order
and in this particular gauge, so that there are no terms quadratic
in $A$ in the right hand side of eq.~(\ref{kinDA}).

The contribution of hard gluons to the induced current can be calculated
from  eq.~(\ref{jg}), which leads to the density
\beq
\label{Jg}
{\cal J}_{{\rm g}\,\mu}^a(k,X)= i\,f^{abc}\Gamma_{\mu\rho\lambda\nu}
\, k^\lambda\,\delta{\cal D}^{\rho\nu}_{bc}(k,X).
\eeq
The separation $\delta {\cal D}=\delta{\cal D}^A+\delta{\cal D}^\psi$
 leads to a corresponding separation in the induced current density,
 ${\cal J}_{\rm g}={\cal J}_{\rm g}^A+{\cal J}_{\rm g}^\psi$.
 We are interested now in transforming
 the previous equations for $\delta{\cal D}^A$ and $\delta{\cal D}^\psi$
into equations for these densities.

Let us  consider  first
the density ${\cal J}_{\rm g}^\psi$. This is obtained from
$\delta{\cal D}^\psi$, given by eq.~(\ref{kinDpsi}) above.
Since the r.h.s. of this equation is antisymmetric with respect to
$simultaneous$ permutations of the color and Minkovski indices, the same
property holds for $\delta{\cal D}^\psi$
\beq
\label{asymD}
\delta{\cal D}^{\psi\,ab}_{\,\,\mu\nu}(k,X)=\,-\,
\delta{\cal D}^{\psi\,ba}_{\,\,\nu\mu}(k,X).
\eeq
Furthermore,  the right hand side of eq.~(\ref{kinDpsi}) is transverse
 with respect to $k$. Indeed, by taking its scalar product with $k^\mu$,
 we find in the r.h.s. combinations like $k\cdot{\cal K}^a$ or $\slashchar{k}
{\cal K}^a_\mu$ which are vanishing in the present approximation,
as shown by eqs.~(\ref{K-prop}) and (\ref{trK}). This implies
\beq
\label{trDpsi}
k^\mu\,\delta{\cal D}^{\psi\,ab}_{\mu\nu}(k,X)=
\delta{\cal D}^{\psi\,ab}_{\mu\nu}(k,X)\,k^\nu\,=\,0.
\eeq
Using this property, together with the  definition (\ref{Jg}) for
the current density ${\cal J}_{\rm g}^\psi$, as well as  the
explicit form of the $\Gamma$ symbol, eq.~(\ref{Gamma}), one obtains
\beq\label{JgD}
{\cal J}^{\psi\,a}_{{\rm g}\,\mu}(k,X)=-i\,k_\mu\,f^{abc}\left (
\delta {\cal D}_{bc}^\psi\right )^\nu_{\nu}(k,X).
\eeq

By  using the transversality property (\ref{trDpsi}), we readily derive from
eq.~(\ref{kinDpsi}) the following equation for $\left (
\delta {\cal D}_{ab}^\psi\right )^\nu_{\nu}$:
\beq
\label{trkinD}
k\cdot\del_X\left (
\delta {\cal D}_{ab}^\psi\right )^\nu_{\nu}\,-\,g(k\cdot A_c)
\left \{f^{acd}\left (\delta {\cal
 D}_{db}^\psi\right )^\nu_{\nu}+
f^{bcd}\left (\delta {\cal D}_{ad}^\psi\right )^\nu_{\nu}\right \}
\nonumber\\
=i\,\frac{g}{2}\,\bar\psi\left \{ t^a{\slashchar{\cal K}}^b-
 t^b{\slashchar{\cal K}}^a\right \}
+i\,\frac{g}{2}\left \{{\slashchar{\cal H}}^b t^a-
{\slashchar{\cal H}}^a t^b\right \}\psi.
\eeq
{}From eq.~(\ref{slashK}), one can show that
 \beq\label{tslK}
t^a{\slashchar{\cal K}}^b-
 t^b{\slashchar{\cal K}}^a\,=\,i\,f^{abc}{\slashchar {\cal K}}^c
\,=\,-T^c\,{\slashchar {\cal K}}^c.\eeq
It then follows from eq.~(\ref{trkinD}) that $\left (\delta {\cal D}_{ab}
^\psi\right )^\nu_{\nu}$ is a color matrix  in the adjoint representation,
that is, we can write
\beq\label{LDpsi}
\left (\delta {\cal D}_{ab}^\psi\right )^\nu_{\nu}\equiv
-(T^c)_{ab}\delta {\cal D}_c^\psi
\equiv - \delta {\cal D}^\psi_{ab},\eeq
and $\delta{\cal D}^\psi$ thus defined satisfies the
{\it gauge covariant} equation \beq\label{psiwig}
\left [ k\cdot\tilde D_X,\, \delta{\cal D}^\psi(k,X)\right ]&=&i\,\frac{g}{2}
\,T^a\left \{\bar\psi(X)\,\slashchar{{\cal K}}^a(k,X)
-\slashchar{{\cal H}}^a(k,X)\,\psi(X)\right \},\eeq
where $[\tilde D_\mu,\,
O]\equiv\del_\mu O+ig[\tilde A_\mu, O]$ for any
 $O(X)\equiv O^a(X)T^a$.

It is now straightforward to write the equation giving the current density
${\cal J}^\psi_{\rm g}$. From eqs.~(\ref{JgD}) and (\ref{LDpsi}) we have
\beq\label{Jgwig}
{\cal J}^{\psi\,a}_{{\rm g}\,\mu}(k,X)\,=\,k_\mu\,{\rm Tr}\,T^a\,
\delta {\cal D}^\psi(k,X)\,=\,N\,k_\mu\delta {\cal D}_a^\psi(k,X).
\eeq
Then, from eq.~(\ref{psiwig}) we deduce the following
gauge covariant equation for ${\cal J}^\psi_{{\rm g}\,\mu}\equiv
t^a\,{\cal J}^{\psi\,a}_{{\rm g}\,\mu}$:
\beq
\label{kJgpsi}
\left [k\cdot D_X,\,{\cal J}^\psi_{{\rm g}\,\mu}(k,X)\right ]^a&=&
i\,\frac {g}{2}\,N\,k_\mu\left\{\bar\psi(X)\,\slashchar{{\cal K}}^a(k,X)
-\slashchar{{\cal H}}^a(k,X)\,\psi(X)\right \}.
\eeq
 There is, of course, a similar equation for
the density ${\cal J}^5_{\rm g}$ of
the axial color current  induced by fermionic fields
on the hard gluons.

Let us consider now the color current induced by the  gauge fields, $i.e.$
 $j_{\rm g}^A$. Its  density, ${\cal J}_{\rm g}^A$, is related to
  $\delta{\cal D}^A$ by eq.~(\ref{Jg}). From eq.~(\ref{kinDA}), we see that
$\delta{\cal D}_{\,\,\mu\nu}^{A\,ab}$ is symmetric in the Minkovski indices
and belongs to the adjoint representation, that is
\beq
\label{rdefD}
\delta{\cal D}_{\,\,\mu\nu}^{A\,ab}\equiv (T^c)_{ab}
\delta{\cal D}_{\mu\nu}^{A\,c},
\eeq
 with the new functions
$\delta{\cal D}^{a}_{\mu\nu}$ satisfying
\beq
\label{DA}
k\cdot\del_X\,\delta{\cal D}^{A\,a}_{\mu\nu}-\frac {g}{4}f^{abc}
A^\rho_b k^\lambda
\left [\Gamma_{\mu\sigma\lambda\rho}\,\delta{\cal D}^
{A\,\sigma }_{c\nu}+
\Gamma_{\nu\sigma\lambda\rho}\,\delta{\cal D}^{A\,\sigma }_{\mu c}
\right ]\nonumber\\
=\frac{g}{2}\,\Gamma_{\mu\nu\rho\lambda}\left
(\del_X^\sigma A^\rho_a\right )\,k^\lambda (\del_\sigma ^k\Delta).
\qquad\qquad
\eeq
In terms of these functions, the density ${\cal J}^A_{\rm g}$ is
\beq
\label{JgA}
{\cal J}^{A\,a}_{{\rm g}\,\mu}(k,X)= N\left \{\delta{\cal D}_{\mu\nu}
^{A\,a}(k,X)
k^\nu\,-\,k_\mu\delta{\cal D}^{A\,\nu}_{a\,\nu}(k,X)\right \},
\eeq
and it depends only on the particular combinations
$\delta{\cal D}_{\mu\nu}^{A\,a}k^\nu$ and
$\delta{\cal D}^{A\,\nu}_{a\,\nu}$.
The equations satisfied by these quantities are easily obtained from
 eq.~(\ref{DA}). They read
\beq
\label{trD}
(k\cdot\del_X)\delta{\cal D}^{a\,\nu}_{\nu}
-g\,f^{abc}(k\cdot A^b)\delta{\cal D}^{c\,\nu}_{\nu}
+g\,f^{abc}A^\mu_b\left (\delta{\cal D}^{c}_{\mu\nu}k^\nu\right )\nonumber\\
\qquad\qquad\qquad =\,g(d-1)k_\mu\left (\del_X^\nu A_a^\mu\right )
(\del_\nu^k\Delta),
\eeq
and
\beq
\label{Dk}
(k\cdot\del_X)\left (\delta{\cal D}_{\mu\nu}^{a}k^\nu\right )
-g\,f^{abc}(k\cdot A^b)\left (\delta{\cal D}_{\mu\nu}^{c}k^\nu
\right )\qquad\qquad\qquad\qquad\nonumber\\
+\frac {g}{4} f^{abc}\left\{k_\mu(A^b\cdot\delta{\cal D}^{c}\cdot k)+
A_\mu^b(k\cdot\delta{\cal D}^{c}\cdot k)+
k^2A^\nu_b \delta{\cal D}_{\nu\mu}^{c}
+(k\cdot A^b)k^\nu \delta{\cal D}_{\nu\mu}^{c} \right \}\nonumber\\
=\frac {g}{2}(k_\mu k_\nu-g_{\mu\nu}k^2)\left (\del_X^\rho A^\nu_a\right )
(\del_\rho^k\Delta).\qquad\qquad\qquad
\eeq
We have dropped the superscript $A$
on $\delta D$ in order to simplify the writing, and this will be done in
the following equations as long as
no confusion is possible.
Furthermore, we have used compact notations  for tensor products, such as
$A^b\cdot\delta{\cal D}^{c}\cdot k\equiv A_b^\rho\delta{\cal D}
^{c}_{\rho\nu} k^\nu$.

We have to solve eq.~(\ref{Dk}) before eq.~(\ref{trD}).
Indeed, its solution, $\delta{\cal D}_{\mu\nu}^{a}
\,k^\nu$,  enters the l.h.s. of eq.~(\ref{trD}) for
$\delta{\cal D}^{a\,\nu}_{\nu}$.
In order to do so, we first replace the term proportional to
 $k^2\delta{\cal D}^{c}_{\nu\mu}$ in the l.h.s. of eq.~(\ref{Dk})
by the expression deduced from eq.~(\ref{k2D}).
Second, we note that the r.h.s. of eq.~(\ref{Dk}) is transverse with respect to
 $k$, so that
\beq
\label{Dtr}
k^\mu\delta{\cal D}_{\mu\nu}^{a}(k,X)\,k^\nu =0.
\eeq
This allows us to eliminate the corresponding term in the l.h.s.
 of eq.~(\ref{Dk}). We are then left with the equation
\beq\label{Dk1}
(k\cdot\del_X)\left (\delta{\cal D}_{\mu\nu}^{a}\,k^\nu\right )
-\frac{3}{4}g\,f^{abc}(k\cdot A^b)\left (\delta{\cal D}_{\mu\nu}^{c}\,
k^\nu\right )
+\frac {g}{4} f^{abc}k_\mu(A^b\cdot\delta{\cal D}^{c}\cdot k)
\nonumber\\
=\frac {g}{2}(k_\mu k_\nu-g_{\mu\nu}k^2)\left (\del_X^\rho A^\nu_a\right )
(\del_\rho^k\Delta)-\frac{3}{4}g^2f^{abc}(k\cdot A^b)A_\mu^c \Delta.
\eeq
The structure of this equation suggests the decomposition
$\delta{\cal D}^{a}_{\mu\nu}k^\nu\equiv gA_\mu^a\Delta+\delta{\cal D}^
a_\mu$, where $\delta{\cal D}^{a}_\mu$ are new functions which satisfy
\beq\label{Dk2}
(k\cdot\del_X)\delta{\cal D}_{\mu}^{a}
-\frac{3}{4}g\,f^{abc}(k\cdot A^b)\delta{\cal D}_{\mu}^{c}
+\frac {g}{4} f^{abc}k_\mu(A^b\cdot\delta{\cal D}^{c})
=\frac {g}{2}k_\mu k_\nu(\del^\rho_X A^\nu_a)(\del_\rho^k \Delta).
\eeq
By comparing this equation with eq.~(\ref{kinGa}) for the ghosts propagator,
we see that $\delta{\cal D}_{\mu}^{a}=-k_\mu\delta{\cal G}^a$.
Therefore eq.~(\ref{Dk1}) has the
following solution
\beq\label{Dk3}
\delta{\cal D}^{a}_{\mu\nu}k^\nu(k,X)&=& gA_\mu^a(X)\Delta(k)-
k_\mu\delta{\cal G}^{a}(k,X),
\eeq
which, not surprisingly, relates ghost and longitudinal gluon degrees
of freedom.
We use this relation in order to rewrite the
induced current density (\ref{JgA}) as
\beq
\label{3Jg}
{\cal J}^{A\,a}_{{\rm g}\,\mu}(k,X)=gN\Delta(k)A^a_\mu(X)-Nk_\mu\left(
\delta{\cal D}^{A\,\nu}_{a\,\nu}(k,X)+\delta{\cal G}^a(k,X)\right),
\eeq
where the  superscript $A$ has been reintroduced.
The relation (\ref{Dk3}) can also be used to transform eq.~(\ref{trD}) for
 $\delta{\cal D}^{A\,\nu}_{a\,\nu}$ into
\beq\label{3trD}
\left [k\cdot D_X,\,\delta{\cal D}^{A\,\nu}_{\nu}(k,X)\right ]^a=
g(d-1)k_\nu (\del^\rho_X A^\nu_a)(\del_\rho^k\Delta)+gf^{abc}(k\cdot A^b)
\delta {\cal G}^c(k,X).
\eeq

We now consider the density ${\cal J}^A_{\rm b}$ of the total color current
carried by gluonic degrees of freedom, $j^A_{\rm b}= j^A_{\rm g}+
j_G+j_\Omega$. By adding
 eqs.~(\ref{3Jg}), (\ref{4jG}) and (\ref{4jt}), we obtain
\beq\label{4Jb}
{\cal J}^{A\,a}_{{\rm b}\,\mu}(k,X)=-gN(d-2)\Delta(k)A^a_\mu(X)-Nk_\mu\left
\{\delta{\cal D}^{A\,\nu}_{a\,\nu}(k,X)+2\delta{\cal G}^a(k,X)\right\}.
\eeq
The first piece in the right hand side is the tadpole contribution to the
induced current density, including only the $N(d-2)$ transverse gluons.
The second piece contains the contributions involving  the 3-gluon vertex.
In this term, the role of the ghosts is clearly seen: the term
$2\delta{\cal G}^a$ compensates for the two spurious degrees of freedom
 contained in  $\delta{\cal D}^{A\,\nu}_{a\,\nu}$.

The equation satisfied by ${\cal J}^A_{\rm b}$ follows from eqs.~(\ref
{kinGa}) and (\ref{3trD})
\beq
\label{kJ8}
\left [k\cdot  D_X,\, {\cal J}_{{\rm b}\,\mu}^A (k,X) \right ]^a
=-gN(d-2) \left \{\left [k\cdot D_X,\, A_\mu\right ]^a
+k_\mu k_\nu \left(\del_X^\rho A_a^\nu \right )\del_\rho^k \right\}\Delta(k).
\eeq
This equation is  not gauge covariant, in spite of the covariant derivative
in the l.h.s. and the presence in the r.h.s. of the explicit factor
$N(d-2)$ indicating that only the transverse gluons
 contribute. This lack of covariance has its origin in
the noncovariant character of the Wigner functions that we have been using.
Recall, however, that the physically relevant quantity is the induced
current $j^A_{\rm b}$, which is obtained from its density after a $k$
integration. This means that  ${\cal J}^A_{\rm b}$ is determined
only up to a total derivative with respect to $k$.
We use this arbitrariness
in order to construct a gauge covariant density. We define,
in analogy with eq.~(\ref{delS}),
\beq
\label{tilJ}
\acute{\cal J}_{{\rm b}\,\mu}^A(k,X)\equiv
{\cal J}_{{\rm b}\,\mu}^A(k,X)+gN(d-2)(A\cdot\del_k)(k_\mu\Delta(k)).
\eeq
This new density gives the same current as ${\cal J}_{\rm b}^A(k,X)$ after the
$k$-integration, and it satisfies the {\it gauge covariant}
 equation, derived from eq.~(\ref{kJ8}),
\beq
\label{3covJg}
 \left [k\cdot D_X,\,\acute{\cal J}^A_{{\rm b}\,\,\mu}(k,X) \right ]^a=
g\,N(d-2)\,k_\mu\,k^\rho F_{\rho\nu}^a\del^\nu\Delta(k).
\eeq
We recognize here a situation similar to that
encountered in Sec.3.5, in relation with $\delta {\cal S}_A$.
By analogy, we can interpret
eq.~(\ref{tilJ}) as expressing the change  between canonical and kinetic
momenta for the $N(d-2)$ transverse gluons.

This may be seen more explicitly by expressing the current
 density ${\cal J}_{\rm b}\equiv
{\cal J}^{\psi}_{\rm g}+\acute{\cal J}^{A}_{{\rm b}}$ in terms of
 a {\it gauge covariant} Wigner
function, denoted by $\acute{\cal D}(k,X)$.
This is a $(N^2-1)\times (N^2-1)$ color matrix
 whose diagonal elements may be interpreted as the
distribution functions of transverse gluons of each color. In equilibrium, it
reduces to
\beq\label{eqbw}
 \acute{\cal D}_0^{ab}(k)&=&\delta^{ab}\,(d-2)\,\Delta(k),\eeq
since there are $(d-2)$ polarization states for gluons of each color.
Off-equilibrium,
\beq\label{bwig}
\acute{\cal  D}(k,X)&=& \acute{\cal D}_0(k)+\delta \acute{\cal D}(k,X),
\eeq
where   $\delta \acute{\cal D}(k,X)$ is related to the induced current
 $j_{\rm b}=j^A_{\rm b}+j^\psi_{\rm b}$ by
\beq\label{jbw}
j^{a}_{{\rm b}\,\mu}(X)&=& g\int\frac{d^4k}{(2\pi)^4}\,k_\mu\,
{\rm tr}\,T^a\,\delta \acute{\cal D}(k,X),
\eeq
with the trace acting in the color space.
Since $\delta \acute{\cal D}(k,X)$
 belongs to the adjoint representation, it is
completely determined by  eq.~(\ref{jbw}). In line with the previous
analysis, we write
\beq\label{indbwig}
\delta \acute{\cal D}(k,X)&=&\delta \acute{\cal D}_a^A(k,X)\,T^a
+\delta{\cal D}_a^\psi(k,X)\,T^a.
\eeq
where  $\delta{\cal D}^\psi$ satisfies the gauge covariant equation
 (\ref{psiwig}), while  $\delta \acute{\cal D}^A$ obeys an equation analogous
to that for $ \acute{\cal J}^A_{\rm b}$, (\ref{3covJg}),
\beq
\label{bweq}
 \left [k\cdot\tilde D_X,\,\delta \acute{\cal D}^A(k,X) \right ]=
g\,(d-2)\,k^\rho\tilde F_{\rho\nu}\del^\nu\Delta(k),
\eeq
with $\tilde F_{\nu\rho}\equiv  F^a_{\nu\rho}T^a$.

In Feynman gauge, the relation between $\delta \acute{\cal D}^A(k,X)$
and the gluon and the ghost propagators is given by (see eq.~(\ref{4Jb}))
\beq
\delta \acute{\cal D}^A(k,X)&=&-\left
\{\delta{\cal D}^{A\,\nu}_{\,\,\nu}(k,X)+2\,\delta{\cal G}(k,X)\right\}
+g(d-2)(\tilde A(X)\cdot\del_k)\Delta(k).\eeq
If we also notice that,
in the same  gauge, $ \acute{\cal D}_0(k)=- D_{0\,\nu}^{<\,\nu}(k)-
2G_0^<(k)$, we see that $ \acute{\cal D}(k,X)$ may be viewed as resulting
from the replacement of the canonical momentum by the kinetic one
 in
\beq
{\cal D}(k,X)\equiv-\left
\{{\cal D}^{\nu}_{\nu}(k,X)+2{\cal G}(k,X)\right\}.\eeq
That is, $ \acute{\cal  D}(k,X)\approx{\cal D}(k+g\tilde A,X)$ in leading
order,
and $ \acute{\cal D}(k,X)$ may be interpreted as the distribution function
for the transverse gluons.
Of course, the separation of $ \acute{\cal D}(k,X)$ in terms of
 ${\cal D}^\nu_\nu$ and ${\cal G}$ depends upon
the choice of the gauge. However,
equations (\ref{kJgpsi}) and (\ref{3covJg}) for the induced currents,
as well as the corresponding equations (\ref{psiwig}) and (\ref{bweq})
 for the covariant Wigner functions, are independent of gauge fixing,
as shown in Appendix B.

\vskip 2cm
\setcounter{equation}{0}

\section{Kinetic equations for hard particles}

In the previous section, we have derived equations for the
Wigner transforms of the off-equilibrum 2-point functions which are needed in
the calculation of the
induced sources for the average fields. These equations describe the dynamics
of the
plasma particles in the presence of the average fields which may be considered
at this stage as
given background fields. We  analyze now the structure of these equations
 and we write their  solutions for retarded conditions, that is for average
fields which vanish adiabatically as
$X_0\to -\infty$. As we shall see, the equations for the densities of induced
sources take the form of
kinetic equations, that is, of equations for generalized  phase space
distribution functions for the
plasma particles.

\subsection{Equations for the densities of the induced sources}

The equations for the densities of the induced sources  have been already
written
in the previous section, or can be
derived in a straightforward way from the  equations
for the corresponding 2-point functions.  Explicitly, we use
eqs.~(\ref{kinK}), (\ref{Vlas}), (\ref{Spsi}), (\ref{kJgpsi})
and (\ref{3covJg}) and obtain
\beq \label{4kinK} \left (k\cdot D_X\right ) {\cal {\slashchar
K}}(k,X) =-i\frac{g}{2}(d-2)C_f(\Delta(k)+\tilde\Delta(k))\slashchar{k}
\psi(X),
\eeq
\beq
\label{4Vlas}
 \left [k\cdot D_X,\,{\cal J}^A_{{\rm f}\, \mu}(k,X) \right ]^a=
2g\,N_f\,k_\mu\,k^\rho F_{\rho\nu}^a(X)\,\del^\nu\tilde\Delta(k),
\eeq
\beq
\label{4Spsi}
\,\left [k\cdot D_X,\,{\cal J}^\psi_{{\rm f}\, \mu}(k,X) \right ]^a&=&ig\,k_\mu
\,\left \{\bar\psi(X)\,t^a\,\slashchar{{\cal K}}(k,X)
-\slashchar{{\cal H}}(k,X)\,t^a\,\psi(X)\right \}\nonumber\\
&-&i\,\frac {g}{2}\,N\,k_\mu\left\{\bar\psi(X)\,\slashchar{{\cal K}}^a(k,X)
-\slashchar{{\cal H}}^a(k,X)\,\psi(X)\right \},
\eeq
\beq
\label{4kJgpsi}
\left [k\cdot\,D_X,\,{\cal J}^\psi_{{\rm b}\, \mu}(k,X)\right ]^a=
\,i\,\frac {g}{2}\,N\,k_\mu\left\{\bar\psi(X)\,\slashchar{{\cal K}}^a(k,X)
-\slashchar{{\cal H}}^a(k,X)\,\psi(X)\right \},
\eeq
\beq
\label{covJg}
 \left [k\cdot D_X,\,\acute{\cal  J}^A_{{\rm b}\,\,\mu}(k,X) \right ]^a=
g\,N(d-2)\,k_\mu\,k^\rho F_{\rho\nu}^a(X)\,\del^\nu\Delta(k),
\eeq
\beq\label{4kB}
\,\qquad (k\cdot \del_X)\,{\cal B}_{\mu}(k,X) &=&ig\,k_\mu
\,\left \{\bar\psi(X)\,\slashchar{{\cal K}}(k,X)
-\slashchar{{\cal H}}(k,X)\,\psi(X)\right \},\eeq
\beq\label{5kB}
\,\qquad (k\cdot \del_X)\,{\cal B}^5_{\mu}(k,X) &=&ig\,k_\mu
\,\left \{\bar\psi(X)\gamma_5\,\slashchar{{\cal K}}(k,X)
+\slashchar{{\cal H}}(k,X)\,\gamma_5\psi(X)\right \}.\eeq
The equations (\ref{4kinK})--(\ref{5kB}) are all first order partial
 differential equations
and, to be specific, we shall solve them in this section with vanishing initial
conditions at $X_0\to -\infty$. They describe the dynamics of the hard
particles of the plasma in the presence of given background fields
$A_\mu(X)$, $\psi(X)$ and $\bar\psi(X)$ which vanish adiabatically
in the remote past.

Only the real hard
particles which exist in the plasma at thermal equilibrum do contribute
to the induced sources. This is manifest in several ways in the previous
equations. First, in the r.h.s. of eqs.~(\ref{4kinK})--(\ref{4kB}),
all possible vacuum contributions cancel.
For instance, we can use the identity
\beq
\label{SUSY}
\tilde\Delta(k)+ \Delta(k)\equiv
\rho_0(k)[N(k_0)+n(k_0)]=2\pi \delta(k^2)[N(\epsilon_k)+n(\epsilon_k)],
\eeq
where $\epsilon_k\equiv |\vec k|$, in order to see that the right hand
side of eq.~(\ref{4kinK}) vanishes at zero temperature; the
zero temperature contributions to the quark and gluon propagators
$S_0^<(k)$ and  $D^<_0(k)$ are mutually compensating in their sum.
Furthermore, the vacuum contribution to eqs.~(\ref{4Vlas})
and (\ref{covJg}) cancels out, as
$k\cdot F\cdot\del_k [(1/k)\delta(k_0\pm \epsilon_k)]=0$. Thus only
thermal fluctuations contribute to ${\slashchar{\cal K}}$ or ${\cal J}$.
Second, there is a factor $(d-2)$ (eventually implicit) in the r.h.s.
 of all the equations involving bosonic degrees of freedom. As
already dicussed, this reflects the fact that only the   transverse
 gluons effectively contribute to these densities.
Third, the equations are independent of gauge fixing, which again reflects
the decoupling of the spurious degrees of freedom.
This is shown explicitly in Appendix B where the equations for the 2-point
functions
are studied for a  covariant gauge with an arbitrary gauge fixing
parameter $\lambda$.

All the previous equations are covariant under local gauge
transformation of  the mean fields $A_\mu$, $\psi$ and $\bar\psi$.
Therefore the induced densities determined from these equations are gauge
covariant quantities. Specifically, $\slashchar{{\cal K}}$ transforms like
$\psi$ as vector in the fundamental representation,
while ${\cal J}_{\rm f}^A$, ${\cal J}_{\rm f}^\psi$, ${\cal J}_{\rm b}^\psi$,
and $\acute{\cal J}_{\rm b}^A$ transform like $F_{\mu\nu}$
 as vectors in the adjoint representation,  and ${\cal B}$ and ${\cal B}^5$
are scalars.

The similarity between the two equations for the vector and axial
baryonic currents densities reflects the helicity conservation in the
elementary interactions.
The same symmetry is, of course, valid
for the corresponding color currents. The densities for the axial color
currents satisfy equations which can be easily obtained by inserting $\gamma_5$
matrices in the  vector equations (\ref{4Spsi}) and
(\ref{4kJgpsi}), in the same way as in going
from eq.~(\ref{4kB}) to (\ref{5kB}).

Quarks and gluons respond
similarly to soft gauge and fermionic fields. This is apparent in the r.h.s.
of eq.~(\ref{4kinK}), and also in the similarity between eqs.~(\ref{4Vlas})
and (\ref{covJg}).
 The different numerical factors in the right hand sides of these two equations
simply count the relevant degrees of freedom.
In fact, by adding together eqs.~(\ref{4Vlas}) and (\ref{covJg}),
one obtains an equation for the total density
 ${\cal J}^A\equiv
{\cal J}^A_{\rm f}+\acute{\cal J}^A_{\rm b}$
\beq
\label{4JA}
\left [k\cdot D_X,\, {\cal J}^A_{\mu}(k,X) \right ]=
g\,\,k_\mu\,k\cdot F(X)\cdot \del\left ( 2 N_{\rm f}\tilde\Delta(k)+
N(d-2)\Delta(k)\right ).
\eeq
This equation generalizes the linearized Vlasov equation to nonabelian plasmas.
Note that previous attempts to derive such an equation were based
on   approximation schemes  mixing
leading and non leading contributions in $g$ and therefore, from the present
point of view,  not entirely consistent \cite{Heinz83,Elze86}.
 One can also combine
 eqs.~(\ref{4Spsi})
and (\ref{4kJgpsi}) into a single equation for  the total current density
induced by the fermionic fields,
  ${\cal J}^\psi\equiv {\cal J}^\psi_{\rm f}+{\cal J}^\psi_{\rm b},$
\beq
\label{4Jpsi}
\left [k\cdot D_X,\,{\cal J}^\psi_\mu(k,X) \right ]=ig\,k_\mu\,t^a
\,\left \{\bar\psi(X)\,t^a\,\slashchar{{\cal K}}(k,X)
-\slashchar{{\cal H}}(k,X)\,t^a\,\psi(X)\right \}.
\eeq
This equation is similar to the corresponding one in the abelian
case \cite{QED}. In doing  the sum of eqs.~(\ref{4Spsi})
 and (\ref{4kJgpsi}), the typical non-abelian effects cancel; these are
contained for example in the second braces in eq.~(\ref{4Spsi}), and involve
the
 3-gluon vertex leading to gauge field insertions on the  hard gluon lines.
 This kind of cancellation was first noted
by Taylor and Wong \cite{Taylor90} in relation with the HTL's for amplitudes
involving one pair of quarks and any number of soft gluons (albeit their
proof is only explicit up to three external gluons).

The right hand sides of the kinetic equations (\ref{4kinK})--(\ref{5kB})
contain a  factor $\delta(k^2)$ showing that the mass shell
condition for the hard particles is not affected by their interactions with the
mean fields.  This
allows us, in the next subsection, to transform the equations for the current
densities into
kinetic equations for generalized distribution functions.

\subsection{Explicit solutions}

 The solution of eq.~(\ref{4kinK}) for
$\slashchar{{\cal K}}$ has the following general structure:
\beq
\label{solK}
\slashchar{{\cal K}}(k,X)=2\pi\delta(k^2)\left\{ \theta(k_0)
\slashchar{\Lambda}^+
(\vec k, X)+\theta(-k_0)\slashchar{\Lambda}^-(-\vec k,X) \right\}.
\eeq
This structure is characteristic to all the
other densities that we compute below. Eq.~(\ref{4kinK}) can be
turned  into  equations for the functions $\slashchar{\Lambda}^\pm$.
It is easily verified that  $\slashchar{\Lambda}^+=
\slashchar{\Lambda}^-\equiv \slashchar{\Lambda}$, where $\slashchar{\Lambda}$
 satisfies
\beq
\label{Lam}
(v\cdot D_X)\slashchar{\Lambda}(\vec k, X)=
-i\frac{g}{2}(d-2)C_f[N(\epsilon_k)+n(\epsilon_k)]\slashchar{v}\psi(X),
\eeq
with the initial condition $\slashchar{\Lambda}(\vec k, X)
\to 0$ for $X_0\to -\infty$.
Here $\slashchar{v}\equiv v^\mu\gamma_\mu$,
with the velocity 4-vector $v^\mu\equiv (1,\,\vec v)$ and $\vec v\equiv
\vec k/\epsilon_k$.  This equation  describes fluctuations where, under the
action of a soft fermionic mean field,
 hard quarks are converted into hard gluons and  vice versa. The quantity
 $\slashchar{\Lambda}(\vec k, X)$ may be viewed as a generalized one body
density matrix mixing
fermionic and bosonic degrees of freedom. The covariant line derivative in the
l.h.s. of
eq.~(\ref{Lam}) is remininscent of the familiar drift term of elementary
kinetic equations; the
presence of a covariant derivative
steems from the fact that  $\slashchar{\Lambda}$ behaves as
$\psi$ under gauge transformations.

The solution to eq.~(\ref{Lam}) can be written by  introducing
the  parallel transporter along a straight line  joining $x$ and $y$  \beq
\label{U}
 U(x,y)=P\exp\{ -ig\int_\gamma dz^\mu A_\mu(z)\},
 \eeq
 where $\gamma$ is the path parametrized by $z^\mu(s)=y^\mu+(x^\mu-y^\mu)s$
with $0\le s\le1$, and $P$ is the path-ordering operator. Thus,
\beq
\label{Ufdt}
U(X,X-vu)=P\exp\{-ig\,u\int_0^1ds \,v\cdot A(X-vu(1-s))\}.
\eeq
In order to solve eq.~(\ref{Lam}), we take $A_\mu\equiv A_\mu^a t^a$
in the fundamental
representation. We have then
\beq
\label{LkX}
\slashchar{\Lambda}(\vec k, X)=-i\frac{g}{2}(d-2)
\,C_f[N(\epsilon_k)+n(\epsilon_k)]
\,\slashchar{v}\int_0^\infty du U(X,X-vu) \psi(X-vu).
\eeq
In order to verify that (\ref{LkX}) is the correct solution to eq.~(\ref{Lam})
 we may use the following formula for the line-derivative of the parallel
transporter
\beq
\label{derU}
(v\cdot D_X)\,U(X,Y)\Big |_{Y=X-vu}=\,0
\eeq
together with the identity $(v\cdot \del_X)f(X-vu)=
-(\del /\del u)f(X-vu)$.

According to eq.~(\ref{4eind}), the fermionic induced source is obtained by
integrating the density $\slashchar{{\cal K}}(k,X)$  over $k$.
By using eqs.~(\ref{solK}) and (\ref{LkX}), we get
\beq
\label{eta-psi} \eta^{ind}(X)&=&
 -i\omega_0^2\int\frac{d\Omega}{4\pi}\,\slashchar{v}
\int_0^\infty duU(X,X-vu)\psi(X-vu),
\eeq
with $\omega_0^2\equiv(d-2)C_fg^2 T^2/16$.

The equation (\ref{4Jpsi}) for  ${\cal J}^\psi$ can be integrated in a similar
fashion. One gets
\beq\label{DJp}
{\cal J}_{\mu\,a}^\psi(k,X)=igv_\mu^\prime\int_0^\infty du\,
\tilde U_{ab}(X,X-v^\prime u)\qquad\qquad\qquad\qquad\qquad\nonumber\\
\left\{
\bar\psi(X-v^\prime u)t^b\slashchar{{\cal K}}(k,X-v^\prime u)
-\slashchar{{\cal H}}(k,X-v^\prime u)t^b\psi(X-v^\prime u)\right\},
\eeq
where $v^\prime_\mu\equiv k_\mu/k_0=(1, \pm \vec v)$ for $k_0=\pm \epsilon_k$.
$\tilde U_{ab}(X,X-v^\prime u)$
is the parallel transporter in the adjoint representation, $i.e.$,
eq.~(\ref{Ufdt}) with $A_\mu\to \tilde A_\mu\equiv A_\mu^a T^a$.
The function  $\slashchar{{\cal H}}$ has a structure similar to (\ref{solK}):
\beq
\label{solH}
\slashchar{{\cal H}}(k,X)\,=\, 2\pi\delta(k^2)\left\{ \theta(k_0)
\bar{\slashchar{\Lambda}}
(\vec k, X)+\theta(-k_0)\bar{\slashchar{\Lambda}}(-\vec k,X) \right\},
\eeq
where $\bar{\slashchar{\Lambda}}(\vec k,X)\equiv {\slashchar{\Lambda}
}^\dagger(\vec k,X)\gamma^0$.

By integrating  eq.~(\ref{DJp}) over $k$ in order to get the induced current
$j_\mu^\psi(X)$, we  obtain, after simple manipulations,
\beq\label{Djp1}
j_{\mu}^{\psi\,a}(X)\,=\,g\int\frac{d^3k}{(2\pi)^3}\,v_\mu\,\left[ \delta
 n_+^{\psi\,a}(\vec k,X)-\delta n_-^{\psi\,a}(\vec k,X)\right],
\eeq
where
\beq\label{npsi}
\delta n_\pm^{\psi\,a}({\vec k},X)\,=\,\pm \,i\frac{g}{2\epsilon_k}
\int_0^\infty du\,
\tilde U_{ab}(X,X-vu)\qquad\qquad\qquad\qquad\qquad\nonumber\\
\left\{
\bar\psi(X-vu)t^b\slashchar{\Lambda}(k,X-vu)
-\bar{\slashchar{\Lambda}}(k,X-vu)t^b\psi(X-vu)\right\}
\eeq
are fluctuations in the quark and antiquark color densities induced by
the fermionic mean fields.
By replacing $\slashchar{\Lambda}$ by its explicit expression (\ref{LkX}),
 we can evaluate the integral over $k\equiv \epsilon_k$ in (\ref{Djp1})
and get \beq\label{b8}
j_\mu^\psi(X)&=&g\,\omega_0^2 t^a\,\int\frac{d\Omega}{4\pi}\,v_\mu
\int_0^\infty  du\int_0^\infty
 du^\prime \tilde U_{ab}(X,X-vu)\nonumber\\
&&\biggl\{\bar\psi(X-vu)t^b\slashchar{v}
U(X-vu,X-v(u+u^\prime))\psi(X-v(u+u^\prime))\nonumber\\
&&+\bar\psi(X-v(u+u^\prime))\slashchar{v}
U(X-v(u+u^\prime),X-vu)t^b\psi(X-vu)\biggr\}.
\eeq
We recognize in this expression the two terms which correspond to the two
possible time
orderings of the scattering processes between the hard particles and the soft
fermionic fields.
One can obtain a more compact expression
by performing simple changes of variables.  Setting $s=u+u^\prime$ and $t=u$
one transforms the first integrals over $u,u^\prime$ into
\beq
\int_0^\infty dt\int_t^\infty ds\,\tilde U_{ab}(X,X-vt)
\bar\psi(X-vt)\slashchar{v} t^b U(X-vt,X-vs)\psi(X-vs).
\eeq
In the second integrals, we set $s=u$ and $t=u+u^\prime$, and obtain
\beq
\int_0^\infty dt \int_0^t ds\,\tilde U_{ab}(X,X-vt)
\bar\psi(X-vt)\slashchar{v} U(X-vt,X-vs)t^b\psi(X-vs).
\eeq
By using the identity
\beq
\tilde U_{ab}(X,X-vt)\,t^b&=&U(X-vt,X)\,t^a\,U(X,X-vt),\eeq
together with the group  property of the parallel transporter,
\beq
U(X-vt,X-vs)=U(X-vt,X)U(X,X-vs),\eeq
we finally obtain
\beq\label{4jpsi}
j_\mu^\psi(X)&=&g\,\omega_0^2\, t^a\int\frac{d\Omega}{4\pi}\,v_\mu\int_0^\infty
dt\int_0^\infty ds\nonumber\\ &&\qquad
\bar\psi(X-vt)\slashchar{v}
U(X-vt,X)\,t^a\,U(X,X-vs)\psi(X-vs).
\eeq

We consider now the equations (\ref{4kB})  and (\ref{5kB}) for the
 baryonic currents. The solution to eq.~(\ref{4kB}) is of the form
  \beq
{\cal B}_{\mu}(k,X)=2\,k_\mu\, (2\pi \delta(k^2))
\left \{\theta (k^0)\delta n_+({\vec k},X) + \theta (-k^0) \delta n_-
(-{\vec k},X)\right \},
\eeq
where  $\delta n_+({\vec k},X)$ and  $\delta n_-({\vec k},X)$
are (color singlet)
fluctuations in the quark and antiquark number densities.
These satisfy
\beq(v\cdot \del_X) \,\delta n_\pm({\vec k},X) =\pm i\frac{g}{2\epsilon_k}
\left [\bar\psi(X){\slashchar \Lambda}({\vec k},X)-\bar
{\slashchar \Lambda}({\vec k},X)\psi(X)\right ].
\eeq
This equation is solved
similarly to the previous one for ${\cal J}^\psi$ (see eq.~(\ref{npsi})).
Then we can calculate the baryonic current
\beq\label{bpsi}
b_\mu(X)\,=\,\int\frac{d^4k}{(2\pi)^4}\,{\cal B}_\mu(k,X)
\,=\,\int\frac{d^3k}{(2\pi)^3}\,v_\mu\,\left[ \delta
 n_+(\vec k,X)-\delta n_-(\vec k,X)\right],
\eeq
and obtain
\beq
\label{4bar}
b_\mu(X)&=&g\,\omega_0^2 \int\frac{d\Omega}{4\pi}\,v_\mu \int_0^\infty
dt\int_0^\infty ds \nonumber\\
&&\qquad\bar\psi(X-vt)\slashchar{v}
U(X-vt,X-vs)\psi(X-vs).
\eeq
This is, of course, similar to $j^\psi_\mu$, eq.~(\ref{4jpsi}), but without the
color
matrices $t^a$. In a complete analoguous way, we
also obtain the axial baryonic current
\beq
\label{5bar}
b_\mu^5(X)&=&g\,\omega_0^2 \int\frac{d\Omega}{4\pi}\,v_\mu \int_0^\infty
dt\int_0^\infty ds \nonumber\\
&&\qquad\bar\psi(X-vt)\gamma_5 \slashchar{v}\,
U(X-vt,X-vs)\psi(X-vs).
\eeq

We turn now to the color currents  induced by soft gauge fields.
Eq.~(\ref{4Vlas}) for ${\cal J}^A_{\rm f}$ implies
  \beq\label{cfluct}
{\cal J}^A_{{\rm f}\,\,\mu}(k,X)=2k_\mu\,N_f\,t^a (2\pi \delta(k^2))
\left \{\theta (k^0)\delta n^{A\,a}_+({\vec k},X) + \theta (-k^0) \delta
 n^{A\,a}_- (-{\vec k},X)\right \},
\eeq
where $\delta n_\pm^A\equiv\delta n^{A\,a}_\pm t^a$ are
fluctuations in the quark color
densities. These fluctuations satisfy (with
$\epsilon_k\equiv |\vec k|$)
\beq\label{clvlas}
\left[ v\cdot D_X,\delta n_\pm^A({\vec k},X)\right]=\mp g\,\vec
v\cdot\vec E(X)\frac{dn(\epsilon_k)}{d\epsilon_k}.
\eeq
Here $\vec E(X)\equiv \vec E^a(X)t^a$ is the average chromoelectric field,
$E_a^i\equiv F^{i0}_a$; possible chromomagnetic contributions cancel
because the equilibrium
distribution $n(\epsilon_k)$ is isotropic.
In the abelian case, this equation coincides with the linearized Vlasov
equation. Here, the gauge field not only modifies the motion of the
particle, through the color electric field that it generates,
but it also induces a ``precession'' of the densities in color space,
through the covariant derivative on the l.h.s.
The solution of eq.~(\ref{clvlas}) is
\beq\label{vlsol}
\delta n_\pm^A(\vec k,X)=\mp g\int_0^\infty du\, U(X,X-vu)
\vec v\cdot\vec E(X-vu)\,U(X-vu,X)\frac{d n(\epsilon_k)}{d\epsilon _k}.
 \eeq

Similarly, eqs.~(\ref{covJg}) or (\ref{bweq})
describe oscillations in the gluon
color densities  $\delta N^A\equiv\delta N^A_a t^a$. They  satisfy an
equation similar to (\ref{clvlas}), viz.
\beq\label{clvl}
\left[ v\cdot D_X,\delta N^A({\vec k},X)\right]=- g\vec
v\cdot\vec E(X)\frac{dN(\epsilon_k)}{d\epsilon_k}.
\eeq
The  total induced current $j^A=j^A_{\rm f}+j^A_{\rm b}$ can be expressed
in terms of these color density fluctuations. Thus
\beq\label{jfa}
j_{\rm f\,\mu}^A(X)=gN_{\rm f}\int\frac{d^3k}{(2\pi)^3}\,v_\mu\,\left[ \delta
 n_+^A(\vec k,X)-\delta n_-^A(\vec k,X)\right],
\eeq
and
\beq\label{jba}
j_{\rm b\,\mu}^A(X)=g(d-2)N\int\frac{d^3k}{(2\pi)^3}\,v_\mu\,\delta
N^A(\vec k,X).
\eeq
Then, by using eq.~(\ref{vlsol}) for $\delta n^A_\pm(\vec k,X)$ and a similar
expression for $\delta N^A(\vec k,X)$, we obtain
\beq
\label{4jA}
j^{A}_\mu(X)&=&3\,\omega^2_p\int\frac{d\Omega}{4\pi}
v_\mu\int_0^\infty du\, U(X,X-vu)\,F_{0\,j}(X-vu)v^j \,U(X-vu,X),\eeq
where $\omega^2_p\equiv((d-2)N+N_{\rm f})g^2
 T^2/18$ is  the plasma frequency.

As a last remark, we note that the final expressions for the induced sources
contain no explicit
 reference to the nature of the hard particles involved. This information is
all contained in the
explicit expressions of
  the two frequencies $\omega_0$ and $\omega_p$, which are obtained as  simple
momentum integrals
involving the statistical occupation factors $N(k)$ and $n(k)$. These
expressions may be easily
generalized to the case of systems with a non
vanishing chemical potential $\mu$:
\beq\label{neww} \omega_0^2=g^2C_f\frac{d-2}{16}\left
(T^2+\frac{\mu^2}{\pi^2}\right),\qquad
\omega^2_p=g^2\left\{\frac{1}{18}\left((d-2)N+{N_f}\right)T^2\,+\,
\frac{N_f}{6}\frac{\mu^2}{\pi^2}\right\}.\eeq

\vskip 2cm
\setcounter{equation}{0}

\section{Mean fields dynamics}

In Sec.3, we have derived consistent equations of motion
for the  2-point functions which describe, at leading order in $g$,
the response of the plasma to weak and slowly varying fields. By solving
these equations in Sec.4, we have expressed the induced sources,
 which enter the mean fields
equations, in terms of the average fields themselves, thereby closing
the system which determines  the dynamics of the mean fields.

The mean field equations,
\beq\label{4avpsi}
i\slashchar{D} \psi(X)=\eta(X)+\eta^{ind}(X),
\eeq
and
\beq
\label{4avA}
\left [\, D^\nu,\, F_{\nu\mu}(X)\,\right ]^a
-g  \bar\psi (X)\gamma_\mu t^a \psi(X)
=\,j_\mu^a(X)+j_\mu^{ind\, a}(X),
\eeq
have a few  remarkable properties, which are worth recalling.

(a) As shown in Sec.4.1, the induced sources are covariant under local
gauge transformations of the mean fields. Specifically, $\eta^{ind}$
transforms like $\psi$, while $j^{ind}$ transforms
 as $F_{\mu\nu}^a$.
It follows that eqs.~(\ref{4avpsi}) and (\ref{4avA}) are gauge covariant
if the external sources transform covariantly. This
covariance is
that of the Euler-Lagrange equations of motion deduced from
the {\sl classical} QCD action, ($ i.e.$, the tree level action, eq.~(\ref
{QCD}), but without the ghosts and the gauge fixing terms).
It is noteworthy to recover gauge covariance at the level of eqs.~(\ref
{4avpsi})--(\ref{4avA}), where  corrections to the tree level dynamics
are taken into account through the induced sources.

(b) The  derivatives of the mean fields in the l.h.s.
of eqs.~(\ref{4avpsi}) and (\ref{4avA})
are of the same order of magnitude as the induced sources in the
r.h.s. That is, $\eta^{ind}\sim gT\psi\sim \del_X\psi$ and
$j^A\sim gTF\sim \del_X F$, as can be readily verified on the
explicit expressions given in eqs.~(\ref{eta-psi}) and (\ref{4jA}).
Thus, the propagation of soft modes is nonperturbatively
renormalized by the medium, reflecting the
collective character of the long wavelength excitations.

(c) For average fields whith maximum strength,
i.e., $F\sim gT^2$ and $\bar\psi\psi\sim gT^3$,
all the terms appearing in eqs.~(\ref{4avpsi})--(\ref{4avA}) are of the
same order of magnitude. The equations  are then nonlinear,
due to the covariant derivatives  in their l.h.s.,
as well as those which enter the structure of the induced sources.
This
is the only source of nonlinearity, showing that the limits on the
 fields strength we are using are equivalent to neglecting all genuine
non linear dynamical effects of the fields.

(d)  Eqs.~(\ref{4avpsi}) and (\ref{4avA})  are nonlocal,
which is a manifestation
of the fact that the ultrarelativistic QCD plasma is a dispersive polarizable
medium.

 For vanishing external sources, the mean field equations describe the
normal modes of the plasma, or equivalently its soft quasiparticles. These
equations have been solved only in the case where the gauge fields are weak,
that is in the linear, abelian, regime. Then the oscillations of the
fermion and gauge fields decouple, and explicit dispersion relations can
be obtained \cite{Kalashnikov80,Weldon82a,BBS92,BO}. In general,
quark and
gluon modes mix, and the non linearity of the equations make them difficult to
solve. Leaving aside this difficult problem, we note that the mean field
equations summarize a lot of information about the soft
quasiparticles, such as their effective propagators, and the effective vertices
which describe their mutual interactions. We show in the next subsections
how these can be  obtained from the induced sources, and derive the simple Ward
identities that they satisfy.

\subsection{Hard thermal loops from induced sources}

As pointed out in Sec.2.2, the knowledge of the induced sources as
functionals of the fields allows us to derive, by functional
differentiation, the one particle irreducible amplitudes for the
corresponding fields. We  use this procedure now, in order to
construct self-energies and proper vertices for soft quasiparticles
valid at leading order in $g$. In doing this, we use boundary conditions such
that the mean fields vanish at $X_0\to -\infty$, as  in the
previous section. The  resulting  amplitudes will be causal with respect to
the time argument in the induced source.

In order to proceed in a systematic way, it is convenient to introduce
retarded and advanced Green's functions for the covariant line derivative
$v\cdot D_X$, denoted as  $G_{ret}(X,Y;v)$ and  $G_{adv}(X,Y;v)$, respectively.
Thus,  $G_{ret}(X,Y;v)$ is defined by
\beq\label{Gret}
i\,(v\cdot D_X)\,G_{ret}(X,Y;v)=\delta^{(4)}(X-Y),
\qquad G_{ret}(X,Y;v)=0\,\,\,{\rm for}\,\,X_0<Y_0,
\eeq
and  $G_{adv}(X,Y;v)$ satisfies a similar equation, but $ G_{adv}(X,Y;v)=0$ for
$X_0>Y_0$. The solution to  eq.~(\ref{Gret}) may be written as
\beq\label{GR}
G_{ret}(X,Y;v)&=&-i\,\theta (X^0-Y^0)\,\delta^{(3)}
\left({\vec X}-{\vec Y}-{\vec v}(X^0-Y^0)
\right )U(X,Y)\nonumber\\
&=&-i\,\int_0^\infty du\,\delta^{(4)}(X-Y-vu)\,U(X,Y),\eeq
where $U(X,Y)$ is the parallel transporter (\ref{U}). The use of this Green's
function is appropriate for the study of the response of the plasma to
perturbations which vanish as $X_0\to -\infty$. One could also consider
perturbations which vanish adiabatically when $X_0\to \infty$. Then, the
induced currents would be expressed in terms of the advanced Green's function,
\beq\label{GA}
G_{adv}(X,Y;v)&=&i\,\int_0^\infty du\,\delta^{(4)}(X-Y+vu)\,U(X,Y),\eeq
which is related to $G_{ret}$ by $G_{adv}(X,Y;v)=G^\dagger_{ret}(Y,X;v)$, where
the Hermitian conjugaison refers to color indices.
For $A_\mu^a=0$, the Fourier components of $G_{ret}$ an $G_{adv}$ are simply
$G_{ret}(P)=(v\cdot P+i\eta)^{-1}=
G_{adv}^{\,\,*}(P)$,
with $\eta\to 0^+$.

According to equations like (\ref{pisib})
or (\ref{Vqg}), $\eta^{ind}$ acts as a
generating functional for all the proper vertices having at least one
pair of external fermion lines and any number $N$ of bosonic lines.
In the present approximation, $\eta^{ind}$ is given by eq.~(\ref{eta-psi}),
which can  also be written as
\beq
\label{5eta} \eta^{ind}(X)&\equiv&  \int d^4Y\delta \Sigma^A(X,Y) \psi(Y).
\eeq
Since $\eta^{ind}$ is linear in $\psi$, the only amplitudes
which do not vanish at the
level of the present approximation involve in fact no more than
 one pair of external fermions.
The kernel $\delta \Sigma^A$  may be identified with the self energy
of a  soft fermion  propagating in  a background gauge field
 (see   eq.~(\ref{pisib})).
In order to get its explicit form, we rewrite  eq.~(\ref{eta-psi}) for
$\eta^{ind}$  as
\beq\label{etaR}
\eta^{ind}(X)&=&\omega_0^2\int\frac{d\Omega}{4\pi}
\,{\slashchar{v}}\int d^4Y\,G_{ret}(X,Y;v)\,\psi(Y).\eeq
Then, obviously,
\beq
\label{SigX}
\delta\Sigma^A(X,Y)&=&\omega_0^2\int\frac{d\Omega}{4\pi}
\,{\slashchar{v}}\,G_{ret}(X,Y;v),
\eeq
which is similar to the corresponding abelian expression \cite{QED}. The only
manifestation of the underlying gauge group is the presence of the
color factor $C_f$ in the expression of $\omega_0$, together
with the non-trivial color structure of $G_{ret}$.
  Eq.~(\ref{GR}) implies $\delta\Sigma^A(X,Y)\,=\,0$
for $X^0<Y^0$, reflecting the retarded boundary conditions in eq.~(\ref{etaR})
for $\eta^{ind}(X)$.
For $A=0$, eq.~(\ref{SigX}) gives the well-known HTL
expression for the soft quark (retarded) self energy \beq \label{Sig}
\delta\Sigma(P) =\omega_0^2\int\frac{d\Omega}{4\pi} \frac{\slashchar{v}}
{v\cdot P+i\eta}.
\eeq
Note that, because $P\sim gT$, the correction (\ref{Sig})
 to the quark inverse propagator is of the same order
of magnitude as the tree-level amplitude ($S_0^{-1}(P)\sim {\slashchar
P}\sim gT$). It can be used to construct the equilibrium propagator
for a soft quark, correct at leading order in $g$: ${}^*S^{-1}(P)=
-{\slashchar P}+\delta\Sigma(P)$ \cite{Braaten90b}.

Because eq.~(\ref{SigX}) is valid for arbitrary (soft) background
gauge fields, we can use it to derive - by successive functional
differentiations - expressions for
 all amplitudes involving one fermion pair and any number
of soft gluons. To be specific, we define the correction to the amplitude
between a quark pair and $N$ soft gluons by
\beq\label{2fNg}
g^N\,\delta \Gamma^{a_1...a_N}_{\mu_1...\mu_N}(X_1,...,X_N;Y_1,Y_2)&=&
\frac {\,\,\,\,\delta^N}{\delta A_{a_N}^{\mu_N}(X_N)...\delta A_{a_1}
^{\mu_1}(X_1)}\,\left(\delta\Sigma^A(Y_1,Y_2)\right).\eeq
In doing these differentiations, we use identities like
\beq\label{derG}
\frac{\delta\,G_{ret}(X_1,X_2;v)}{\delta A_{a}^\mu(Y)}\,=\,g\,v_\mu\,
G_{ret}(X_1,Y;v)\,t^a\,G_{ret}(Y,X_2;v),\eeq
and similarly for $G_{adv}$, which follow from eqs.~(\ref{GR}) and (\ref{U}).
The normalization we choose for the amplitudes (\ref{2fNg}) is such that
$\delta\Gamma$  depends on $g$ only through $\omega_0^2$.
For all these amplitudes, $Y_1^0$ is the largest time, while $Y_2^0$ is
the smallest one. The relative chronological ordering of the $N$ gluon
lines is arbitrary, and, in fact, the amplitudes are totally symmetric
under their permutations. Up to minor changes due to the color algebra, all
the amplitudes obtained in this way coincide with the corresponding abelian
amplitudes \cite{QED}.

For the sake of illustration, we give now the explicit expressions for the
amplitudes involving  a quark pair and one or two gluons. The first case
corresponds to
the quark-gluon vertex correction in the presence of a background gauge
field $A_\mu$ (see eq.~(\ref{Vqg}))
\beq
\label{delG}
\delta\Gamma_{\,\,\mu}^{A\,a}(X;Y_1,Y_2)&=&\omega_0^2\gamma^\nu
\int\frac{d\Omega}{4\pi}\,v_\mu
v_\nu\,G_{ret}(Y_1,X;v)\,t^a\,G_{ret}(X,Y_2;v),\eeq
where the time arguments satisfy $Y_1^0\ge X^0\ge Y_2^0$. For $A=0$,
 eq.~(\ref{delG}) gives the HTL correction to the quark-gluon vertex
\cite{Frenkel90}, with appropriate boundary conditions.
Defining the Fourier transform of  $\delta\Gamma^a_\mu$ by
\beq\label{GMF}
\lefteqn{(2\pi)^4\,\delta^{(4)}(P+K_1+K_2)\,\delta\Gamma^a_{\mu}(P;K_1,K_2)\equiv
\qquad\qquad}\nonumber\\
& &\int d^4X d^4Y_1 d^4Y_2 \,e^{i(P\cdot X+ K_1\cdot Y_1+ K_2\cdot Y_2)}
\delta\Gamma^{a}_{\mu}(X;Y_1,Y_2),\eeq
 we get
\beq\label{4gamma}
\delta\Gamma_{\mu}^a(P;K_1,K_2)\,=\,-t^a
\omega_0^2\gamma^\nu \int\frac{d\Omega}{4\pi}\frac{v_\mu\,
v_\nu}{(v\cdot K_1+i\eta)(v\cdot K_2-i\eta)}\equiv t^a\delta\Gamma_\mu
(P;K_1,K_2).\nonumber\\\eeq
As the external  momenta are  all  of order $gT$,
 $g\delta \Gamma_\mu$ is of the same order as the bare vertex $g\gamma_\mu$.
Thus, the effective quark-gluon vertex at leading order in $g$ is
$g\,t^a\,{}^* \Gamma_\mu$, where ${}^* \Gamma_\mu\equiv \gamma_\mu + \delta
 \Gamma_\mu$ \cite{Braaten90b}.

The second example is that of the vertex between a quark pair and two gluons,
$g^2\,{}^* \Gamma_{\mu\nu}^{ab}$.
This vertex does not exist at tree level, and in leading order
it arises entirely from the hard thermal loop,
${}^* \Gamma_{\mu\nu}^{ab}=\delta\Gamma_{\mu\nu}^{ab}$, with
\newpage
 \beq\label{2f2g}
\delta\Gamma_{\mu\nu}^{ab}(X_1,X_2;Y_1,Y_2)&=&\omega_0^2\gamma^\rho
\int\frac{d\Omega}{4\pi}\,v_\mu v_\nu v_\rho\nonumber\\
&&\biggl
\{G_{ret}(Y_1,X_1;v)\,t^a\,G_{ret}(X_1,X_2;v)\,t^b\,G_{ret}(X_2,Y_2;v)\nonumber\\
&&\,\,+G_{ret}(Y_1,X_2;v)\,t^b\,G_{ret}(X_2,X_1;v)\,t^a\,G_{ret}(X_1,Y_2;v)
 \biggr \}.\nonumber\\\eeq
 For $A=0$, defining the Fourier transform as in eq.~(\ref{GMF}), we obtain
\beq\label{2f2gG}
\delta\Gamma_{\mu\nu}^{ab}(P_1,P_2;K_1,K_2)&=&
-\omega_0^2\gamma^\rho \int\frac{d\Omega}{4\pi}\frac{v_\mu
\,v_\nu\,v_\rho }{(v\cdot K_1+i\eta)(v\cdot K_2-i\eta)}\nonumber\\
&&\left\{ \frac {t^at^b}{v\cdot (K_1+P_1)+i\eta}+
 \frac {t^bt^a}{v\cdot (K_1+P_2)+i\eta}\right\},\eeq
which, when summed over the color indices, coincides with the expressions
given
in refs.\cite{Braaten90b,Braaten92}.

Alternatively, we can derive the amplitudes (\ref{2fNg})
from the expression (\ref{4jpsi}) for the induced current
$j^\psi(X)\equiv j^\psi_{\rm f}(X)+j^\psi_{\rm b}(X)$, but,
this time, the boundary conditions will be different, as the time
argument of $j^\psi(X)$ is now the largest one. We first use
the definitions (\ref{GR}) and (\ref{GA}) to rewrite $j^\psi(X)$,
eq.~(\ref{4jpsi}), as
\beq\label{jfRA}
j_\mu^{\psi}(X)&=&g t^a\,\omega_0^2 \int\frac{d\Omega}{4\pi}\,v_\mu
\int d^4Y_1d^4Y_2\nonumber\\&&
 \bar\psi(Y_1) \,{\slashchar{v}}\,G_{adv}(Y_1,X;v)\,t^a\,
G_{ret}(X,Y_2;v)\,\psi(Y_2).\eeq
The resulting correction to the quark-gluon vertex is (recall the
second equality in eq.~(\ref{Vqg}))
\beq\label{GRA}
\delta\Gamma_{\,\,\mu}^{A\,a}(X;Y_1,Y_2)&=&\omega_0^2\gamma^\nu
\int\frac{d\Omega}{4\pi}\,v_\mu
v_\nu\,G_{adv}(Y_1,X;v)\,t^a\,G_{ret}(X,Y_2;v),\eeq
where now the time arguments satisfy $X^0\ge max(Y_1^0,Y_2^0)$, the
chronological order of $Y_1$ and $Y_2$ being arbitrary (compare, in this
respect, with eq.~(\ref{delG}) above).
 For $A=0$, we have
\beq\label{4gamma2}
\delta\Gamma_{\mu}^a(P;K_1,K_2)&=&-t^a
\omega_0^2\gamma^\nu \int\frac{d\Omega}{4\pi}\frac{v_\mu
\,v_\nu}{(v\cdot K_1-i\eta)(v\cdot K_2-i\eta)},\eeq
which differs from (\ref{4gamma}) solely by the $i\eta$'s in the denominators
reflecting the respective boundary conditions. If we further differentiate the
$\delta\Gamma_{\mu}^{A\,a}(X_1;Y_1,Y_2)$ of
 eq.~(\ref{GRA}) with respect to $A_\mu$,
we generate amplitudes of the type (\ref{2fNg}), in which $X_1^0$ is  the
largest time.

All the amplitudes involving only soft gluons may be derived from the
induced current $j^A$. In the present approximation, $j^A$ is given
by eq.~(\ref{4jA}), which, after some algebraic
manipulations, can be cast into the form
\beq\label{4ja}j^{A\,a}_\mu(X)
&=&-3\,\omega^2_p A_0^a(X)g_{0\mu}+
3\,i\,\omega^2_p\int\frac{d\Omega}{4\pi}\,v_\mu
\int d^4Y\, \tilde G_{ret}^{ab}(X,Y)(v\cdot\dot A^b(Y)),\eeq
where $\tilde G_{ret}$ is the retarded Green's function (\ref{GR}) written in
the adjoint representation ($i.e.$, with $U\to \tilde U$),
while  $\dot A\equiv \del_0 A$. By successive differentiations
 with respect to $A$ in $j^A_\mu(X)$, eq.~(\ref{4ja}), we derive equilibrium
amplitudes with any number of soft gluon external  lines,
which are causal with respect to $X_0$.
 For example, by using eq.~(\ref{pisia}) one gets,
after a simple calculation, the well-known expression of the (retarded)
soft gluon polarization tensor:
\beq
\label{htlpi}\delta\Pi^{ab}_{\mu\nu}(P)=
3\,\omega_p^2\,\delta^{ab}
\left \{-\delta^0_\mu\delta^0_\nu \,+\,P^0 \int\frac{d\Omega}{4\pi}
\frac{v_\mu\, v_\nu} {v\cdot P+i\eta}\right\}.
\eeq
 As  $P\sim gT$, this is of the same order as the tree-level inverse
propagator, $D_0^{-1}(P)\sim P^2\sim g^2T^2$. Thus, in leading
order, the effective propagator for soft gluons is ${}^*D_{\mu\nu}^{-1}(P)
\equiv D_{0\,\mu\nu}^{-1}(P)\,+\,\delta\Pi_{\mu\nu}(P)$
 \cite{Pisarski89,Braaten90b}. Note that the constant term in eq.~(\ref{htlpi})
stems from the first
 term (linear in $A_0$) of the  expression eq.~(\ref{4ja}) for $j^A$;
this term does not contribute
 to amplitudes with $N\ge 3$ external lines.
As a further example, we write
 the expression for the three-gluon vertex
correction which follows by a second differentiation with respect to $A$
(as in eq.~(\ref{V3g})). Using the same definition as in (\ref{GMF}) for the
Fourier transform,  we obtain from eq.~(\ref{4ja})
\beq \label{htl3}
\delta\Gamma^{abc}_{\mu\nu\rho}(P_1,P_2,P_3)
&=&i f^{abc}\,3\,\omega^2_p\int
\frac{d\Omega}{4\pi}\frac {v_\mu v_\nu v_\rho}{v\cdot P_1 +i\eta}\nonumber\\
&&\left \{\frac{P_3^0}{v\cdot P_3-i\eta}\,-\,\frac{P_2^0}{v\cdot  P_2-i\eta}
\right \},\eeq
where  the imaginary parts in the denominators correspond to the  time
orderings $X_1^0\ge X_2^0\ge X_3^0$ for the first term inside the parantheses,
and, respectively, $X_1^0\ge X_3^0\ge X_2^0$ for the second term.
We can rewrite this more symmetrically as $\delta\Gamma^{abc}_{\mu\nu\rho}
\equiv if^{abc}\delta\Gamma_{\mu\nu\rho}$ with \cite{Frenkel90}
\beq \label{htl3g}
\lefteqn{
\delta\Gamma_{\mu\nu\rho}(P_1,P_2,P_3)
\,=\,\omega^2_p\int
\frac{d\Omega}{4\pi}\,v_\mu v_\nu v_\rho
\,\Biggl\{\frac{P_1^0-P_2^0}{(v\cdot P_1+i\eta)(v\cdot
 P_2-i\eta)} }\nonumber\\
& &\mbox{}+\frac{P_2^0-P_3^0}{(v\cdot P_2-i\eta)(v\cdot
 P_3-i\eta)}+\frac{P_3^0-P_1^0}{(v\cdot P_3-i\eta) (v\cdot
 P_1+i\eta)}\Biggr\}.\eeq

Generally, all the amplitudes with  soft external
gluons which are derived in this way coincide with the corresponding
HTL's of the diagrammatic approach  \cite{Frenkel90,Braaten90b},
with boundary conditions  such that  $X_1^0$ is the
largest time, while all the other external lines enter symmetrically.

\subsection{Conservation laws and Ward identities}
The induced currents satisfy covariant conservation laws related to the
symmetries of the QCD lagrangian. These
 can be easily derived
from the  equations of motion for the corresponding densities.
These conservation laws imply, in particular, simple Ward identities
relating the hard thermal loops.

By acting on eq.~(\ref
{4Jpsi}) with the covariant derivative $D_X^\mu$, we get
\beq\label{1der}
\left [D_X^\mu,\,\left [k\cdot D_X,\,{\cal J}^\psi_\mu(k,X) \right ]\right ]
=\left [k\cdot D_X,\,igt^a
\left (\bar\psi(X)t^a\slashchar{{\cal K}}(k,X)
-\slashchar{{\cal H}}(k,X)t^a\psi(X)\right )\right ].
\eeq
In the left hand side, we use the Jacobi identity to write
\beq\label{4Jac}
\left [D_X^\mu,\,\left [k\cdot D_X,\,{\cal J}^\psi_\mu \right ]\right ]=
\left [k\cdot D_X,\,\left [D_X^\mu,\,{\cal J}^\psi_\mu \right ]\right ]-
ig\left [{\cal J}^\psi_\mu,\,F^{\mu\nu}k_\nu\right ],
\eeq
where the definition of the field tensor  $igF_{\mu\nu}\equiv [D_\mu,D_\nu]$
has been used. Since ${\cal J}^\psi_\mu$ is proportional
to $k_\mu$, as shown by eq.~(\ref{4Jpsi}), the last term cancels
in the right hand side of eq.~(\ref{4Jac}). Then eq.~(\ref{1der}) leads to
\beq
\left [D_X^\mu,\,{\cal J}^\psi_\mu(k,X) \right ]=ig\,t^a
\left (\bar\psi(X)\,t^a\,\slashchar{{\cal K}}(k,X)
-\slashchar{{\cal H}}(k,X)\,t^a\,\psi(X)\right ).
\eeq
After integrating over $k$, this becomes
a continuity equation for $j^\psi$:
\beq\label{cjp}
\left[D^\mu,\,j_\mu^\psi(X)\right]=ig\,t^a\left(\bar\psi(X) t^a\eta^{ind}(X)-
\bar\eta^{ind}(X)t^a\psi(X)\right).
\eeq
Furthermore, the equations of motion for
the fermionic fields $\psi$ and $\bar\psi$, that is, eq.~(\ref{4avpsi})
and its Hermitian conjugate, may be used to show that
\beq\label{c1jp}
\left[D^\mu,\,g t^a \bar\psi\gamma_\mu t^a \psi\right]=
-ig\,t^a\left(\bar\psi t^a\eta -
\bar\eta t^a\psi\right)-ig\,t^a\left(\bar\psi t^a\eta^{ind}-
\bar\eta^{ind}t^a\psi\right).\eeq
By adding eqs.~(\ref{cjp}) and (\ref{c1jp}), we finally obtain
the continuity equation satisfied by the total current
$g  \bar\psi\gamma_\mu t^a \psi+ j^{\psi\,a}_\mu$  associated
 to the fermionic mean
fields:
\beq\label{c2jp}
\left[D^\mu,\, g t^a\bar\psi\gamma_\mu t^a \psi+
j^{\psi}_\mu\right]=-igt^a\left(\bar\psi t^a\eta-
\bar\eta t^a\psi\right).
\eeq
Thus the current $g  \bar\psi\gamma_\mu t^a \psi+ j^{\psi\,a}_\mu$
is conserved in the absence of external fermionic sources.

Similar conservation laws hold for the axial color current
and for the  baryonic currents (\ref{4bar}) and (\ref{5bar}).

Repeating the same steps as above, but starting with the Vlasov equation
 (\ref{4JA}), we obtain the covariant  conservation law for the induced
current $j^A$:
\beq\label{cjA}
\left[D^\mu,\,j_\mu^A(X)\right]=0.
\eeq

Finally, by using the conservation laws for the induced currents,
eqs.~(\ref{c2jp})--(\ref{cjA}),
 together with the Jacobi identity $\left [D^\mu,
 [ D^ \nu, F_{\nu\mu}]\right ]=0$, we see that, in order for
 the mean fields equations (\ref{4avpsi}) and (\ref{4avA}) to be
consistent, the external sources must
satisfy
\beq
\left[D^\mu,\,j_\mu\right]=igt^a\left(\bar\psi t^a\eta-
\bar\eta t^a\psi\right).
\eeq

 By differentiating the  equations (\ref{cjA}) and (\ref{cjp}) with respect to
the
fields, one obtains relations between HTL refered as  ``QED-like'' (or
``tree-level like'') Ward identities
 in Refs.~\cite{Frenkel90,Braaten90b}. In particular,
the transversality of the polarization tensor follows from eq.~(\ref{cjA}),
after one differentiation with respect to $A_\mu$:
\beq
\label{W2}
\delta\Pi_{\mu\nu}P^\nu=P^\mu\delta\Pi_{\mu\nu}=0.
\eeq
Of course, this property can be easily verified on the explicit expression
for $\delta\Pi_{\mu\nu}$
given in eq.~(\ref{htlpi}). If we differentiate once again eq.~(\ref{cjA}),
we derive the simple identity
\beq
\label{W3}
P_1^\mu\delta\Gamma_{\mu\nu\lambda}(P_1,P_2,P_3)=\delta\Pi_{\nu\lambda}
(P_3)-\delta\Pi_{\nu\lambda}(P_2),
\eeq
which also  follows from eqs.~(\ref{htl3}) and (\ref{htlpi}) above.

For amplitudes involving a quark pair,  Ward identities
result from differentiating eq.~(\ref{cjp}). For instance,
 by differentiating eq.~(\ref{cjp}) with respect to
both $\psi$ and $\bar\psi$, and using eqs.~(\ref{pisib})
and (\ref{Vqg}),
we recognize the QED-like Ward identity
\beq
\label{W1}
P^\mu\delta\Gamma_\mu(P;K_1,K_2)=\delta\Sigma(K_1)-\delta \Sigma(K_1+P),
\eeq
which can also be deduced from the explicit expressions
 eqs.~(\ref{4gamma}), (\ref{4gamma2}) and (\ref{Sig}).
As a final example, consider the following identity, which results
from eq.~(\ref{cjp}) after an extra differentiation with respect
 to $A_\nu^b$,
\beq\label{W22}
\lefteqn{
P_1^\mu\,\delta\Gamma_{\mu\nu}^{ab}(P_1,P_2;K_1,K_2)\,=\, if^{abc}
\delta\Gamma_\nu^c(P_1+P_2;K_1,K_2) }\nonumber\\
& &\mbox{}+\delta\Gamma^b_\nu(P_2;K_1,K_2+P_1)\,t^a\,-\,
t^a\,\delta\Gamma^b_\nu(P_2;K_1+P_1,K_2),\eeq
and can also be  obtained directly from eqs.~(\ref{4gamma}) and (\ref{2f2gG}).

\subsection{Effective action for soft fields}

Using the results of Sec.5.1, one can rewrite the equations of motion
for the average  fields as
\beq\label{5avpsi}
i\slashchar{D} \psi(X) -\omega_0^2\int\frac{d\Omega}{4\pi}
\,{\slashchar{v}}\int d^4Y\,G_{ret}(X,Y;v)\,\psi(Y)=\,\eta(X),
\eeq
and
\beq
\label{5avA}
\left [\, D^\nu,\, F_{\nu\mu}(X)\,\right ]^a -
3\,i\,\omega^2_p\int\frac{d\Omega}{4\pi}\,v_\mu
\int d^4Y\, \tilde G_{ret}^{ab}(X,Y)\,\vec v\cdot\vec E^b(Y)
\nonumber\\
=\,j^a_\mu(X)+g\,\bar\psi (X)\gamma_\mu t^a \psi(X) +j_\mu^{\psi\,a}(X),
\eeq
where $j_\mu^{\psi}(X)$ is the current induced by fermionic
mean fields, given in eq.~(\ref{jfRA}).
In these equations, the fields are assumed to vanish adiabatically for
$X_0\to -\infty$, which correspond to the plasma being initially in
equilibrium. Of course, similar equations may be written to describe
perturbations which  vanish adiabatically in the remote future; then,
the induced currents would be expressed in terms of the advanced
Green's function.

In order to  make contact with previous works, we shall now restrict
the fields $\psi(X)$, $\bar\psi(X)$ and $A_\mu^a(X)$ to belong to a
 space ${\cal R}$ in which the  operators
 $v\cdot D_X$, $v\cdot\tilde D_X$ and $(v\cdot \tilde D_X)^2$ never vanish.
Specifying the boundary conditions becomes then unnecessary, and it is possible
to deduce the equations of motion (\ref{5avpsi}) and (\ref{5avA})
 by applying the minimum  action principle to an effective action
which can be written in a simple compact form.
The space ${\cal R}$ is spaned by the fields which vanish exponentially
for $|X_0|\to\infty$ and satisfy the conditions
\beq\label{cond1}
\int_{-\infty}^\infty du\,U(X,X-vu)\psi(X-vu)&=&0,\eeq
\beq\label{cond2}
\int_{-\infty}^\infty du\,\tilde U_{ab}(X,X-vu) F^b_{\mu\nu}(X-vu)&=&0,\eeq
\beq\label{cond3}
\int_{-\infty}^\infty du\,u\,\tilde U_{ab}(X,X-vu) F^b_{\mu\nu}(X-vu)&=&0.\eeq
These conditions guarantee  that  the fields have no components in the
kernel of  the covariant line derivatives. The left hand sides of
eqs.~(\ref{cond1})--(\ref{cond3}) represent such components.
For instance,
\beq
(v\cdot D_X)\, \int_{-\infty}^\infty du\,U(X,X-vu)\psi(X-vu)\,=\,0, \eeq
as one can easily verify (see e.g. eq.~(\ref{derU})).
Physically, the conditions
above imply that the fields do not suffer Landau damping. This is seen most
simply in case of weak fields, where we can ignore the parallel transporter.
Then the Fourier transforms of eq.~(\ref{cond1}) and (\ref{cond2}) result in
\beq \qquad \psi(P^0=\vec v\cdot\vec P, \,\vec P)=0,\qquad
and \qquad\,
F_{\mu\nu}^a(P^0=\vec v\cdot\vec P, \,\vec P)=0.\eeq
Within ${\cal R}$, the inverse operators $1/v\cdot D$, $1/v\cdot\tilde D$
and $1/(v\cdot \tilde D)^2$ are nonambiguous and represent indifferently the
retarded or advanced Green's functions, which are  equivalent on ${\cal R}$.

We write the effective action as $S_{eff}=S_0+S_{ind}$, where
 $S_0$ is the classical QCD action, (as given in eq.~(\ref{QCD}), but
without the ghosts and the gauge fixing term),
 while $S_{ind}$ contains the effects
of the interactions between the soft fields and the hard particles.
In order that $S_{eff}$ generates the mean field equations of motion
(\ref{5avpsi}) and (\ref{5avA}), one requires \beq
\delta S_{ind}/\delta \bar\psi(X)=\eta^{ind}(X),\qquad\qquad
\delta S_{ind}/\delta A^\mu_{a}(X)=j^{ind\, a}_\mu(X).\eeq
As proven in Appendix C,
these conditions are satisfied by $S_{ind}\equiv S_{\rm f}+S_{\rm b}$, with
\beq
\label{Sf}
S_{\rm f}&=& \omega_0^2\int\frac{d\Omega}{4\pi}\int
 d^4 X\int d^4 Y\,
\bar\psi(X)\langle X|\frac
{\slashchar{v}} {i(v\cdot D)}|Y\rangle \psi(Y),
\eeq and \beq
\label{Sb}
\qquad S_{\rm b}&=&\frac{3}{2}\,\omega^2_p\int \frac{d\Omega}{4\pi}\int d^4 X
\int d^4 Y \,{\rm tr} \left [ F_{\mu\lambda}(X)
\langle X|\frac {v^\mu v_\nu}{(v\cdot \tilde D)^2}|Y\rangle
 F^{\nu\lambda}(Y)\right ],
\eeq
where the trace acts on color indices only.
The action $S_{\rm f}$ accounts for all the polarization
effects induced by soft fermionic mean fields, that is
\beq\label {delSf}
\delta S_{\rm f}/\delta \bar\psi(X)=\eta^{ind}(X),\qquad\qquad
\delta S_{\rm f}/{\delta A^\mu_{a}(X)}=j_\mu^{\psi\, a}(X),
\eeq
while $S_{\rm b}$ satisfies
\beq \label{delSb}\delta S_{\rm b}/{\delta A^\mu_{a}(X)}&=&j_\mu^{A\,
a}(X).\eeq
The effective action (\ref{Sf})--(\ref{Sb})
 coincides with the generating functional for HTL's first derived
in \cite{Taylor90,Braaten91,Frenkel91} on the basis of gauge invariance.
One sees from our analysis that $S_{eff}$ may be given the physical
interpretation of the classical action which describes long wavelength
excitations of the hot quark-gluon
plasma, at leading order in the coupling $g$, but only as long as
 Landau damping
is inoperative.
One should emphasize that, indeed,
the functional derivatives of $S_{ind}$ generate properly the induced
currents only on ${\cal R}$ (see the discussion in Appendix 3).
 In cases where the boundary conditions matter,
 we have to recourse to the kinetic equations written in Sec.4.1 to solve
for the induced sources appropriate to the given physical conditions
(as done e.g. in Sec.4.2 for retarded conditions.)

\vskip 2cm
\setcounter{equation}{0}

\section{Conclusions}

We have presented a consistent description of soft collective
excitations of quark-gluon plasmas. This description relies on the
hierarchy of Dyson-Schwinger equations, which in the leading order
that we have considered, truncate at the level of 2-point
functions.  Our resulting equations take then the form of
a coupled set of mean field and kinetic equations. The mean fields
describe the soft excitations and have many
features of classical degrees of
freedom. The kinetic equations describe the dynamics of the hard
plasma particles, which respond collectively to the action of the
mean fields. This separation between mean fields and kinetic
equations reflects the distinct roles played by the two types, hard and soft,
degrees of freedom of the plasma, and also the fact that the dominant
interactions at leading order are those between the hard particles and the mean
fields.

As hard particles behave in a remarkably similar fashion,
irrespective of whether they are quarks or gluons, they
provide sources not only for average gauge fields, but also for
average fermionic fields. And indeed the quark-gluon plasma
sustains collective excitations with fermionic quantum numbers.
That is, there exists excitations in which the soft fermion
degrees of freedom oscillate over a long time scale. These
oscillations are maintained by the hard particles coherently
changing their quantum numbers from bosonic to
fermionic and vice versa, and this over long space/time scales \cite{BO}.

The excitations that we have studied are related to oscillations
in the conserved currents associated with the QCD Lagrangian.
There are thus oscillations in the baryon currents
which result from the action of a fermionic mean field.
There are also oscillations of color currents induced either by a
fermionic field or by a gauge field. All these oscillations can be
described in terms of fluctuations  of density matrices which are
generally non diagonal either in color space or in a space
containing fermionic and bosonic degrees of freedom. The diagonal
matrix elements, when they are non vanishing may be given the
interpretation of classical distribution functions for particles
of a given color.

The simple physical picture that we have just outlined is of course
fully justified by the long technical developments presented in this
paper, whose main features we now summarize.

The cornerstone of the whole approach is
that
collective modes develop at a particular energy scale, $gT$. The
emergence of this scale  is seen most clearly on the mean field
equations which, in leading order, become universal equations, once
length are measured in units of $1/gT$ and energies in units of
$gT$. It is the existence of this particular scale which allows us
to organize our analysis of the  Dyson-Schwinger equations, by controlling
consistently at the same time the strength of the coupling, the wavelength and
the amplitude of the collective oscillations,  to eventually arrive at a
consistent set of equations of motion which incorporate all leading order
effects.

One outcome of our work is a set of consistent and gauge covariant kinetic
equations for the plasma particles moving in background fields.
 As another by-product, we recover in a simple way all hard
thermal loops which have been analyzed first in a diagrammatic
approach. These are simply obtained from compact expressions for
the induced sources, which may be regarded as generating
functionals for the HTL's. Thus our approach provide a new and
independent derivation of such generating functionals. It has also
the advantage of clearly exhibiting the role of boundary
conditions. In particular our induced currents are easily obtained
with the retarded boundary conditions appropriate to the
calculation of the causal response functions. The standard form of the HTL
generating functional is recovered by restricting the fields to
obey special conditions. The particular Ward identities which
relate  HTL's  follow simply from conservation laws obeyed by the induced
currents. Finally, the mean field equations provide expressions for
the resummed soft propagators and vertices.\newpage

\vskip 2cm
\setcounter{equation}{0}
\renewcommand{\theequation}{A.\arabic{equation}}

\appendix{\noindent{\LARGE Appendix A}}

We prove here that the propagation of soft mean fields with ghost
quantum numbers does not give rise to polarization effects at leading
order in $g$. That is, in this order, the induced sources can be
ignored in the r.h.s. of eqs.~(\ref{avz}) and (\ref{avbz})
 for the ghost mean fields.
(In the diagrammatic language, this is equivalent to say that
there are no ``hard thermal loops'' with ghost external lines \cite
{Braaten90b}.) It follows that the ghost fields vanish in the absence of
external sources and they can be omitted in the equations of motion for
the 2-point functions (see Section 3).
For brevity, we present the explicit proof only in Feynman gauge
and for vanishing gauge fields. The general case can be treated along
the lines described at length in Appendix B.

The induced sources
 $C^{ind}_a$ and $\bar C^{ind}_a$ are related to the diagonal matrix
elements of the ghost-gluon off-equilibrum propagators, $O^{ab}_{\,\,\nu}$
and ${\bar O}^{ab}_{\nu}$, by  eqs.~(\ref{Ci}) and (\ref{bCi}), which imply
\beq
\label{CiA}
C^{ind\,a}(X) = gf^{abc}\left[\del^\nu_X O^{bc\,<}_{\,\,\nu} (s,X)\right]_{s=0}
=gf^{abc}\int\frac{d^4k}{(2\pi)^4}  \del_X^\nu {\cal{O}}^{bc}_{\,\,\nu} (k,X),
\eeq
and
\beq
\label{bCiA}
\bar C^{ind\,a}(X) = gf^{abc}\left[\del^\nu_s
{\bar  O}^{bc\,<}_\nu (s,X)\right]_{s=0}
=-i\,g\,f^{abc}\int\frac{d^4k}{(2\pi)^4}  k^\nu{\bar {\cal O}}^{bc}_\nu
(k,X).
\eeq
Note that in going from eq.~(\ref{bCi}) to eq.~(\ref{bCiA}) we have neglected
the soft derivative $\del_X$ in comparaison with $\del_s$, whereas (\ref{CiA})
is an exact rewriting of eq.~(\ref{Ci}).


The equations for
$O^{ab}_{\,\,\nu}$ and ${\bar O}^{ab}_{\nu}$ are obtained by
differentiating the mean fields equations with respect to the external
sources, as explained in Sec.3.3.
First, we differentiate the gauge field equation
(\ref{avA}) with respect to
$\bar C^b$ and $C^b$, and get, respectively,
\beq
\label{O1}
{\cal L}_{\mu\nu}^{ac}(y)\,O_{bc}^{\,\,\nu}(x,y)=\,-gf^{acd}\left
[\del_\mu^y\,G^{bc}(x,y)\right ]\langle\zeta^d(y)\rangle,
\eeq
and
\beq
\label{bO1}
{\cal L}_{\mu\nu}^{ac}(x)\,\bar O_{cb}^{\nu}(x,y)=\,gf^{acd}\,G^{cb}(x,y)
\del_\mu\langle\bar\zeta^d(x)\rangle.
\eeq
Then, we differentiate  eqs.~(\ref{avz}) and (\ref{avbz}) for the ghost fields
with respect $j_b^\nu$, and obtain
\beq
\label{O2}
\del_x^2\,O_{\,\,\nu}^{ab}(x,y)-gf^{acd}\langle A_\mu^c(x)\rangle\,\del_x^\mu
O_{\,\,\nu}^{db}(x,y)\qquad\qquad\nonumber\\
=gf^{acd}\,\del_x^\mu\left
[\,D_{\mu\nu}^{cb}(x,y)\,\langle\zeta^d(x)\rangle\,\right ],
\eeq
and
\beq
\label{bO2}
\del_y^2\,\bar O_{\nu}^{ba}(x,y)-gf^{acd}\langle A_\mu^c(x)\rangle\,\del_y
^\mu\bar O_{\nu}^{bd}(x,y)\qquad\qquad\nonumber\\
=gf^{acd}\,D_{\nu\mu}^{bc}(x,y)\,\del^\mu\langle\bar\zeta^d(y)\rangle
\eeq
Let us
 consider first $C^{ind}_a$. This will be obtained after solving
 eqs.~(\ref{O1}) and (\ref{O2}) for $O^{ab}_{\,\,\nu}$.
As mentioned earlier, we choose $\langle A_\mu^a\rangle = 0$, in order to
simplify the discussion.
Then, after doing approximations similar to those described at length in
 Section  3, we obtain for  $O^{ab}_{\,\,\nu}$ the following two equations:
\beq
\label{O1A}
\del_y^2 O^{ab\,<}_{\,\,\nu}(x,y)=gf^{abc}\left (\del_\nu^y \Delta(x-y)\right )
\zeta^c(y),\eeq
and
\beq
\label{O2A}
\del_x^2\,O_{\,\,\nu}^{ab\,<}(x,y)
=-gf^{abc}\del^x_\nu\left [\Delta(x-y)\zeta^c(x)\right ].
\eeq
The ghost mean fields
are simply denoted here by $\zeta^a$ and $\bar\zeta^a$.
 These  equations are further transformed into
 an equation for the Wigner function ${\cal{O}}^{ab}_{\,\,\nu}(k,X)$
\beq\label{kinO}
(k\cdot\del_X){\cal{O}}^{ab}_{\,\,\nu}(k,X)=i\,\frac{g}{2} \,
f^{abc}k_\nu(\del^\mu_X
\zeta^c)(\del^k_\mu \Delta).
\eeq
This equation includes consistently all the terms at leading order in $g$.
It shows that ${\cal O}(k,X)\sim g(\zeta/T)\Delta$. One deduces then from
eq.~(\ref{CiA}) that $C^{ind}\sim g^3T^2\zeta$, which is $g$ times smaller
than the l.h.s. of eq.~(\ref{avz}) for $\zeta$, $\del^2\zeta\sim g^2T^2\zeta$.
In fact, $C^{ind}$ is even more suppressed than this estimate suggests.

The corresponding proof for the anti-ghost mean field - $i.e.$, for $\bar
 C^{ind}$ and $\bar O^{ab}_{\nu}$ - is slightly different, owing to the
disymmetry with which the two kinds of ghost fields enter the QCD action
 (\ref{QCD}). Now it is the abnormal propagator
$\bar O^{ab}_{\nu}$ itself which  vanishes at leading order. This
can be easily verified by isolating the dominant terms
in eqs.~(\ref{bO1}) and (\ref{bO2}) for $\bar O^{ab}_{\nu}$.

\vskip 2cm
\setcounter{equation}{0}
\renewcommand{\theequation}{B.\arabic{equation}}

\appendix{\noindent {\LARGE Appendix B}}

In the derivation of the equations of motion in the main text,
we have used systematically the Feynman
gauge ($\lambda = 1$) in order to simplify the algebra. We prove now
that this particular choice does not affect the final results: the
 equations that we have obtained  for the mean fields and
their induced sources preserve their form
in any covariant gauge. As a consequence, the HTL's are independent of
the gauge fixing parameter $\lambda$ for arbitrary (soft) external momenta.

The gauge fixing parameter $\lambda$ enters our calculations through the
 gluonic propagator, that is, either through the
differential operator ${\cal L}^{ab}_{\mu\nu}(x)$, eq.~(\ref{L}), or
through $D^{<\,ab}_{0\,\mu\nu}(k)\equiv \delta^{ab}\rho_{\mu\nu}(k)N(k_0)$.
Here $\rho_{\mu\nu}(k)$ is the spectral density in a general covariant gauge
 \cite{Landsman87}
\beq\label{rho}
\rho_{\mu\nu}(k)=\rho_{\mu\nu}^F(k)+\left(1-\lambda ^{-1}\right)
\rho_{\mu\nu}^\lambda(k),
\eeq
with  $\rho_{\mu\nu}^F(k)=-g_{\mu\nu}\rho_0(k)$  the spectral density
in Feynman gauge ($\rho_0(k)
\equiv 2\pi\epsilon(k_0)\delta(k^2)$), while $\rho^\lambda_{\mu\nu}(k)
\equiv -2\pi\epsilon(k_0) k_\mu k_\nu \delta '(k^2)$,  $\delta '(k^2)$
 denoting the derivative of $\delta(k^2)$ with respect to $k^2$.
 We also define $\Delta'(k)\equiv2\pi\epsilon(k_0) \delta '(k^2)N(k_0)$,
by analogy with  $\Delta(k) =\rho_0(k)N(k_0)$.
 By using $k^2\delta'(k^2)=-
\delta(k^2)$, we verify that
\beq\label{Delprim}
k^2\Delta'(k)=-\Delta(k),
\eeq
and consequently the gluon propagator can also be written as
\beq\label{Dol}
D^<_{0 \,\mu\nu}(k)=[g_{\mu\nu} k^2-(1-\lambda^{-1})k_\mu k_\nu]\Delta'(k).
\eeq
We shall need later the following identity:
\beq\label{delDo}
k^\mu D^<_{0\,\mu\nu}(k)=-\lambda^{-1}\,k_\nu\Delta(k).
\eeq

After these preparations, let us formulate the proposition
 to be proved below:
{\it in a general covariant gauge parametrized by $\lambda$,
the $k$-space densities for the induced sources, $i.e.$ ${\slashchar{\cal
K}}(k,X)$,  ${\cal J}^\psi_{\rm f}(k,X)$,  ${\cal J}^\psi_{\rm b}(k,X)$,
  ${\cal J}^A_{\rm f}(k,X)$ and
 ${\cal J}^A_{\rm b}(k,X)$, are independent of $\lambda$ when
 consistently computed at leading order in $g$.}

Note first that, since the gluon propagator is not involved
in the equations for $\delta{\cal S}^A$, ${\cal J}^A_{\rm f}$
 is trivially gauge independent. Furthermore, by examining the derivation
of the equations satisfied by   ${\cal J}^\psi_{\rm f}$
and  ${\cal J}^\psi_{\rm b}$, one sees that these may depend on $\lambda$
only through  $\slashchar{\cal K}$ and $\slashchar{\cal H}$.
Thus, we only need to prove that $\slashchar{\cal K}$ and
${\cal J}^A_{\rm b}$ are independent of $\lambda$.

We start  with  $\slashchar{\cal K}$.
By following the same steps as in  Sec.3.3, we derive from
eq.~(\ref{K4}) the equation which generalizes
 eq.~(\ref{K6}) to arbitrary $\lambda$
\beq\label{1DKX}
\left [ k^2 +i k\cdot\left (\del_X+2igA\right )\right ]
{\cal K}^a_\mu
\approx- {g}\Delta(k)\left[\slashchar{k}\gamma_\mu
 -(1-\lambda^{-1})k_\mu\right] t^a \psi,
\eeq
where we have preserved only the terms of relevant order of magnitude.
Similarly, starting from eq.~(\ref{K3}) we get
\beq\label{2DKX}
(k^2-i k\cdot\del_X){\cal K}^a_\mu-i{g}f^{abc}\Gamma_
{\mu\nu\rho\lambda} A_c^\rho k^\lambda {\cal K}_b^\nu+i(1-\lambda)
k_\mu\tilde{\cal K}_a
\approx {g}\tilde\Delta(k) \slashchar{k}\gamma_\mu t^a\psi .
\eeq
We have introduced here the function $\tilde{\cal K}_a(k,X)$,
 the Wigner transform of
\beq\label{tildeK}
\tilde K_a(x,y)\equiv\del_y^\nu K_{\nu\,a}^<(x,y).
\eeq
We now expand
 ${\cal K} = {\cal K}^{(0)}+{\cal K}^{(1)}+...$, with ${\cal K}^{(1)}
\sim g{\cal K}^{(0)}$, etc.
At leading order,
we recover from the previous equation
the consistency conditions (\ref{ms}) and (\ref{trK}), that is,
 $k^2{\cal K}_\mu^{(
0)}=0$ and $k^\mu{\cal K}_\mu^{(0)}=0$.
At next to leading order, we subtract the two equations and obtain
the generalization of eq.~(\ref{KXk}) to arbitrary $\lambda$
\beq
\label{DKX}
 k\cdot\left (\del_X+igA\right ){\cal K}^a_\mu+ \frac{g}{2}f^{abc}\Gamma_
{\mu\nu\rho\lambda} A_c^\rho k^\lambda {\cal K}_b^\nu-\frac{1-\lambda}
{2}k_\mu\tilde{\cal K}_a\nonumber\\
=i\frac {g}{2}[\Delta(k)+\tilde\Delta(k)]
\slashchar{k}\gamma_\mu t^a\psi -i\frac{g}{2}(1-\lambda^{-1})\Delta(k)
k_\mu t^a \psi,
\eeq
 where ${\cal K}\equiv{\cal K}^{(0)}$.
Because $k^\mu{\cal K}_\mu^{(0)}=0$,
the leading order contribution to $\tilde{\cal K}_a$ (which enters
 eq.~(\ref{DKX})) is $\sim gT{\cal K}^{(0)}$:
\beq
\tilde{\cal K}_a(k,X)=ik^\nu{\cal K}_\nu^{(1)\,a}(k,X)+\frac{1}{2}\del^\nu_X
{\cal K}_\nu^{(0)\,a}(k,X).
\eeq
Thus eq.~(\ref{DKX}) is consistent, $i.e.$, all terms are of the same order
of magnitude, $\sim gT^2{\cal K}^{(0)}$.
The unknown function ${\cal K}^{(1)}$, as well as the $\lambda$ dependence
disappear in the equation for ${\slashchar{\cal K}}^a
\equiv\gamma^\mu{\cal K}^{(0)}_{\mu\,a}$.
These cancellations  result from the identity
\beq\label{id}
{\slashchar k}\tilde{\cal K}_a(k,X)=-i\frac{g}{\lambda}{\slashchar k}
\Delta(k) t^a\psi(X),
\eeq
which is proved by first differentiating eq.~(\ref{K4}) with respect to
 $\del^\nu_y$, and then  using the identity (\ref{delDo}), to get
\beq
{\slashchar D}_x \tilde K_a(x,y)=-i\frac{g}{\lambda}\gamma^\mu t^a\psi(x)
\del_\mu\Delta(x-y),
\eeq
which, in leading order in $g$, gives precisely
eq.~(\ref{id}). Thus, we recover for
  ${\slashchar{\cal K}}^a$ the Feynman gauge equation (\ref{slashK}).

Consider now the gluonic current ${\cal J}^A_{\rm b}$. As discussed in Sec.3.7,
 ${\cal J}^A_{\rm b}\equiv {\cal J}^A_{\rm g}+{\cal J}_G+{\cal J}_\Omega$.
Since, by construction, ${\cal J}_G$ is independent of $\lambda$,
we need to prove
that a similar property holds for the sum ${\cal J}^A_{\rm g}+{\cal J}_\Omega$.

{}From eq.~(\ref{TP1})
\beq
\label{Djt}
j^a_{\Omega\,\mu}(X)=g^2\hat
\Gamma_{\mu\nu\rho\lambda}^{abcd}\,A_b^\nu(X)D_{0\,cd}^{<\,\rho\lambda}(0)\,=\,Ng^2\left[g_{\mu\nu}(D_0^<)^\rho_
{.\,\,\rho}(0)-(D_0^<)_{\mu\nu}(0)\right]A^\nu_a(X).
\eeq
By using eq.~(\ref{Dol}), this can be written as a integral in $k$-space
of ${\cal J}_\Omega\equiv {\cal J}_\Omega^F+
(1-\lambda^{-1}){\cal J}^\lambda$, where
\beq\label{JtF}
{\cal J}^{F\,a}_{\Omega\,\mu}=-gN(d-1)\,\Delta(k)\,A_\mu^a(X),
\eeq
is the Feynman gauge expression, eq.~(\ref{4jt}), while
\beq\label{Jtl}
{\cal J}_{\,
\mu}^{\lambda\,a}(k,X)=-gN(g_{\mu\nu}k^2-k_\mu k_\nu)\Delta'
(k)A_a^\nu(X)\eeq
is a supplementary term for $\lambda\not =1$.

As indicated in eq.~(\ref{JgA}), ${\cal J}^A_{\rm g}$ is determined by the
 quantities
$\delta {\cal D}^a_{\mu\nu}k^\nu$ and $\delta {\cal D}^{a\,\nu}_\nu$.
(To simplify the writing, we omit, here and below,  the
 $A$-superscripts.)
These quantities will be determined by following the same steps as in  Sec.3.7.
We start  from eqs.~(\ref{D2}) and (\ref{D3}),
written for arbitrary $\lambda$ and $\psi=\bar\psi=0$.
 By using eq.~(\ref{L})
 for ${\cal L}^{ab}_{\mu\nu}$, we write eq.~(\ref{D2}) as
\beq
\label{DD1}
\left (\del_x^2 g_\mu^\rho -(1-\lambda)\del_\mu^x\del^\rho_x\right )
D^{ab\,<}_{\rho\nu}(x,y)
&=& -gf^{adc}\left
[\Gamma_{\mu\sigma\rho\lambda} A_c^\rho(x)\del_x^\lambda-
\Gamma_{\mu\rho\lambda\sigma} (\del_x^\lambda A_c^\rho(x))\right ]D^{\sigma b\,
<}_{d\nu}(x,y)\nonumber\\
&+&g^2\hat \Gamma_{\mu\sigma\rho\lambda}^{adce}A_c^\rho(x)
 A_e^\lambda(x)D^{\sigma b\,<}_{d\nu}(x,y).
\eeq
We shall refer to this equation as (I).
We have a similar equation deduced from (\ref{D3}) and  to which we
refer below as  (II). It is obtained from (I) by exchanging, respectively,
$x$ and $y$, $a$ and $b$, and $\mu$ and $\nu$.
We separate the induced part of $D$ by setting $D^<\equiv D^<_0+\delta D$,
with $\delta D\sim gD_0$ for $A\sim T$.
Note that $D_0^<$
does not contribute to the l.h.s. of (I) and (II), because
it satisfies  $D_0^{-1}\,D_0^<=0$; ($D_0^{-1}$ is the free inverse
propagator, eq.~(\ref{D})). We expand
$\delta D =\delta D^{(0)}+\delta D^{(1)}+...$, where $\delta D^{(1)}
\sim g\delta D^{(0)}$, etc...   At leading order,  (I) and (II)
lead to the consistency condition which generalizes eq.~(\ref{k2D}) for
$\lambda\not =1$
\beq\label{Dk2D}
k^2\delta{\cal D}^a_{\mu\nu}-(1-\lambda)k_\mu k^\rho
\delta{\cal D}^{a}_{\rho\nu}=g\Gamma_{\mu\sigma\rho\lambda}
A^\rho_c(X)k^\lambda D_{0\,\nu}^{<\,\sigma}(k),
\eeq
where we have set (see eq.~(\ref{rdefD}))
\beq
\delta{\cal D}^a_{\mu\nu}(k,X)\equiv(i/N)f^{abc}\delta{\cal D}^{(0)\,bc}
_{\mu\nu}(k,X).\eeq

 At next-to-leading order, we encounter terms involving the unknown
function  $\delta D^{(1)}$.
 For instance, $(k\cdot\del_X)\delta D^{(0)}_{\mu\nu}\sim
k^2\delta D^{(1)}_{\mu\nu}\sim k_\mu
k^\rho\delta D^{(1)}_{\rho\nu}$ in the l.h.s. of eq.~(I).
Contrary to what happened in the Feynman gauge, now
such terms do not  disappear totally in the difference of eqs.~(I) and (II).
Nor do the terms
proportional to $\hat\Gamma$ in the r.h.s.'s of Eqs.(I) and (II). However,
things simplify in the equations for the functions of interest,
 $\delta {\cal D}^a_{\mu\nu}k^\nu$ and $\delta {\cal D}^{a\,\nu}_\nu$.

Consider first the equations for  $\delta {\cal D}^{a\,\nu}_\nu$.
When traced over  space-time indices, the terms depending
on $\delta D^{(1)}$, as well as those coming from the four gluons interactions,
become identical in eqs.~(I) and (II) (at the order of interest).
Then, by taking the difference of these two equations, we get
\beq\label{DtrD}
\lefteqn{ \left[k\cdot D_X, \delta{\cal D}^\nu_{\nu}\right]^a
+g\,f^{abc}\left(A^\mu_b\delta{\cal D}^{c}_{\mu\nu}k^\nu\right )
-(1-\lambda)\left(\del^\mu_X\delta{\cal D}^a_{\mu\nu}k^\nu\right) }
\nonumber\\& &=g(d-1)k^\nu (\del_\mu A^a_\nu)(\del^\mu\Delta)
+g(1-\lambda^{-1}) (\del_\mu A_\nu^a)(k^\mu k^\nu-g^{\mu\nu}k^2)\Delta'(k),
\eeq
which generalizes eq.~(\ref{trD}) for arbitrary $\lambda$.

Similarly, we shall prove shortly that   $\delta {\cal D}^a_{\mu\nu}k^\nu$
satisfies
\beq\label{DDk1}
(k\cdot\del_X)\left (\delta{\cal D}_{\mu\nu}^{a}\,k^\nu\right )
-\frac{3}{4}g\,f^{abc}(k\cdot A^b)\left (\delta{\cal D}_{\mu\nu}^{c}\,
k^\nu\right )
+\frac {g}{4} f^{abc}k_\mu(A^b\cdot\delta{\cal D}^{c}\cdot k)
\nonumber\\
=\frac {g}{2\lambda}(k_\mu k_\nu-g_{\mu\nu}k^2)\left (\del^\rho A^\nu_a
\right )
(\del_\rho\Delta)-\frac{3}{4\lambda}g^2f^{abc}(k\cdot A^b)A_\mu^c \Delta.
\eeq
\parindent 30pt
Thus, $\lambda \delta {\cal D}^a_{\mu\nu}k^\nu$ satisfies the same equation as
$\delta {\cal D}^a_{\mu\nu}k^\nu$ does in Feynman gauge (see eq.~(\ref{Dk1})),
and, according to eq.~(\ref{Dk3}),
\beq\label{DDk3}
\lambda\delta{\cal D}^{A\,a}_{\mu\nu}k^\nu(k,X)&=& gA_\mu^a(X)\Delta(k)-
k_\mu\delta{\cal G}^{a}(k,X).
\eeq
\parindent 30pt We can use  this relation  to eliminate the terms dependent on
$\delta {\cal D}^a_{\mu\nu}k^\nu$ in eq.~(\ref{DtrD}), which
becomes then
\beq\label{1DtrD}
\lefteqn{ \left [k\cdot D_X,\,\,\delta{\cal D}^{\nu}_{\nu}
-(1-\lambda^{-1})\left(\delta{\cal G}+g(k\cdot A)\Delta'(k)\right )\right]^a
\qquad } \nonumber\\ & &
\qquad=g(d-1)k_\nu (\del^\rho A^\nu_a)(\del_\rho^k\Delta)+gf^{abc}(k\cdot A^b)
\delta {\cal G}^c.
\eeq
This equation shows that $\delta{\cal D}^{\nu}_{\nu}-(1-\lambda^{-1})
\left(\delta{\cal G}+g(k\cdot A)\Delta'(k)\right) $
is independent of $\lambda$; for $\lambda=1$, (\ref{1DtrD})
 reduces to eq.~(\ref{3trD}).

By  using the expression  (\ref{DDk3}) in  eq.~(\ref{JgA}), we obtain the
following expression for ${\cal J}_{\rm g}\equiv{\cal J}_{\rm g}^A$:
\beq\label{DJg}
{\cal J}^a_{{\rm g}\,\mu}(k,X)=\frac{g}{\lambda}N\Delta(k)A^a_\mu(X)-
Nk_\mu\left(\delta{\cal D}^{a\,\nu}_\nu(k,X)+\frac{1}{\lambda}
\delta{\cal G}^a(k,X)\right).
\eeq

It is now straightforward to verify that ${\cal J}_{\rm g}=
{\cal J}_{\rm g}^F-(1-\lambda^{-1}){\cal J}^\lambda$, where ${\cal J}_{\rm g}
^F$ is the expression obtained for ${\cal J}_{\rm g}^A$ in Feynman gauge,
eq.~(\ref{3Jg}), while ${\cal J}^\lambda$ is given by eq.~(\ref{Jtl}).
Indeed, from eqs.~(\ref{Jtl}) and (\ref{DJg}), we see that\beq
\lefteqn{ {\cal J}^a_{{\rm g}\,\mu}+(1-\lambda^{-1}){\cal J}^{\lambda\,a}_\mu=
gN\Delta(k)A^a_\mu }\nonumber\\ & &
-Nk_\mu\left\{\delta{\cal D}^{a\,\nu}_\nu
-(1-\lambda^{-1})\left(\delta{\cal G}^a+g(k\cdot A^a)\Delta'(k)\right )
+\delta{\cal G}^a\right\},
\eeq
where the r.h.s. is, according to  eq.~(\ref{1DtrD}), independent of
$\lambda$, and equal to ${\cal J}_{\rm g}^F$ (compare, in this respect,
 eqs.~(\ref{1DtrD}) and (\ref{3trD})).
Thus, ${\cal J}_{\rm g}+{\cal J}_\Omega={\cal J}_{\rm g}^F+{\cal J}_\Omega^F$,
as claimed.

 To complete our proof, we still have to establish
 eq.~(\ref{DDk1}) for $\delta {\cal D}^a_{\mu\nu}k^\nu$.
 The technical difficulty is the elimination of the terms
involving the next-to-leading order correction, $\delta D^{(1)}\sim g\delta
D^{(0)}$.
In order to do this, we consider the equation satisfied by the derivatives
of $\delta D(x,y)$. First,  we obtain an equation for
 $\del^\nu_y\delta D^{ab}_{\mu\nu}$ by  differentiating Eqs.(I) and (II)
 with respect to $\del^\nu_y$  and subtracting them. The l.h.s. of the
 resulting equation is
\beq\label{dy}
(\del_x^2-\del_y^2)(\del^\nu_y\delta D^{ab}_{\mu\nu})-(1-\lambda)
\del_\mu^x(\del^\rho_x \del^\lambda_y\delta D^{ab}_{\rho\lambda}).
\eeq
At leading order,  $\delta D^{(1)}$ appears only in the second term:
 $\del^s_\mu(\del_s^\rho\del_s^\lambda\delta
D^{(1)}_{\rho\lambda})\sim g^2 T^3 D_0$, $i.e.$, of the order of
$(\del_x^2-\del_y^2)(\del^\nu_y\delta D^{ab}_{\mu\nu})\approx
-2(\del_s\cdot\del_X)(\del^\nu_s\delta D_{\mu\nu}^{(0)\,ab})\sim
g^2 T^3 D_0$ (recall that $(\del_s^\rho\del_s^\lambda\delta
D^{(0)}_{\rho\lambda})=0$).
By the same procedure, we get an equation $\del^\nu_x\delta D^{ab}_{\nu\mu}$,
whose l.h.s. is
\beq\label{dx}
(\del_x^2-\del_y^2)(\del^\nu_x\delta D^{ab}_{\nu\mu})+(1-\lambda)
\del_\mu^y(\del^\rho_x \del^\lambda_y\delta D^{ab}_{\rho\lambda}).
\eeq
Then,  eq.~(\ref {DDk1}) for $\delta {\cal D}^a_{\mu\nu}k^\nu$ is obtained
by  subtracting the equations corresponding to (\ref{dx}) and (\ref{dy}).
The l.h.s. is, at the relevant order,
\beq\label{ds}
4(\del_s\cdot\del_X)(\del_s^\nu\delta  D^{(0)\,ab}_{\mu\nu})
+(1-\lambda)
\del_\mu^X(\del^\rho_x \del^\lambda_y\delta D^{ab}_{\rho\lambda}).
\eeq
The second term in eq.~(\ref{ds}) is now negligible
at this order, as it is $\sim g^3 T^3 D_0$. Getting the r.h.s. of this
equation presents no difficulty. A Fourier transform with respect to
 $s$ then yields eq.~(\ref {DDk1}) for $\delta {\cal D}^a_{\mu\nu}k^\nu$.

\vskip 2cm
\setcounter{equation}{0}
\renewcommand{\theequation}{C.\arabic{equation}}

\appendix{\noindent{\LARGE Appendix C}}

We prove here that the induced sources $\eta^{ind}$, eq.~(\ref
{eta-psi}), and $j_\mu^{ind}=j_\mu^\psi+j_\mu^A$, eqs.~(\ref{4jpsi}) and
(\ref{4jA}), are generated formally by $S_{ind}=S_{\rm f}+S_{\rm b}$, with
$S_{\rm f}$ and $S_{\rm b}$ given, respectively, by eqs.~(\ref{Sf}) and
(\ref{Sb}), {\it if the average fields $\psi(X)$ and $A_\mu^a(X)$ belong
to the  space ${\cal R}$}, defined by eqs.~(\ref{cond1})--(\ref
{cond3}).

Recall first that, within ${\cal R}$, the inverse operators $1/v\cdot D$
and $1/(v\cdot \tilde D)^2$ which appear in eqs.~(\ref{Sf}) and
 (\ref{Sb}) are unambiguous and represent indifferently the
retarded or advanced Green's functions, which are
 equivalent on ${\cal R}$. For instance, we can  easily verify, using
 eqs.~(\ref{GR}) and (\ref{GA}), that
\beq\int d^4Y\left ( G_{ret}(X,Y;v)- G_{adv}(X,Y;v)\right )\psi(Y)=
-i\int_{-\infty}^\infty duU(X,X-vu)\psi(X-vu)=0,\nonumber\\
\eeq
where the last equality follows from eq.~(\ref{cond1}). This allows us,
when doing formal operations in ${\cal R}$, to replace
 $1/v\cdot D$ by the retarded Green's function, which has a simple
expression in terms of the parallel transporter:
\beq \int d^4Y\langle X|\frac{1}{v\cdot D}|Y\rangle\psi(Y)=
\int_0^\infty du \,U(X,X-vu)\psi(X-vu).\eeq
This is convenient as it allows us to use nice identities satisfied by
 the parallel transporters.
Similarly, we can identify on ${\cal R}$ the operators $1/(v\cdot \tilde D)^2$
and
\beq \tilde G^{(2)}_{ret}(X,Y;v)&\equiv&-\int d^4Z\,
\tilde G_{ret}(X,Z;v)\tilde G_{ret}(Z,Y;v)\nonumber\\&=&
\theta (X^0-Y^0)(X^0-Y^0)\,\delta^{(3)}\left({\vec X}-{\vec Y}-{\vec
v}(X^0-Y^0)
\right )\tilde U(X,Y).\eeq
Thus, on ${\cal R}$,
\beq \int d^4Y\langle X|\frac{1}{(v\cdot\tilde D)^2}|Y\rangle \,F_{\mu\nu}(Y)=
\int_0^\infty du \,u\,U(X,X-vu) F_{\mu\nu}(X-vu) U(X-vu,X).\eeq
We conclude that the previous expressions for $S_{\rm f}$, eq.~(\ref{Sf}), and
 for $S_{\rm b}$, eq.~(\ref{Sb}), are equivalent respectively to
\beq
\label{CSf}
S_{\rm f}&=& -i\omega_0^2\int\frac{d\Omega}{4\pi}\int
d^4X\int_0^\infty du\bar\psi(X)\slashchar{v}U(X,X-vu)\psi(X-vu),
\eeq and \beq
\label{CSb}
S_{\rm b}&=&\frac{3}{2}\,\omega^2_p\int \frac{d\Omega}{4\pi}\int d^4 X
\int_0^\infty du\,u\nonumber\\
&&\qquad\,{\rm tr} \left\{v^\mu F_{\mu\lambda}(X)\,U(X,X-vu)\,
v_\nu F^{\nu\lambda}(X-vu)\,U(X-vu, X)\right \},
\eeq
where the trace acts on color indices only.

Let us prove now that, when the conditions (\ref{cond1})--(\ref{cond3}) hold,
the induced sources are indeed generated from the effective action,
 according to eqs.~(\ref{delSf})--(\ref{delSb}). The first equality
in eq.~(\ref{delSf}) follows immediatly
from eqs.~(\ref{CSf}) for $S_{\rm f}$ and (\ref{eta-psi})
for $\eta^{ind}$. To prove the second identity, which relates $S_{\rm f}$ to
$j_\mu^\psi$, we differentiate eq.~(\ref{Sf}) with respect to $A^\mu_a(X)$,
 using \beq
\frac{\delta\qquad} {\delta A^\mu_a(X)}
\,\langle Y_1|\frac{1}{i(v\cdot D)}|Y_2\rangle\,=
\,gv_\mu\langle Y_1|\frac{1}{i(v\cdot D)}|X\rangle
\,t^a\,\langle X|\frac{1}{i(v\cdot D)}|Y_2\rangle,\eeq
and obtain
\beq
\frac{\delta S_{\rm f}}{\delta A^\mu_a(X)}&=&g\omega_0^2
\int\frac{d\Omega}{4\pi}\,v_\mu\int d^4Y_1\int d^4 Y_2\nonumber\\
&&\qquad\left\{\bar\psi(Y_1)
\langle Y_1|\frac{1}{i(v\cdot D)}|X\rangle
\,{\slashchar v}t^a\,\langle X|\frac{1}{i(v\cdot D)}|Y_2\rangle\psi(Y_2)
\right \}.\eeq
This is precisely the expression (\ref{jfRA}) for $j_\mu^\psi$
(recall that, on ${\cal R}$, $G_{adv}(X,Y;v)=G_{ret}(X,Y;v)\\
=\langle X|1/i(v\cdot D)|Y\rangle$).

We now turn to the verification of eq.~(\ref{delSb}), which proceeds in three
steps.  In order to do this, it is  convenient to use the expression
of $S_{\rm b}$ in terms of the parallel transporter.

{\bf i)} We  differentiate functionally $S_{\rm b}$, eq.~(\ref{CSb}), with
respect to $A^a_\mu(X)$ and write the result of this operation as
\beq\label{CdS}
\delta S_{\rm b}\equiv \delta S_{\rm b}^{(1)}+\delta S_{\rm b}^{(2)}.
\eeq The first piece comes from differentiating the $F_{\mu
\nu}$'s in eq.~(\ref{CSb}),
\beq\label{CdS1}
\lefteqn{
\delta S_{\rm b}^{(1)}\,=\,\frac{3}{4}\,\omega^2_p\int \frac{d\Omega}{4\pi}
\int d^4 X \int_0^\infty du\,u \,v^\mu \delta F_{\mu\lambda}^a(X) \qquad
}\nonumber\\
& &\left \{\tilde U_{ab}(X,X-vu)\,v_\nu F^{\nu\lambda}_b(X-vu)+
\tilde U_{ab}(X,X+vu)\,v_\nu F^{\nu\lambda}_b(X+vu)\right \}.
\eeq
The second term, $\delta S_{\rm b}^{(2)}$, comes from the parallel transporter,
\beq\label{CdS2}
\delta S_{\rm b}^{(2)}=\frac{3}{4}\,\omega^2_p\int \frac{d\Omega}{4\pi}
\int d^4 X \int_0^\infty du\,u\,
v^\mu F_{\mu\lambda}^a(X)\left [\delta\tilde U_{ab}
(X,X-vu)\right ]v_\nu F^{\nu\lambda}_b(X-vu).
\eeq
{}From eq.~(\ref{Ufdt}), we derive \beq\label {CdelU1}
\delta\tilde U_{ab}(X,X-vu)&=&\int d^4Y\, \frac {\delta\tilde U_{ab}(X,X-vu)}
{\delta A^a_\rho(Y)}\,\delta A^a_\rho(Y)\nonumber\\
&=&-igu\int_0^1ds \left[\tilde U (X,Z)\,v\cdot\delta \tilde A(Z)\,
\tilde U (Z,X-vu)\right]_{ab},\eeq
with $\tilde A_\mu\equiv T^a A^a_\mu$ and $Z^\mu(s)=X^\mu-u(1-s)v^\mu$.
We introduce (\ref{CdelU1}) into (\ref{CdS2}) and derive, after
elementary changes of integration variables,
    \beq\label{CdS3}
\lefteqn{
\delta S_{\rm b}^{(2)}\,=\,
-ig\,\frac{3}{4}\,\omega^2_p\int \frac{d\Omega}{4\pi}
\int d^4 X \int_0^\infty du\int _0^\infty dt\,(u+t)\qquad}\nonumber\\
& &\left\{v^\mu F_{\mu\lambda}^a(X+vt)\left [\tilde U
(X+vt,X)\,v\cdot\delta\tilde A(X)\,\tilde U(X,X-vu)\right ]_{ab}
v_\nu F^{\nu\lambda}_b(X-vu)\right \}.
\eeq

{\bf ii)} We  now derive a new expression for the induced current
 $j^A_\mu$,
which will make easier the comparaison with $\delta S_{\rm b}/\delta
A^\mu$. We note that  eq.~(\ref{4JA}) can be  solved in the form
 \beq\label{CJA}
{\cal J}^{A\,a}_\mu(k,X)=g\,v_\mu^\prime k_\nu\,\del_\rho
\Delta_{eff}(k)\int_0^\infty du
\,\tilde U_{ab}(X,X-v^\prime u)F^{\nu\rho}_b(X-v^\prime u), \eeq
whith $v_\mu^\prime\equiv k_\mu/k_0$ and $\Delta_{eff}(k)\equiv
2N_f\tilde \Delta(k)+(d-2)N\Delta(k)$.
The  expression  (\ref{4jA}) of the induced current $j^A_\mu$ is obtained
from (\ref{CJA}) after a  direct integration over  $k$.
Alternatively, one can  integrate by parts with respect to
$k$ and derive, after some elementary calculations,
\beq\label{Cj1}
j^{A\,a}_\mu(X)\,=\,-\frac{3}{2}\,\omega_p^2\int\frac{d\Omega}
{4\pi}\,v_\nu\,\int_0^\infty du\,\frac{\del\,\,}{\del v^\rho}\left [
v_\mu\tilde U_{ab}(X,X-vu)F^{\nu\rho}_b(X-vu)\right ]. \eeq
Here, $v_\mu\equiv (1,\vec k/\epsilon_k)$,
but the four components have to be considered as independent variables
before taking the derivative $\del/\del v^\rho$. The integrand
in eq.~(\ref{Cj1}) is further transformed by writing
\beq\label{Cj2}
\frac{\del\,\,}{\del v^\rho}\left [
\tilde U_{ab}(X,X-vu)F^{\nu\rho}_b(X-vu)\right ]\,=\,
-u\frac{\del\,\,}{\del X^\rho}\left [
\tilde U_{ab}(X,X-vu)F^{\nu\rho}_b(X-vu)\right ]\nonumber\\
+u\left [\frac{\del\,\,}{\del X^\rho}
\tilde U_{ab}(X,Y)\right ]_{Y=X-vu}F^{\nu\rho}_b(X-vu), \eeq
and the second term in the r.h.s. is evaluated by using \cite{Elze86}
\beq\label{CdelU}
\tilde D_X^\rho \tilde U(X,Y)\Big |_{Y=X-vu}\,=\,
igu\int_0^1ds\,s \left[\tilde U (X,Z)\,v_\lambda  \tilde F^{\lambda\rho}
(Z)\,\tilde U (Z,X-vu)\right],\eeq
with $\tilde F^{\lambda\rho} \equiv T^a  F_a^{\lambda\rho}$ and
$Z^\mu(s)=X^\mu-u(1-s)v^\mu$.
Introducing (\ref{Cj2}) and (\ref{CdelU}) into (\ref{Cj1})
and using the identity
\beq
\int _0^\infty du\left [1-u(v\cdot\tilde D_X)\right ]_{ab}
\tilde U_{bc}(X,X-vu)F^{\nu\rho}_c(X-vu)\,=\,0,\eeq
we derive the following expression for the induced current:
\beq\label {Cj3} j^A_\mu(X)&=& j_\mu^{(1)}(X)+ j_\mu^{(2)}(X), \eeq
with
\beq\label{CjA1}
 \lefteqn{
 j_\mu^{(1)\,a}(X)\,=\,-
\frac{3}{2}\,\omega^2_p\int \frac{d\Omega}{4\pi}
\int_0^\infty du\,u\qquad\qquad}\nonumber\\
& &\left [g_{\mu\rho}(v\cdot \tilde D_X)-v_\mu\tilde D_\rho\right ]_{ab}
\tilde U_{bc}(X,X-vu)v_\nu F^{\nu\rho}_c(X-vu), \eeq
and\beq\label{CjA2}
\lefteqn{
j_\mu^{(2)\,a}(X)\,=\,-ig\,
\frac{3}{2}\,\omega^2_p\int \frac{d\Omega}{4\pi}\,v_\mu\,\int_0^\infty du\,u^2
\int_0^1ds\,s\qquad}\nonumber\\
& & \left[\tilde U (X, Z)\,v^\lambda  \tilde F_{\lambda\rho}
(Z)\,\tilde U (Z,X-vu)\right]_{ab}v_\nu F^{\nu\rho}_b(X-vu).\eeq
After using the identity
\beq\tilde U(X,Z)\,T^a\,\tilde U(Z,X)&=&\tilde U_{ab}(Z,X)T^b,\eeq
we make a change of integration variables in eq.~(\ref{CjA2}) and
 finally obtain
   \beq\label{CjA3}
 \lefteqn{
 j_\mu^{(2)\,a}(X)\,=\,ig\,
\frac{3}{4}\,\omega^2_p\int \frac{d\Omega}{4\pi}\,v_\mu\,
\int_0^\infty du\int _0^\infty dt\,(u-t)\qquad}\nonumber\\
& &\left\{v^\rho F_{\rho\lambda}^b(X-vt)\left [\tilde U
(X-vt,X)\,T^a\,\tilde U(X,X-vu)\right ]_{bc}
v_\nu F^{\nu\lambda}_c(X-vu)\right \}.
\eeq

{\bf iii)} It is now straightforward to verify that, for gauge fields
 $A_\mu^a$ in ${\cal R}$, the following
equalities are true
\beq \label{CdSb}
\delta S^{(1)}_{\rm b}/{\delta A^\mu_a(X)}=j^{(1)\,a}_\mu(X),\qquad
\delta S^{(2)}_{\rm b}/{\delta A^\mu_a(X)}=j^{(2)\,a}_\mu(X).\eeq
The first equality results from eqs.~(\ref{CdS1}) and (\ref{CjA1}), by
recalling that
\beq \delta F_{\mu\nu}(X)&=&\tilde D_\mu\,\delta A_\nu(X)-
\tilde D_\nu\,\delta A_\mu(X),\eeq
and by using the identity
\beq\label{Ccond3}
\int_0^\infty du\,u\,\tilde U_{ab}(X,X-vu) F^b_{\mu\nu}(X-vu)=
\int_0^\infty du\,u\,\tilde U_{ab}(X,X+vu) F^b_{\mu\nu}(X+vu),\eeq
which follows from the condition (\ref{cond3}).
As for the second equality (\ref{CdSb}), we note that
 (\ref{CdS3}) and (\ref{CjA3}) imply
\beq
\delta S^{(2)}_{\rm b}/{\delta A^\mu_a(X)}-j^{(2)\,a}_\mu(X)
\,=\,
-ig\,\frac{3}{4}\,\omega^2_p\int \frac{d\Omega}{4\pi}\,v_\mu
 \int_{-\infty}^\infty dt\int _0^\infty du\,(u+t)\qquad\nonumber\\
\left\{v^\rho F_{\rho\lambda}^b(X+vt)\left [\tilde U
(X+vt,X)\,T^a\,\tilde U(X,X-vu)\right ]_{bc}
v_\nu F^{\nu\lambda}_c(X-vu)\right \}.
\eeq
Again, this vanishes because of the conditions
(\ref{cond2}) and (\ref{cond3}). This completes our proof.

\setcounter{equation}{0}

\newpage


\centerline{\bf FIGURE CAPTIONS}
\vskip\baselineskip
{\parindent 0pt
\parskip\baselineskip

\noindent{\bf Fig.1:}\par\noindent
Diagrammatic illlustration of the equation of motion
(2.18) for the average gauge field $\langle
A_\mu(x)\rangle$. The thick line represents the gluon
propagator in the presence of the background field $\langle
A_\mu\rangle$, i.e. with all possible mean field
insertions included. The cross represents the external
current $j_\mu(x)$, while the circled cross represents
the induced current $j_\mu^{ind}(x)$ whose detailed
contributions in terms of 2-point and 3-point functions
are displayed in (b) (see eq.(2.22)). The contributions of the ghost mean
fields are not drawn. Quark,
gluon and ghost propagators are represented respectively by straight, wavy and
dashed lines.

\noindent{\bf Fig.2:}\par\noindent
Diagrammatic illustration of eqs. (2.19) for the fermionic
field $\langle\psi(x)\rangle$  (a), and its
conjugate $\langle\bar\psi(x)\rangle$ (b). The thick lines
represent quark propagators in the presence of the
background gauge field $\langle
A_\mu\rangle$. The crosses stand for the external
sources $\eta$ and $\bar\eta$, and the induced sources
are represented explicitly by the appropriate 2-point
functions (see eq. (2.23)).

\noindent{\bf Fig.3:}\par\noindent
 Lowest  order contribution to the off equilibrium fermion
propagator (a). The blob represents a gauge field
insertion and carry a four momentum $P\sim gT$. Typical
radiative corrections are shown in (b). The contributions
in (b) are negligible compared to (a) when $k\sim T$, but
are of the same order as (a) when $k\sim gT$.

\noindent{\bf Fig.4:}\par\noindent
Illustration of the approximate equations of motion for
$K_\nu^b(x,y)$, eq. (3.22) (a), and  eq. (3.23) (b). Thick
lines represent propagators in the presence of the
background gauge field $\langle
A_\mu\rangle$ (see Fig. 5a for the quark propagator and Fig.6a for the gluon
one).

\noindent{\bf Fig.5:}\par\noindent
Illustration of the approximate equation of motion (3.24) for $S(x,y)$.
(a): quark propagator with gauge field insertions;  (b): corrections
induced by the fermionic fields;
 the thick lines represent propagators in the presence of the background
gauge field $\langle A_\mu\rangle$.

\noindent{\bf Fig.6:}\par\noindent
Illustration of the approximate equation of motion
(3.27) for $D(x,y)$.
(a):gluon propagator with gauge field insertions;  (b): corrections
induced by the fermionic fields (only the contribution due to the $K$ function
is shown); the thick lines represent propagators in the presence of the
background gauge field $\langle A_\mu\rangle$.

\noindent{\bf Fig.7:}\par\noindent
The induced current $j^{ind}$ calculated from the 2-point
functions obtained in section 3.3. In the first two lines are
represented the various contributions to
$j^A=j_{\rm f}^A+j^A_{\rm g}+j^A_G+j^A_\Omega$. In the last two
 lines are represented, respectively, the
contributions $j_{\rm f}^\psi$ and $j_{\rm g}^\psi$
to $j^\psi=j_{\rm f}^\psi +j_{\rm g}^\psi$, as obtained by joining
the end points x and y in Figs.5
and 6. The ghost mean field  contributions are not drawn.

\noindent{\bf Fig.8:}\par\noindent
The induced fermionic source $\eta^{ind}$ calculated from
the 2-point functions obtained in section 3.3, that is,  by joining the end
 points x and y in the diagrams in Fig.4.

\end{document}